\documentclass[12pt]{article}
\usepackage{amsmath}
\usepackage{graphicx}
\usepackage{makecell} 
\usepackage{enumerate}
\usepackage{natbib}
\usepackage{tikz}          
\usepackage{ulem}
\usepackage{url} 
\usepackage{amsthm, bm}
\usepackage[a4paper]{geometry}
\usepackage{graphicx}
\usepackage{natbib}
\usepackage{booktabs} 
\usepackage{thmtools}
\declaretheorem[style=definition]{example}
\usepackage{bookmark}
\usepackage{caption}
\usepackage{subcaption}
\usepackage{siunitx}
\usepackage{float}
\usepackage{enumerate}
\usepackage{soul}
\usepackage{amsmath, amssymb} 
\usepackage{mathrsfs}
\usepackage{mhequ}
\usepackage{xcolor} 
\usepackage{bbold} 
\usepackage{mathtools} 

\newtheorem{theorem}{Theorem}
\newtheorem{lemma}{Lemma}

\newtheorem{proposition}{Proposition}  
\newtheorem{assumption}{Assumption} 
 
\newtheorem{definition}{Definition}
\usepackage{setspace} 
\usepackage{tcolorbox} 

\usepackage{mdframed}   

\newmdenv[
    backgroundcolor=yellow!20,
    linecolor=yellow!50!black,
    linewidth=1pt,
    roundcorner=5pt,
    skipabove=10pt,
    skipbelow=10pt,
    nobreak=true,
    frametitlebackgroundcolor=yellow!40,
]{revise}
\newcommand{\blind}{1}

 \soulregister\cite7
 \soulregister\citep7
\soulregister\ref7
\soulregister\pageref7
\soulregister\eqref7
\begin{document}

\renewcommand\thmcontinues[1]{Continued}
\def\spacingset#1{\renewcommand{\baselinestretch}%
{#1}\small\normalsize} \spacingset{1}


\if1\blind
{
  \title{\bf Identifiable and interpretable nonparametric factor analysis}
  \author{Maoran Xu\hspace{.2cm}\\
    Department of Statistics, Indiana University Bloomington\\ 
    Steven Winter\hspace{.2cm} \\ 
    Department of Statistical Science, Duke University\\
    Amy H. Herring \\
    Department of Statistical Science, Global Health,\\ and Biostatistics \& Bioinformatics, Duke University \\
    and \\
    David B. Dunson \\
    Departments of Statistical Science and Mathematics, Duke University\\}
    
      \maketitle
} \fi

\if0\blind
{
  \bigskip
  \bigskip
  \bigskip
  \begin{center}
    {\LARGE\bf Identifiable and interpretable nonparametric factor analysis}
\end{center}
  \medskip
} \fi

\bigskip
\begin{abstract}
Factor models are widely used to reduce dimensionality in modeling high-dimensional data. However, there remains a need for models that can be reliably fit in modest sample sizes and are identifiable, interpretable, and flexible. To address this gap, we propose a NIFTY model that uses a linear factor structure with Gaussian residuals, but with a novel latent variable modeling structure. In particular, we model each latent variable as a one-dimensional nonlinear mapping of a uniform latent location. A key innovation is allowing different latent variables to be transformations of the same latent locations, accommodating intrinsic lower-dimensional nonlinear structures. Leveraging on pre-trained data obtained by diffusion maps and post-processing of MCMC samples, we obtain model identifiability. In addition, we softly constrain the empirical distribution of the latent locations to be close to uniform to address a latent posterior shift problem, which is common in factor models and can lead to substantial bias in parameter inferences, predictions, and generative modeling. We show good performance in density estimation and data visualization in simulations, and apply NIFTY to bird song data in an environmental monitoring application.
\end{abstract}

\vfill
\newpage
\section{Introduction} 
Factor analysis \citep{fruchter1954introduction,harman1976modern} is a fundamental tool in various disciplines, including psychology, sociology, and medical research, where datasets are often high-dimensional and features exhibit strong dependence. Factor models uncover a set of low-dimensional latent variables that summarize the observed data, which is useful for exploratory analysis and understanding dependence structures. Given observed data \(\bm x_1,\ldots,\bm x_N \in \mathbb{R}^P\), a general factor model expresses each observation through a latent-to-observed mapping:
\begin{equation}\label{eq: general factor model}
\bm x_i = \bm g(\bm\eta_i) + \bm\varepsilon_i, \quad i=1,\ldots,N,
\end{equation}
where \(\bm\eta_i \in \mathbb{R}^H\) denotes low-dimensional latent factors (with \(H \le P\)), \(\bm g: \mathbb{R}^H \to \mathbb{R}^P\) is a possibly nonlinear mapping, and \(\bm\varepsilon_i\) is a residual term.

When the latent factors \(\bm\eta_i\) are drawn from a specified distribution \(F\), one can define the likelihood for the model in~\eqref{eq: general factor model} hierarchically. This setup enables both density estimation and generative modeling, where the latent distribution \(F\) is part of the likelihood specification. For example, \citet{kundu2014latent} constructed flexible univariate density models by applying random nonlinear transformations to uniform latent variables. Modern approaches such as variational autoencoders (VAEs; \citealp{kingma2014stochastic}) and diffusion models \citep{ho2020denoising} adopt a similar framework: they assume a simple latent distribution like \(\bm\eta_i \sim N(0, \bm{I_H})\), and learn a flexible decoder \(\bm{g}\) to map from latent to observed space. The synthetic data are then generated by drawing \(\bm\eta_i^* \sim F\) and evaluating \(\bm{x}_i^* = \bm{g}(\bm{\eta}_i^*)+\bm\varepsilon_i\).

A classical special case is the \emph{Gaussian linear factor model}:
\begin{equation}\label{eq: ppca}
\bm{x}_i = \bm\Lambda \bm \eta_i + \bm\varepsilon_i, \quad \bm\eta_i \sim N_H(\bm{0}, \bm{I}_H), \quad \bm\varepsilon_i \sim N_P(\bm{0}, \bm\Sigma), \quad i = 1,\ldots,N,
\end{equation}
where the data are mean-centered so that \(\mathbb{E}(\bm{x}_i) = 0\). Here, \(\bm\Lambda\) is a \(P \times H\) loading matrix, and \(\bm\Sigma\) is a diagonal residual covariance matrix. The marginal likelihood of the data obtained by integrating out the latent factors has the form: \(\bm{x}_i \sim N_P(0, \bm\Lambda\bm\Lambda^\top + \bm\Sigma)\).
Model \eqref{eq: ppca} is widely used due to its simplicity, computational tractability, and a tendency to produce a good fit in-sample to a variety of datasets.

\begin{definition}
Let $\hat{F}_n$ denote the estimated distribution of the latent variables in-sample, corresponding to the posterior expectation of the empirical distribution of $\{\bm\eta_i, i=1,\ldots,N\}$. \label{def1}  
\end{definition}

\begin{figure}[ht]
    \centering
    \begin{tikzpicture}

        \node (fig11) at (0, 0) {
            \begin{subfigure}[t]{0.25\textwidth}                 \includegraphics[width=.6\textwidth]{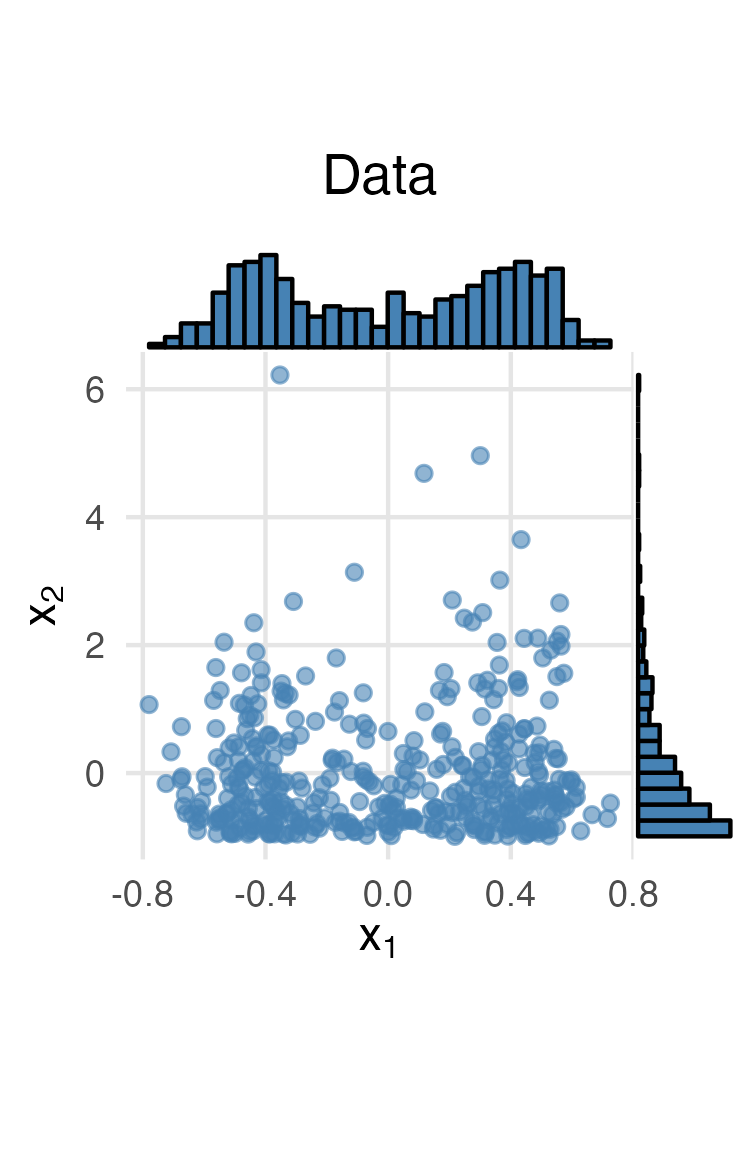} 
                \label{fig:11}
            \end{subfigure}
        };

        \node (fig12) at (4, 0) {
            \begin{subfigure}[t]{0.25\textwidth}
                \includegraphics[width=.6\textwidth]{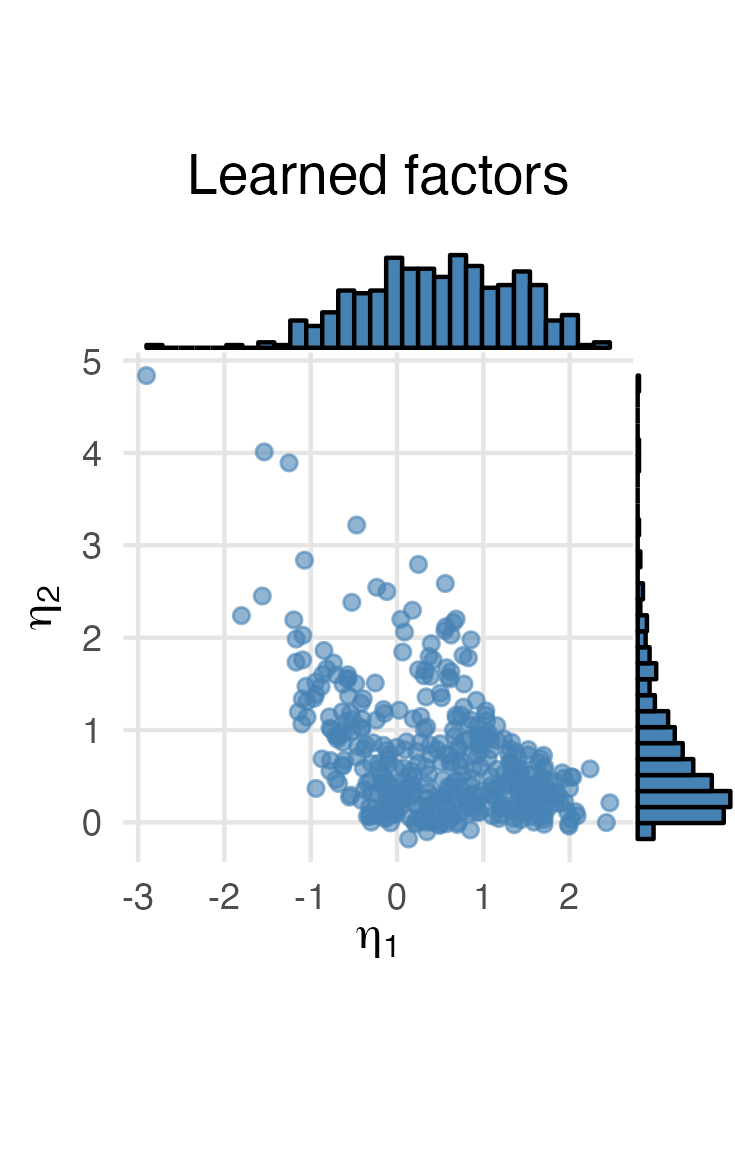} 
            \end{subfigure}
        };

        \node (fig13) at (8, 0) {
            \begin{subfigure}[t]{0.25\textwidth} \includegraphics[width=.6\textwidth]{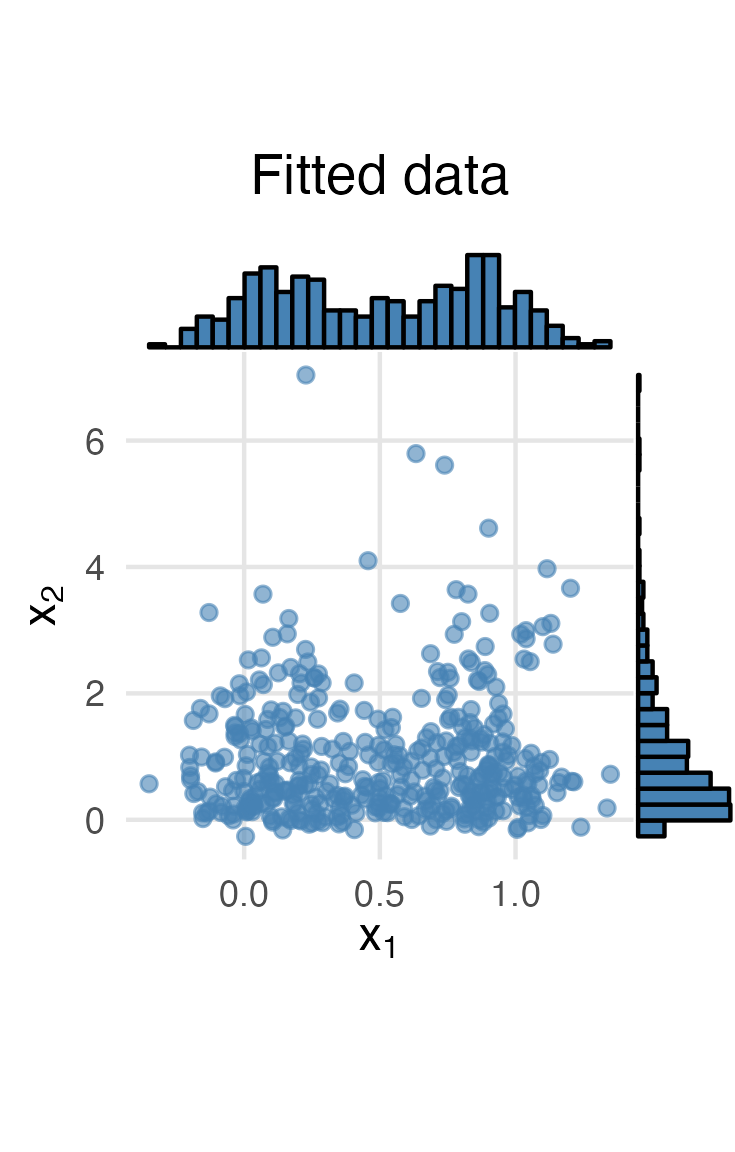} 
            \end{subfigure}
        };

     \node[anchor=west, align=left, font=\scriptsize] (row2) at (-2, -4) {
    Generate new data \\ 
    by drawing factors \\ 
    from the assumed dist. \\
    $\eta_{ih}\sim N(0,1)$ and \\
    apply the trained model
};

        \node (fig22) at (4, -4) {
            \begin{subfigure}[t]{0.25\textwidth}
                \includegraphics[width=.6\textwidth]{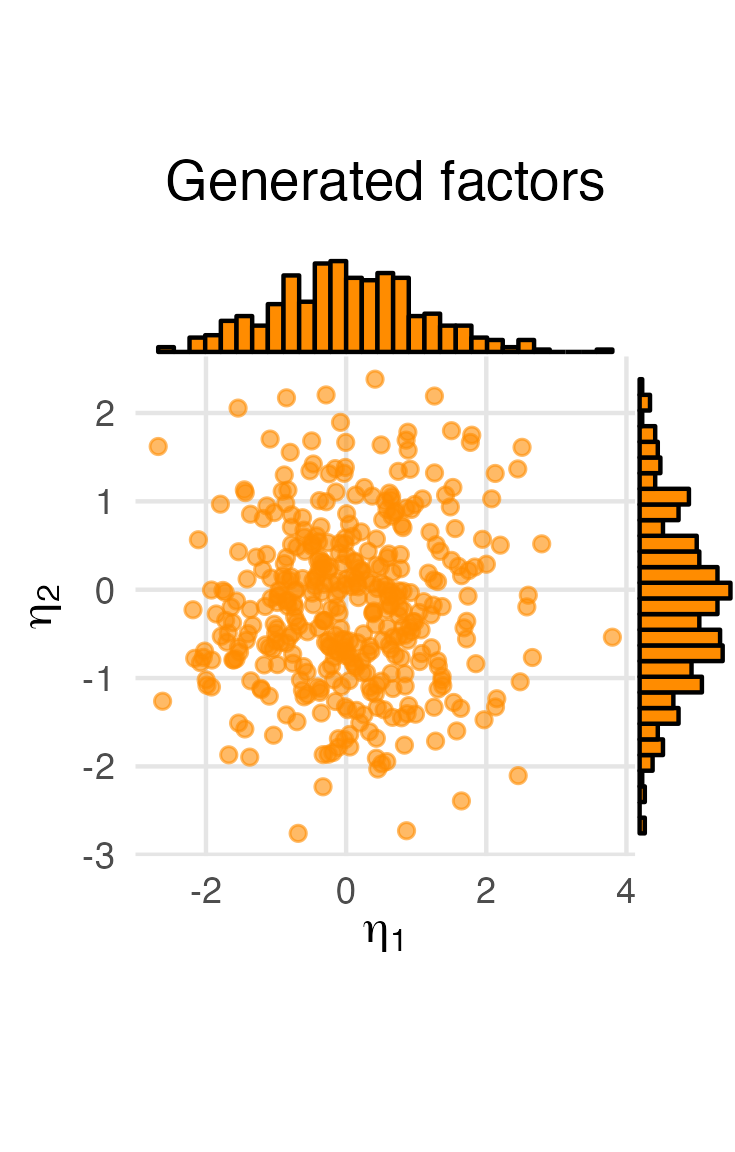} 
            \end{subfigure}
        };

        \node (fig23) at (8, -4) {
            \begin{subfigure}[t]{0.25\textwidth}
                \includegraphics[width=.6\textwidth]{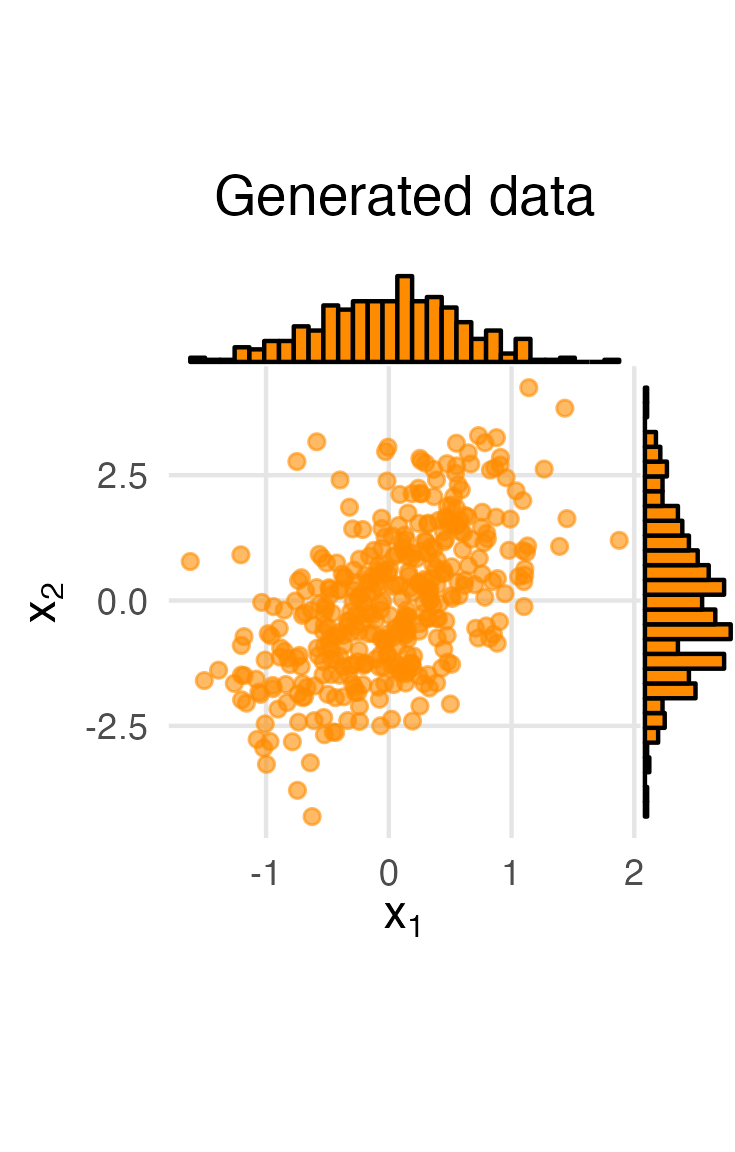} 
            \end{subfigure}
        };

        \draw[->, thick] (.75, -.35) -- (1.75, -.35) node[midway, above, align=left,font=\scriptsize] {Training};  
         \draw[->, thick] (4.5, -.35) -- (5.5, -.35) node[midway, above,font=\scriptsize] {$\bm{\hat\Lambda}\bm{\hat \eta} $}; 
     \draw[->, thick] (4.5, -4.35) -- (5.5, -4.35)node[midway, above,font=\scriptsize] {$\hat{\bm\Lambda}\bm\eta^*$}; 
    \end{tikzpicture} 
    \vspace{-.4cm}
    \caption{\label{fig: ppca}An example of latent posterior shift problem: Gaussian linear factor models for data violating normality. The top row shows the raw data, inferred factors having distribution $F_n$ deviating substantially from $F$, and the fitted data. The bottom row illustrates that even with good in-sample fit the distribution of the generated data is highly biased. The setting description is in example \ref{eg1}.} 
\end{figure}

However, for non-Gaussian data, the good in-sample fit of \(\hat{\bm x}_i = \hat{\bm\Lambda} \hat{\bm\eta}_i\) can disguise problems caused by the
distribution of the inferred factors $\hat{F}_n$, formally defined in Definition \ref{def1}, deviating substantially from the assumed Gaussian latent distribution $F \equiv N_H(\bm{0}, \bm{I}_H)$. This problem, which we refer to as {\em latent posterior shift}, leads to bias in the estimated parameters, so
\(\hat{\bm\Lambda}\hat{\bm\Lambda}^\top + \hat{\bm\Sigma}\)
does not approximate the covariance of the data. In addition, as illustrated in Figure~\ref{fig: ppca}, if we generate new data from the fitted distribution by sampling $\bm\eta_i \sim N_H(\bm{0}, \bm{I}_H)$ and letting $\bm x_i = \hat{\bm\Lambda}\bm\eta_i + \bm\varepsilon_i,$ with $\bm\varepsilon_i \sim N_P(\bm{0}, \hat{\bm\Sigma})$, then the distribution of these samples can be far from the true data distribution. 

There has been abundant recent interest in probabilistic generative models, which provide flexible black-box approaches for generating new data from approximately the same distribution as the training data.
These include normalizing flows \citep{kobyzev2020normalizing}, generative adversarial networks (GANs; \citealp{creswell2018generative}), variational autoencoders (VAEs; \citealp{kingma2014stochastic, rezende2014stochastic}), and Gaussian process latent variable models (GP-LVMs; \citealp{li2016review, titsias2010bayesian}). 
The most popular methods are based on deep neural networks and face three challenges: (i) lack of identifiability and reproducibility, often leading to posterior collapse or mismatch between training and generation \citep{wang2021posterior}; (ii) black-box mappings that limit interpretability of latent dimensions; and (iii) poor performance with small sample sizes due to the data-hungry nature of neural networks. Recent extensions such as VampPrior \citep{tomczak2018vae}, sparse nonlinear ICA \citep{zheng2022identifiability}, and sparseVAE \citep{moran2022identifiable} attempt to improve identifiability, but the remaining issues are intrinsic to deep architectures and remain difficult to resolve.

In contrast to these complex models, a rich literature has developed more structured extensions of linear factor models that aim to improve flexibility while retaining interpretability. One class of such extensions introduces nonlinearities via parametric transformations of the latent factors, such as polynomials or exponentials \citep{yalcin2001nonlinear, arminger1998bayesian}. Another important line includes copula factor models (Murray et al., 2013), which allow
for non-Gaussian marginals while preserving linear dependence. Mixtures of factor analyzers
(MFA; \cite{ghahramani1996algorithm, mclachlan2003modelling}) offer another approach by modeling
heterogeneous data as arising from a mixture of Gaussian factor models. Alternatively, \cite{chandra2023escaping} models factors as mixtures of low-dimensional Gaussian variables, providing dimension-reduced clustering. However, these methods can be brittle in the presence of smooth nonlinear structures without distinct clusters.

To bridge the gap between inflexible Gaussian models and overparametrized deep generative models, we propose NIFTY: a nonparametric linear factor analysis framework that (i) flexibly models the latent factor distributions, (ii) eliminates latent distribution shift while supporting efficient generative modeling, (iii) retains the interpretability of linear factor models, and (iv) admits provable identifiability and consistency results. NIFTY retains the linear form in \eqref{eq: ppca}, but instead of Gaussian latent factors, we allow each \(\eta_{ih} \sim F_h\) to follow a flexible univariate distribution. Specifically, we express each factor as \(\eta_{ih} = g_h(u_{ih})\), where \(u_{ih} \sim \text{Uniform}(0,1)\) and \(g_h\) is a non-decreasing spline function, interpretable as the inverse CDF of \(F_h\). This formulation enables a nonparametric specification of the latent distribution. A key innovation in NIFTY is to allow some \(u_{ih}\) to be shared across dimensions, inducing dependence between latent factors. We establish identifiability under mild conditions and develop a hybrid Hamiltonian-Gibbs sampler for efficient posterior computation. Empirical results on simulated and real acoustic monitoring data demonstrate that NIFTY recovers meaningful latent structure and provides accurate uncertainty quantification.

 The remainder of the paper is structured as follows. Section 2 introduces the NIFTY model, including its formulation, inference procedure, and approach to address latent distribution shift. Section 3 describes the posterior sampling algorithm. Section 4 establishes identifiability and consistency results. Section 5 presents simulation studies comparing NIFTY to alternative methods. Section 6 illustrates the method’s performance on a bird song dataset. Section 7 concludes with a discussion of implications and directions for future work.
 
\section{The NIFTY Framework} 
\subsection{Linear factor model with nonparametric factors} 

Suppose $\bm x_1,\ldots,\bm x_N\in \mathbb R^P$ are drawn from an unknown $P$-dimensional density $f\in\mathcal F$, with $\mathcal F$ the set of densities with respect to Lebesgue measure on $\mathbb R^P$. Consider the generative model:     
\begin{equation}\label{eq: nifty+model}
    \begin{aligned}\bm x_i &={\bm\Lambda}\bm \eta_i+\bm\varepsilon_i, \quad\bm\varepsilon_i\sim N_P(\bm 0,{\bm\Sigma}), \quad i=1,\ldots,N,\\
      \eta_{ih}&= g_{h}(u_{ik_h}),\quad h=1\ldots, H,\\
   u_{ik}&\stackrel{\text{iid}}{\sim} \text{U}(0,1),\quad k =1,\ldots,K, \quad K\le H.    \end{aligned}
\end{equation} 

Each latent factor $\eta_{ih}$ is a transformation of a \emph{latent location} $u_{ik_h}$ through a \emph{latent mapping} $g_h$ which is a non-decreasing $[0,1]\to\mathbb R$ function. The subscript $k_h$ creates a surjective mapping from $\{1,\ldots,H\}$ to $\{1,\ldots,K\}$. For each $k$, the latent factors mapped from $u_{ik}$ are indicated by the subscript $k_h$, and we let $\bm\eta_{i}^k:=\{ \eta_{ih}: k_h = k\}.$ We further let $H_k$ denote the number of latent mappings from $u_{ik}$ and $H=\sum_{k=1}^KH_k$. Applying $g_h:=F_h^{-1}$ to a uniform variable, one gets $\eta_{ih}$ with marginal CDF $F_h$. 
We assume diagonal residual covariance ${\bm\Sigma}=\text{diag}(\sigma_1^2,\ldots,\sigma_P^2)$. For simplicity of computation, we use monotone piecewise linear functions to model the 1-dimensional mappings: Let $0=s_0,s_1,\ldots,s_{L-1},s_L=1$ denote a segmentation of the [0,1] interval and $\alpha_{lh}$ denote the slope for the $l$-th piece. Then we represent $g_h$ as 
\[g_h(u) = \alpha_{0h} + \alpha_{1h}(u-s_0)1\{u\in [s_0,s_1)\}+\cdots+\alpha_{Lh}(u-s_{L-1})1\{u\in [s_{L-1},s_L]\}.\] By simplifying $\bm g$ to consist of monotone piecewise linear functions, we reduce inference of nonlinear mappings to a simpler problem of estimating slope parameters $\{\alpha_{lh}\}$. In the rest of the paper,  we slightly abuse $\bm g$ to denote the nonlinear mapping and the slope parameters.

The above model induces a multivariate density $\bm x_i \sim f$ while maintaining the interpretability of a linear latent factor model. Slightly abusing notation, let $\bm g$ denote the vector-valued function $(g_1,\ldots,g_H)$ and $\bm g(\bm u_i)$ denote the vector $[g_1(u_{ik_1}),\ldots,g_H(u_{ik_H})]$. The density $f$ is a deterministic function of ${\bm\Lambda}, \bm g,{\bm\Sigma}$, after marginalizing out the latent variable $\bm u_i$ as
\[
f_{{\bm\Lambda},\bm g,{\bm\Sigma}} :=f(\bm x_i;{\bm\Lambda}, \bm g, {\bm\Sigma})  = \int_{[0,1]^H} \phi_{\bm\Sigma}[\bm x_i-{\bm\Lambda}\bm g(\bm u_i)] d\bm u_i,
\] 
where $\phi_{\bm\Sigma}(\bm x)$ refers to the density of a multivariate Gaussian with mean zero and covariance matrix $\bm\Sigma$ evaluated at $\bm x$.
By assigning priors (to be specified in later sections)
	${\bm\Lambda}\sim \Pi_{{\bm\Lambda}}, g_h \sim \Pi_{{g_h}}, {\bm\Sigma}\sim \Pi_{{\bm\Sigma}},$
we induce a prior for $f$ with $f\sim \Pi_{\bm f}$.  
We refer to $f_{{\bm\Lambda},\bm g,{\bm\Sigma}}$ as the NIFTY model.

\subsection{Learning identifiable factors and loadings} 
A central question for latent variable models like \eqref{eq: nifty+model} is whether the latent locations and their associated parameters---\((\{u_{ik}\}, \{g_h\}, \bm{\Lambda})\)---are identifiable, i.e., uniquely determined by the likelihood. The answer is no: the model exhibits two types of ambiguity.
 First, it is well-known that linear factor models exhibit rotational ambiguity. Specifically, the parameterization \(\bm{\Lambda}\bm{g}(\bm u_i)\) is equivalent to \((\bm{\Lambda} \bm{R}^\top)(\bm{R}\bm{g}'(\bm u_i))\) for any rotation matrix \(\bm{R}\) satisfying \(\bm{R}^\top\bm{R} = \bm{I}\). Second, the unknown number of latent locations introduces additional non-identifiability. For example, each uniform variable \(u\) can be represented as \(\Phi\big(\sqrt{-2\log (v_1)}\cos(2\pi v_2)\big)\), where \(v_1\) and \(v_2\) are two independent \(\text{U}(0,1)\) variables, and \(\Phi\) is the CDF of a standard normal distribution. 

To address the ambiguities in \eqref{eq: nifty+model}, we adopt a two-step strategy: (i) pre-training the dataset to estimate the number of latent locations and (ii) post-processing the MCMC output to resolve rotational and label-switching ambiguities, as commonly done in linear factor analysis. Theorem \ref{thm: main_identify} shows that pre-training produces identifiable factors and loadings up to rotation and scaling, which are then addressed through post-processing. 
For tasks focused solely on generative modeling or density estimation, post-processing may be skipped without affecting the results. We now detail the pre-training step and defer post-processing to Section~\ref{sec: compute}.
   
Suppose that we have \(n\) observed data points drawn from the model \eqref{eq: nifty+model} generated by \(K\) independent latent locations. If there are \(K\) dimensions in the data set \(x_{ik_1},\ldots,x_{ik_K}\), which depend only on \(u_{i1},\ldots,u_{iK}\), respectively, then Theorem \ref{thm: main_identify} establishes identifiability. These \(K\) variables are called \emph{anchor data}. Although anchor feature assumptions are common in factor models \citep{bing2020adaptive, arora2013practical, moran2022identifiable}, in practice, the subset of anchor features is typically unknown. To address this, we propose inferring anchor features and determining the latent dimensionality via a pre-training step.

We use the \(p\)-dimensional observed data vectors to compute the lower-dimensional anchor features \(x_{i1}^*,\ldots,x_{iK}^*\). We recommend using diffusion maps (DM) \citep{coifman2006diffusion} as the default nonlinear dimension reduction method due to their robust theoretical and practical performance in capturing intrinsic geometric structures in data. Other dimensional reduction methods may also be applied effectively. For example, we demonstrate the use of t-SNE later in the article, motivated by its utility in data visualization. Appendix \ref{sec: review-DM} includes a brief review of DMs and an algorithm to infer the dimensionality \(K\).

The anchor features are given the following model: 
\begin{equation}\label{eq: augment_nifty}
    \begin{aligned}
        x_{ik}^* &= \eta^*_{ik}+\varepsilon^*_{ik}, \quad \varepsilon^*_{ik}\sim N(0,(\sigma_k^*)^2), \\
        \eta^*_{ik} &= g_k^*(u_{ik}),\quad k=1,\ldots, K. 
    \end{aligned}
\end{equation}
We then fit a combined model, incorporating models \eqref{eq: nifty+model} and \eqref{eq: augment_nifty}, which can be expressed as 
\[(\bm x_i, \bm x_i^*) = \begin{bmatrix}
\bm\Lambda & \bm 0 \\
\bm 0 & \bm I_k
\end{bmatrix}(\bm\eta_i, \bm \eta_i^*) + (\bm \varepsilon_i, \bm \varepsilon_i^*).\]
Essentially, the model \eqref{eq: augment_nifty} augments the observed data set with the anchor data of dimension \(K\) and enforces that the latent factors are exclusive to their respective anchor dimensions. This augmented dataset and model \eqref{eq: augment_nifty} ensure identifiable learning of latent locations.

When the true generating distribution of $\bm x_i$ follows \eqref{eq: nifty+model}, DMs consistently recover the true low-dimensional manifold homeomorphic to $u_{i1},\ldots u_{iK}$ \citep{el2016graph, shen2022scalability}. Therefore, it is reasonable to treat the inferred $\bm x_i^*$s as data close to the true manifold.
With this motivation, we fix \((\sigma_k^*)^2\) by fitting splines to each anchor feature and estimating \((\sigma_k^*)^2\) as the mean squared error of the MLE spline fits (details described in Appendix \ref{sec: review-DM}). The residual variances \((\sigma_1^2,\ldots,\sigma_P^2)\) are treated as unknown and updated within our MCMC algorithm.

Although we recommend DMs as the default choice for pre-training, alternative dimensionality reduction algorithms may be preferred in certain settings. For example, t-SNE and UMAP were developed to visualize high-dimensional data in 2-3 dimensions, though they can sometimes fail to accurately represent the global intrinsic geometry of the data. When data visualization is the focus, one may use these approaches instead of DMs in inferring the $\bm x_i^*$s, as we demonstrate in Section \ref{sec: data-vis}. 

\subsection{Inferring the number of factors and latent mappings}   

In applying DMs, or an alternative dimensionality reduction algorithm in pre-training, we also infer the number of latent dimensions. The number of latent locations $K$ in NIFTY is then fixed at this number, but it remains to infer the number of latent factors $H \ge K$. Recall that for each uniform latent location $u_{ik}$, $H_k$ denotes the number of latent mappings from $u_{ik}$. NIFTY allows uncertainty in $H_k$ within a Bayesian framework, but if we fix $H_k=1$, then Bayesian independent component analysis (ICA) can be recovered as a special case.

\begin{example}\label{eg1}(Bayesian ICA) 
When $\eta_{ik}=g_k(u_{ik})$ for $k=1,\ldots,K$ in NIFTY, the latent factors are independent and we obtain a Bayesian ICA model. ICA models \citep{comon1994independent} are widely used in signal processing to infer additive and independent latent subcomponents. Consider a 10-dimensional random vector $\bm x_i=\bm\Lambda \bm\eta_{i}+\bm\varepsilon_i$ with two independent factors $\eta_{i1} \sim \text{Beta}(0.4,0.4)$, $\eta_{i2}\sim \text{Gamma}(1,1)$, and $\varepsilon_{ij} \sim N(0,0.1^2)$.
For a better visual effect, we set $\Lambda_{11}=1,\Lambda_{12}=0,\Lambda_{21}=0,\Lambda_{22}=1$, so that the first two dimensions $x_1$ and $x_2$ are the two factors plus noise.  Figure \ref{fig: ppca} shows that the Gaussian linear factor model does not accommodate skewness and bimodality in marginal distributions. In contrast, NIFTY (depicted in Figure \ref{fig: independent example}) recovers the two non-Gaussian factors as transformations of two independent uniform latent variables. 
	\begin{figure}[H]
		\centering \includegraphics[width=.9\textwidth]{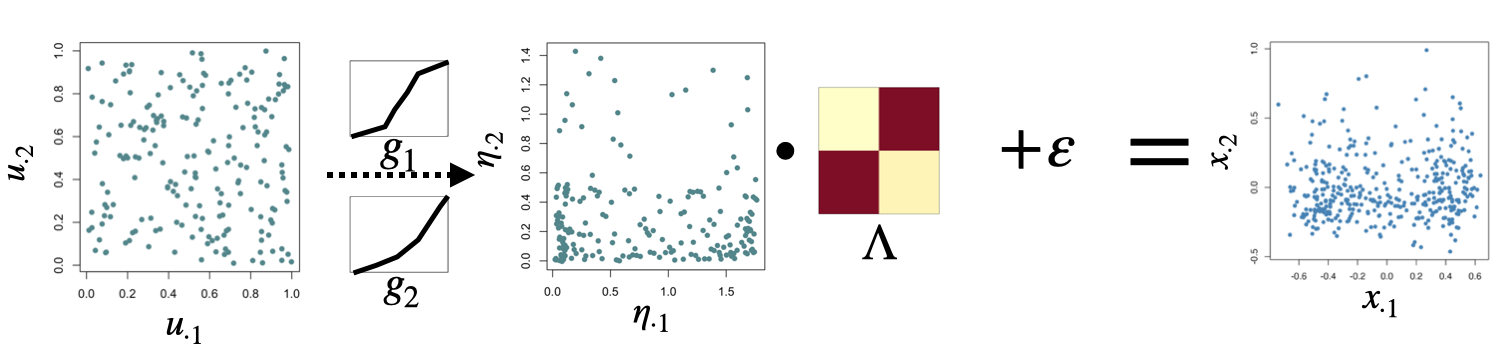}
		\caption{\label{fig: independent example}Illustration of NIFTY in the ICA special case. For data in example \ref{eg1}, NIFTY learns two mutually independent uniform variables $u_{\cdot 1}$ and $u_{\cdot 2}$, mapped to $\eta_{\cdot 1}$ and $\eta_{\cdot 2}$ via $g_1$ and $g_2$. } 
	\end{figure}
\end{example}

\begin{figure}[H]
        \centering  
        \begin{subfigure}[t]{0.45\textwidth}
                \centering \includegraphics[width=.7\textwidth]{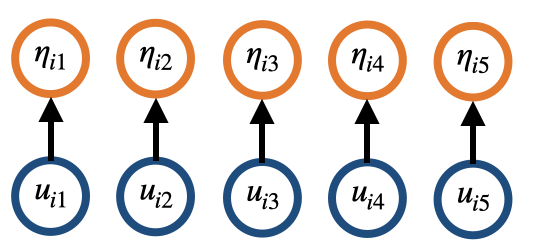}  
                \caption{ICA models.}
        \end{subfigure}  \quad 
        \begin{subfigure}[t]{0.45\textwidth}
                \centering \includegraphics[width=.92\textwidth]{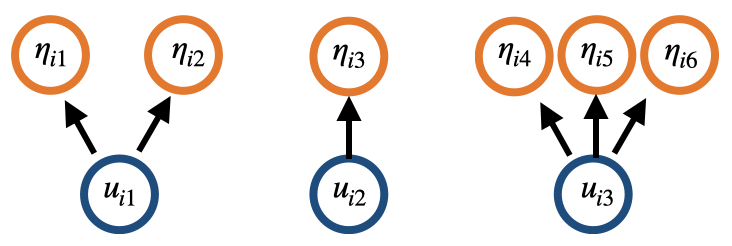}    
                \caption{Nonlinearly dependent factors.}
        \end{subfigure}  
        \caption{\label{fig: diagram} Mapping from uniform latent locations to latent factors in (a) special case of ICA model and (b) a general NIFTY model with nonlinearly dependent factors.}  
\end{figure}

ICA models are unable to characterize nonlinear dependence among the dimensions. For example, when $x_{i1}=p(x_{i2})$ with $p(\cdot)$ a polynomial function, ICA performs poorly in inferring the components. We solve this limitation by a novel model construction: by allowing $\eta_{ih}$'s mapped from the same uniform location $u_{ik_h}$, we introduce dependence among the factors. If $k_{h_1}=k_{h_2}$, then $\eta_{ih_1}$ is a nonlinear transformation from $\eta_{ih_2}$, otherwise $\eta_{ih_1}$ and $\eta_{ih_2}$ are independent. Thus, we broaden the class of densities characterized by an ICA model to accommodate wider nonlinear dependence. We illustrate in Figure \ref{fig: diagram} the difference between the general NIFTY model and the special case of Bayesian ICA.

\noindent\textbf{Inferring unknown $H_k$s.} 
In Bayesian factor modeling, it is popular to choose an initial upper bound on the number of factors and then shrink the loadings to effectively remove the factors that are not needed. We propose such an approach to infer the $H_k$s in NIFTY, letting $\bar H$ be the upper bound on $H_k$ for $k=1,\ldots,K$. We choose a prior distribution for $\bm\Lambda$, which allows effective deletion of columns and therefore additional factors, by adapting horseshoe priors \citep{carvalho2009handling, xu2016bayesian}: $\lambda_{jh}\sim N(0,\tau^2\gamma_h^2\sigma_j^2), \gamma_h\sim C^+(0,1), \tau\sim C^+(0,1)$ with
$C^+(0,1)$ denoting a half-Cauchy distribution. Example 2 illustrates the flexibility of this approach.
 
\begin{example}[label=exa: cont](Two curves.) We generate data $\bm x_i$ from $N_{10}[(2z_{i1},2z_{i1}^2,2z_{i2},2z_{i2}^2,0,\ldots,0),\sigma^2 \bm I_{10}]$ with $\sigma^2=0.01$ and $z_{ij}\stackrel{\text{iid}}\sim \mbox{Beta}(0.5,0.5)$ for $j=1,2$. The pairwise plot and marginal histograms are shown in Figure \ref{fig: 2-factor}(a). NIFTY learns the generative process from two independent uniform variables $u_{\cdot 1}$ and $u_{\cdot 2}$ mapped to $\eta_{\cdot 1},\ldots,\eta_{\cdot 4}$ via  $g_1,\ldots, g_4$, and linearly mapped to the data space. According to Figure \ref{fig: 2-factor}(e), the number of latent factors is indicated by the nonzero columns in the sparse loading matrix, which is 4 in this example. We postpone the details of this numerical experiment along with a comparison with other methods to Section \ref{sec: numerical}.
\begin{figure}
        \centering  
        \begin{subfigure}[t]{0.45\textwidth}
                \centering \includegraphics[width=.9\textwidth]{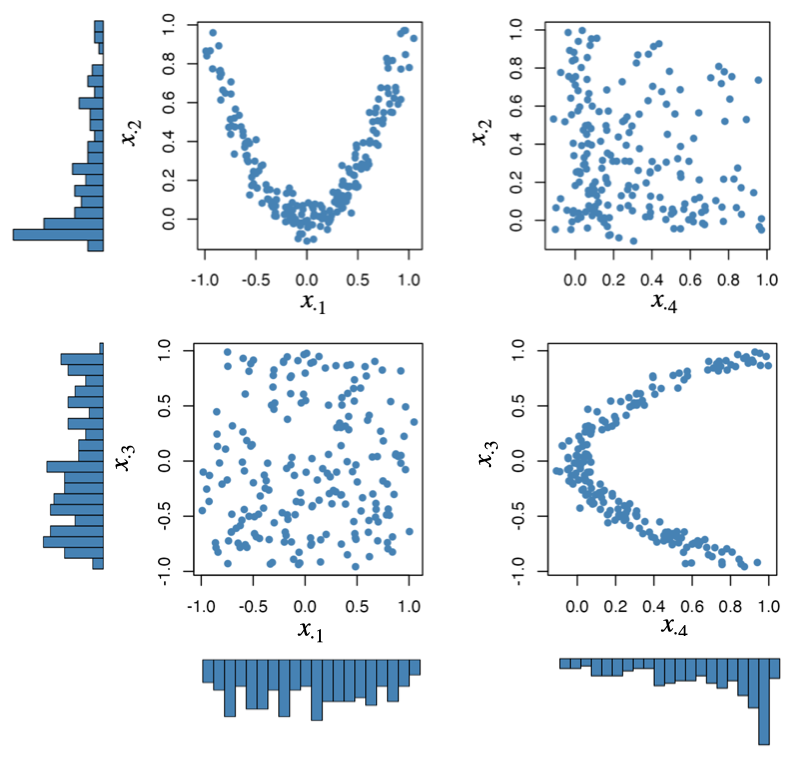}  
                \caption{The first four dimensions of $\bm x_i$ generated from two latent curves.}
        \end{subfigure}  
        \begin{subfigure}[t]{0.45\textwidth}
                \centering \includegraphics[width=.92\textwidth]{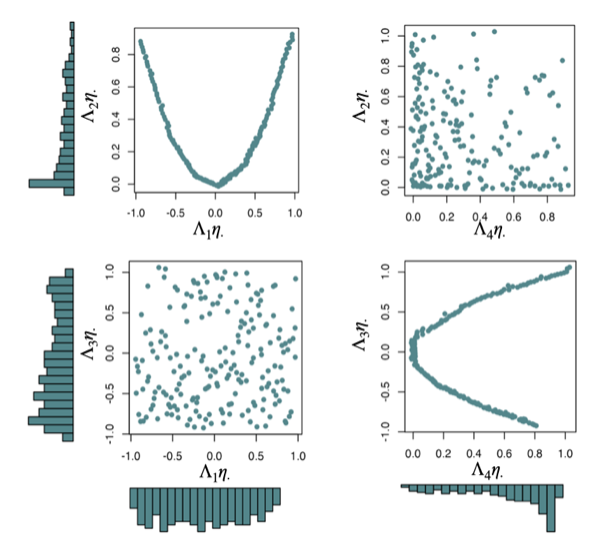}    
                \caption{The fitted components $\bm\Lambda\bm\eta_i$ generated from two latent curves.}
        \end{subfigure}  \\
        \begin{subfigure}[t]{0.2\textwidth}
                \centering \includegraphics[width=.8\textwidth]{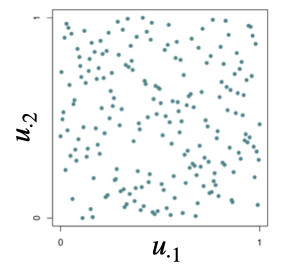}
                \caption{Uniform latent locations.}
        \end{subfigure} 
        \begin{subfigure}[t]{0.58\textwidth}
                \centering \includegraphics[width=1\textwidth]{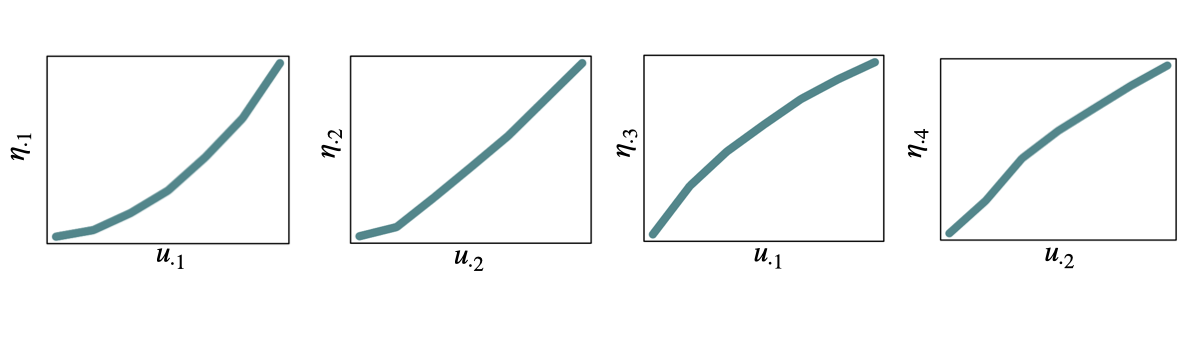}  
                \caption{Latent mappings $g_1,g_2,g_3$ and $g_4$.}
        \end{subfigure}   
        \begin{subfigure}[t]{0.2\textwidth}
                \centering \includegraphics[width=.8\textwidth, height = .8\textwidth]{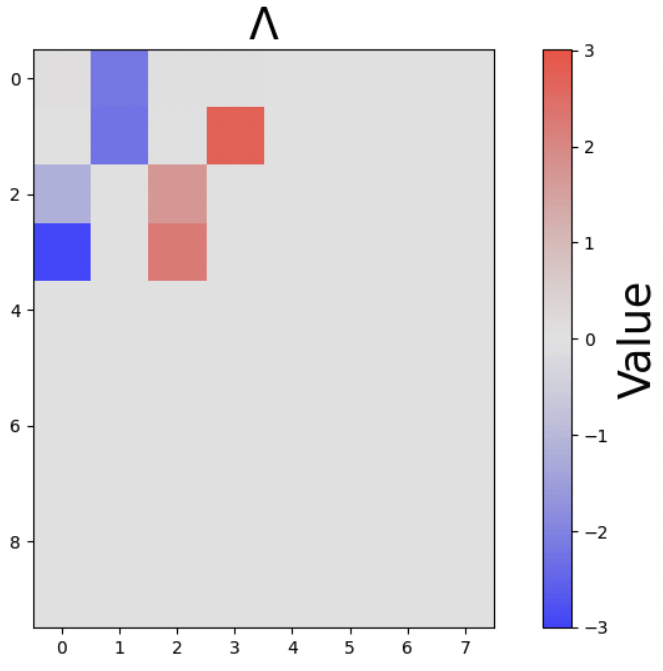} 
                \caption{Loading matrix.}
        \end{subfigure}  
        \caption{\label{fig: 2-factor} An illustration of the NIFTY+ framework on 10-dimensional data arising from two latent curves. Starting from the independent latent locations (c), the framework learns four mappings (d) and transforms the uniform latent variables to latent factors $\eta_1,\ldots,\eta_4$, along with a loading matrix (e). The fitted components accurately infer the two independent latent curves. 
}  
\end{figure}   
 \end{example}  

\subsection{Addressing latent posterior shift} 

In latent factor models, the likelihood of the data $\bm x_i$ is obtained by marginalizing the conditional likelihood of $\bm x_i$ given $\bm \eta_i$ over the distribution of the latent factors $F$. Hence, $F$ is a component of the likelihood. As motivated in Section 1, when the model provides a poor fit to the observed data, there is a tendency for $\hat{F}_n$, the inferred distribution of $\{ \bm\eta_i, i=1,\ldots,n\}$ (refer to Definition \ref{def1}), to differ substantially from $F$. This {\em latent posterior shift} problem leads to bias in parameter estimation, prediction, and generative modeling. Example~1 provides an illustration for Gaussian linear factor models, with further results on biased covariance estimation provided in 
Section~1 of the supplementary materials. 

This issue has been acknowledged in the VAE literature \citep{hoffman2016elbo}, and solutions include: (i) 
Using learnable pseudo-data to align $\hat{F_n}$ and $F$ \citep{tomczak2018vae};
    (ii) penalizing correlations among factors to achieve disentanglement \citep{kim2018disentangling}; and  
    (iii) enforcing the posterior to be Gaussian, for example, using the diffusion process to sequentially transform data into i.i.d.\@ Gaussian latent variables \citep{ho2020denoising}.    

We propose an alternative solution for NIFTY inspired by the constraint relaxation approach of \cite{duan2020bayesian}. 
We propose to enforce $\hat{F}_n \approx F$ by constraining the empirical distribution of $\bm u_{\cdot k}=\{ u_{ik}, i=1,\ldots,n\}$ to be very close to uniform by letting 
\begin{equation}\label{eq: prior_u}
	\Pi_{\bm u_{\cdot k}}( u_{1k},\ldots, u_{Nk}) = \prod_{i=1}^N 1(u_{ik}\in[0,1]) \exp(-\nu\mathcal W_2(U_{k}, U)),\quad k=1,\ldots,K,
\end{equation}
where $\mathcal W_2(U_k,U)$ denotes the Wasserstein-2 distance between the empirical distribution of $\bm u_{\cdot k}$ and a $U(0,1)$ variable. The exponential term
introduces a continuous relaxation of the uniform distributional constraint, with $\nu>0$ controlling closeness to uniform.  
The Wasserstein distance measures the optimal transport distance \citep{kolouri2017optimal} between two distributions. The Wasserstein-2 distance between univariate random variables with CDFs $F$ and $G$ is 
$\mathcal W_2(F,G) = \left(\int_0^1\left|F^{-1}(z)-G ^{-1}(z)\right|^p dz\right)^{1 / p}$. 
As an approximation of $\mathcal W_2(U_k,U)$, let 
\[\hat{\mathcal W}_2(U_k, U)=\left(\sum_{i=1}^N\left\|u_{(i)k}-u^0_{(i)}\right\|^2\right)^{1 / 2},\]
where $u_{(i)k}$ is the $i$th order statistic of $\{u_{1k},\ldots,u_{Nk}\}$
and $u^0_{(i)}=i/N$. 
\begin{proposition}\label{proposition: CoRe}
	Suppose $u_{1 k},\ldots,u_{N k}$ are generated from \eqref{eq: prior_u}. Then the empirical distribution of $\bm u_{\cdot k}$ converges to $U(0,1)$, as $\nu\to\infty$ and $N\to\infty$. 
\end{proposition}
 
Refer to the appendix for proof of the proposition and Section 3 of the Supplementary Materials for a simulation experiment shedding light on how to choose $\nu$ in practice. We obtained the best empirical results when $\nu \approx 10^2 - 10^3$. Large values strongly constrain the empirical distribution to be close to uniform, but huge $\nu$ can lead to slow mixing. The following example provides an illustration.
 
\begin{example}[continues=eg1]
Continuing the example of non-Gaussian marginals, we compare NIFTY with and without addressing latent posterior shift.  From panel (a) in Figure \ref{fig: CoRe}, without the constraint, the distribution of the latent $(u_{i1},u_{i2})$ deviates from the 2-dimensional independent uniform distribution, and shifts towards the data distribution. As a result, $g_1$ and $g_2$ are almost identity maps. Adding the distributional constraint enables us to reliably fit and interpret the nonparametric factors.

\begin{figure}[H]
        \centering  
        \begin{subfigure}[t]{0.45\textwidth}
                \centering \includegraphics[width=.6\textwidth]{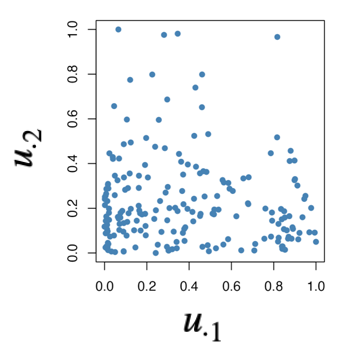}  
                \caption{One posterior sample of the latent locations without the constraint.}
        \end{subfigure}  \quad 
        \begin{subfigure}[t]{0.45\textwidth}
                \centering \includegraphics[width=.6\textwidth]{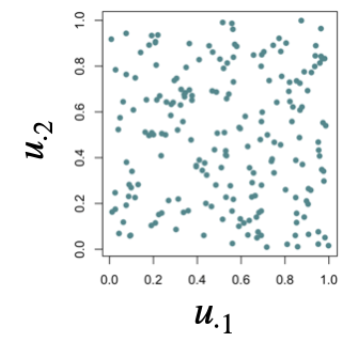}    
                \caption{One posterior sample of the latent locations with the constraint ($\nu=1000$).}
        \end{subfigure}  
        \caption{\label{fig: CoRe}Comparing NIFTY results with and without the (soft) distributional constraint in Example \ref{eg1}. Without the constraint, latent locations deviate from the 2-dimensional independent uniform distribution and shift toward the data distribution.}  
\end{figure}   
\end{example}

\section{Posterior computation\label{sec: compute}} 
\subsection{MALA-within-Gibbs posterior sampling} 
In this section, we describe an algorithm to obtain posterior samples. We sample the parameters $({\bm\Lambda},  \alpha_{lh}, u_{ik}, \sigma^2_{j})$ via block updates: sampling each block of parameters from their joint distribution conditioned on all other blocks. Conjugate updates are available for ${\bm\Lambda},  \alpha_{lh}$ and $\sigma^2_{j}$, and we use a Metropolis-adjusted Langevin algorithm (MALA) \citep{roberts1996exponential,roberts2002langevin} to update the $u_{ik}$'s. The sampling steps are as follows. 



\noindent {\bf Step 1}: The rows of ${\bm\Lambda}$ have conditionally independent conjugate Gaussian distributions. Update the $j$-th row $(j=1,\ldots,P)$ of ${\bm\Lambda}$ using
${\bm\Lambda}_j \sim N_H(\bm V_j \sum_{i=1}^n \bm\eta_i x_{ij}\sigma_j^{-2}, \bm V_j),$
where $\bm V_j=\text{diag}[(\tau\gamma_{j1})^{-1},\ldots,(\tau\gamma_{jH})^{-1}]+\sigma_j^{-2}\bm\eta\bm\eta^T.$

\noindent {\bf Step 2}: Sample $\sigma_1^2,\ldots,\sigma_P^2$ from conditionally independent inverse-Gamma distributions,
$$\sigma_j^{-2} \sim \text{Gamma}\bigg(a_\sigma+N/2, b_\sigma+\frac{1}{2}\sum_{i=1}^n (x_{ij}-{\bm\Lambda}_j^T \bm\eta_i)^2\bigg).$$

\noindent {\bf Step 3}: Let $\bm \alpha$ denote the vector of all the $\alpha_{lh}$s and $u_{ikl}=\min[\max(u_{ik}-s_{l-1},0), s_{l}-s_{l-1}]$. Sample from the conditional posterior $\bm \alpha \sim N_{LH}(\bm m^\alpha, {\bm\Sigma}^{\bm\alpha}),$ where   
$\bm m^\alpha = {\bm\Sigma}^{\bm\alpha} \bm \mu^{\bm\alpha},$
$\mu^{\bm\alpha}_{lh} = \sum_{i=1}^N\sum_{j=1}^P (x_{ij}-\bm\Lambda_j\bm\eta_i + \lambda_{jh}\alpha_{lh}u_{ik_hl})/\sigma^2_j,$ and
$\bm\Sigma^{\bm\alpha}_{lh,l'h'} = \text{Cov}(\alpha_{lh},\alpha_{l'h'}) = \sum_{i=1}^N\sum_{j=1}^P \lambda_{jh} \lambda_{jh'} u_{ik_hl} u_{i'k_h'l'}/\sigma_j^2.$

\noindent {\bf Step 4}: Update $u_{ik}$'s using MALA. Let $\pi(u\mid -)$ denote the conditional posterior density. At step $t$, propose $\tilde u^{t+1}$ using the Langevin diffusion
$\tilde {\bm u}^{t+1} = \bm u^t + \epsilon\nabla \log\pi(\bm u^t \mid -)+\sqrt{2\epsilon} \bm v,$
where $\epsilon>0$ is a small step size, and each entry in $\bm v$ is generated independently from a standard normal distribution. This proposal is accepted with probability
$$p_{\text{accept}} =\min \left\{1,\frac{\pi(\tilde {\bm u}^{t+1})q(\bm u^t\mid\tilde {\bm u}^{t+1})} {\pi(\bm u^{t})q(\tilde {\bm u}^{t+1}\mid \bm u^t)}\right\},$$
where $q(\bm u'\mid \bm u) \propto \exp\left(-\frac{1}{4\varepsilon}\|\bm u'-\bm u-\epsilon\nabla\log\pi(\bm u)\|_2^2\right)$ and 
\[\log \pi(\bm u^t\mid -) = - \sum_{i=1}^N\sum_{j=1}^P\sigma_j^{-2}\bigg(x_{ij} - \sum_{h=1}^H\lambda_{jh}\sum_{l=1}^L\alpha_{lh}u_{ik_hl}\bigg)^2-\nu \sum_{i=1}^N\|u_{(i)k}-u^0_{(i)}\|_2^2.\]  
 
\noindent {\bf Step 5}: Update the hyperparameters $(\tau^2, \gamma_h^2)$ according to \cite{makalic2015simple}.

\subsection{Posterior alignment for scale, rotation and label ambiguity}
From MCMC one obtains $M$ posterior samples  \[(\bm u^{(1)}, \bm \Lambda^{(1)},\bm g^{(1)},\bm\Sigma^{(1)}),\ldots,(\bm u^{(M)}, \bm \Lambda^{(M)},\bm g^{(M)},\bm\Sigma^{(M)}).\] As discussed above, there could be scale and rotational ambiguity between $\bm \Lambda^{(m)}$ and $\bm g^{(m)}$. We describe a post-processing algorithm to align the samples.
We first divide $(\bm \Lambda^{(m)}, \bm g^{(m)})$ into $K$ blocks $(\bm \Lambda^{(m)1}, \bm g^{(m)1}),\ldots, (\bm \Lambda^{(m)K}, \bm g^{(m)K})$,  such that $g^{(m)k} = \{g^{(m)}_h\}_{h:k_h=k}$ and $\bm \Lambda^{(m)k} = (\Lambda^{(m)}_{\cdot,h})_{h:k_h=k}$. We then apply the \texttt{MatchAlign} algorithm proposed by \cite{poworoznek2021efficiently}, which first orthogonalizes the samples using Varimax and then fixes label and sign switching with a matching algorithm. The post-processing procedure is summarized in Algorithm \ref{algo: matchalign}, and we show in the theory section that this step leads to strict parameter identifiability.
\begin{center} \label{algo: matchalign}
\begin{tabular}{l}
\toprule
\textbf{Algorithm 1: post-processing the posterior samples to solve ambiguity}\\
\hline 
\def\arraystretch{1.9}%
\textbf{Input:} $(\bm \Lambda^{(1)},\bm g^{(1)},\bm\Sigma^{(1)}),\ldots,(\bm \Lambda^{(M)},\bm g^{(M)},\bm\Sigma^{(M)})$\\  
\textbf{for} $m=1,\ldots,M$ \textbf{do}\\
\quad\quad\quad\quad 
 \textbf{for}  $k=1,\ldots,K$ \textbf{do}\\
\quad\quad\quad\quad\quad\quad Orthogonalize the $k$th partition ${\bm\Lambda}^{(m)k}$ and address label and \\sign switching with the \texttt{MatchAlign} algorithm \citep{poworoznek2021efficiently};\\
\quad\quad\quad\quad\quad\quad Obtain rotation matrix $\bm R^{(m)k}$ by performing Varimax optimization ;\\
\quad\quad\quad\quad\quad\quad
Rotate ${\bm\Lambda}_h^{(m)k}$ and set ${\bm\Lambda}_h^{(m)k}\leftarrow {\bm\Lambda}^{(m)k} \bm R^{(m)k}$;\\
\quad\quad\quad\quad\quad\quad Rotate ${\bm g}^{(m)k}$ and set ${\bm g}^{(m)k} \leftarrow (R^{(m)k})^T{\bm g}^{(m)k}.$ \\
\quad\quad\quad\quad   \textbf{for} $h=1,\ldots,H$ \textbf{do}\\
\quad\quad\quad\quad\quad\quad $\bm\Lambda_h^{(m)}\leftarrow \bm\Lambda_h^{(m)}/\|\bm\Lambda_h^{(m)}\|_2$;\\
\quad\quad\quad\quad\quad\quad  $\bm g^{(m)}\leftarrow \bm g^{(m)}\|\bm\Lambda_h^{(m)}\|_2;$\\
 \bottomrule
\end{tabular} 
\end{center} 

In summary, NIFTY fitting involves the following steps: (1) pre-train anchor data 
and infer the dimension of the latent space $K$ (Section 2.2); (2) generate posterior samples using the MALA-within-Gibbs sampler (Section 3.1); (3) post-process these posterior samples to address rotational alignment, scale ambiguity and label switching (Section 3.2); and (4) calculate posterior and predictive summaries of interest and/or generate new data based on the post-processed samples. Step (3) can be excluded when the focus is only on generative modeling or inference on the density of $\bm x$.
 
\section{Theoretical Properties} 
Identifiability is a key property for a latent variable model to provide reliable, reproducible, and interpretable latent representations. There are two notions of identifiability for factor models (introduced by \cite{allman2009identifiability}),  \emph{strict identifiability} and \emph{generic identifiability}. Strict identifiability ensures that the mapping from the observed data to the parameters is one-to-one, and generic identifiability only requires the mapping to be one-to-one except on a Lebesgue measure zero subset of the parameter space. 

In this section, we establish strict identifiability for $\bm \Lambda$, $\bm \Sigma$ and the parameters that characterize $g_1,\ldots,g_h$ in two steps. First, under a mild anchor dimension assumption, we show that the parameters and factors are unique up to a simple transformation, leading to generic identifiability. Second, we show that post-processing removes ambiguity in parameters yielding an equivalent likelihood, leading to strict identifiability and posterior consistency.

\begin{definition}
	 (Strict Identifiability). Latent factor model \eqref{eq: nifty+model}  is strictly identifiable if   
	 $$\text{pr}(\bm x_i = \bm a\mid{\bm\Lambda}, \bm g, \bm u_i,  {\bm\Sigma}) = \text{pr}(\bm x_i = \bm a\mid {\bm\Lambda}', \bm g',\bm u_i',  {\bm\Sigma}') \quad \forall\bm a\in \mathbb R^P$$
	 holds if and only if $({\bm\Lambda}, \bm g,\bm u_i,  {\bm\Sigma})$ and $({\bm\Lambda}', \bm g', \bm u_i', {\bm\Sigma}')$ are identical up to a latent class permutation.
\end{definition}

\begin{definition}
	 (Generic Identifiability).  Latent factor model \eqref{eq: nifty+model} is generically identifiable if the set $\mathcal S_{({\bm\Lambda}, \bm g,  {\bm\Sigma})}:= \{({\bm\Lambda}', \bm g',  {\bm\Sigma}'): \text{pr}(\bm x_i = \bm a\mid{\bm\Lambda}, \bm g, {\bm\Sigma}) = \text{pr}(\bm x_i = \bm a\mid {\bm\Lambda}', \bm g', {\bm\Sigma}'), \forall\bm a\in \mathbb R^P\}$ has Lebesgue measure zero.
\end{definition}

\subsection{Generic identifiability with anchor dimension assumption}

Our identifiability results build on related theory for factor models, including \cite{bing2020adaptive, arora2013practical, moran2022identifiable}. These works assume the existence of ``anchor features'' (or ``pure variables''), which refers to certain dimensions of data that depend on only one factor. In practice, we use pre-training to add additional anchor dimensions to the observed dataset before NIFTY model fitting, as described in Section 2.2.
 \begin{definition}(Anchor dimension)
 	A dimension $j$ is said to be an anchor dimension for the latent location $u_{ik}$ if $x_{ij}$ depends only on that variable. That is, there exists an index $h$ with $k_h=k$, such that
 	$E(x_{ij}\mid \bm u_i) = \lambda_{jh}g_h(u_{ik})$ for all $i$. The $x_{ij}$ is known as the anchor feature.
 \end{definition}
 \begin{assumption}\label{assp: anchor}(Anchor feature) For each factor $u_{ik}$ in model \eqref{eq: nifty+model}, there exists at least one anchor feature.  \end{assumption}

An equivalent likelihood can be obtained
replacing $(\lambda_{jh}, g_h)$ with $(a\lambda_{jh}, \frac{1}{a}g_h)$ for any scalar $a>0$. To remove such scale ambiguity, we assume that the columns of $\bm\Lambda$ are scaled and the loading matrix is of full column rank.
\begin{assumption}\label{assp: loadings }(Scaled and full-ranked loadings) The loading matrix $\bm\Lambda$ is of full column rank and each column of $\bm\Lambda$ satisfies $\|\bm\Lambda_h\|_2 = 1$. \end{assumption}
  
We let ${j_1}, \ldots, j_K$ denote the anchor dimensions of $u_{i1},\ldots,u_{iK}$. When the data and model satisfy the Assumptions \ref{assp: anchor}-\ref{assp: loadings }, we can express each anchor dimension in the data as 
\begin{equation}\label{eq: 1danchor_dimension}
    x_{ij_k} = \lambda_{j_kh}g_h(u_{ik}) + \varepsilon_{ij_k},\quad  \varepsilon_{ij_k}\sim N(0,\sigma^2_{ij_k}). 
\end{equation} 
Without constraining the loading matrix, it is typically impossible to identify the diagonal residual covariances. For example, in a one-dimensional factor model, \( x_i = \eta_i + \varepsilon_i \) can be equivalently written as \( x_i = (\eta_i + \varepsilon_i/2) + \varepsilon_i/2 \), with \( \gamma_i = \eta_i + \varepsilon_i/2 \) interpreted as the latent factor. When applying maximum likelihood estimation (MLE) to factor models, this leads to the \emph{Heywood problem} \cite{martin1975bayesian,joreskog1967some}, where some residual variances are estimated as zero. A similar issue arises in Bayesian settings \cite{fruhwirth2024sparse}, where the posterior for \( \sigma_j^2 \) can have a mode at zero.

The following assumption helps resolve the ambiguity and avoid Heywood cases. Although this assumption may appear restrictive, in Section 2.2 we propose a pre-training procedure that infers the anchor features and their variances from the data and then fixes them in the subsequent analyses to avoid Heywood cases. 

\begin{assumption}[Residual variance of the anchor features]\label{assp: sigma}
    The residual variance of each anchor feature is a known fixed value.
\end{assumption}

We now state the generic identifiability of the NIFTY model. Recall the model setting in \eqref{eq: nifty+model} and that the $H$ factors are mapped from $K$ uniform latent locations. We introduce a partition according to the mapped latent location: Let $\bm{\eta}_i^1,\ldots,\bm{\eta}_{i}^K$ be a partition of $\bm{\eta}_{i}$, such that each $\bm{\eta}_i^k$ consists of factors mapped from uniform location $u_{ik}$. Let $\bm\Lambda^1,\ldots, \bm\Lambda^K$ be the partition in the columns of $\bm\Lambda$ and $\bm g^1,\ldots, \bm g^K$ be the partition of $\bm g$, according to the same logic. 

\begin{theorem}\label{thm: main_identify}Suppose the data and model \eqref{eq: nifty+model} satisfy Assumptions 1-3, and the parameters are partitioned into groups according to the latent locations. Then we have
\begin{itemize}
    \item (generic identifiability of $\bm \Lambda$ and $\bm g$)
	$$\text{pr}(\bm x_i = \bm a\mid{\bm\Lambda}, \bm g) = \text{pr}(\bm x_i = \bm a\mid {\bm\Lambda}',\bm g') \quad \forall\bm a\in \mathbb R^P$$ holds if and only if there exists rotation matrices  $\bm R^1,\ldots,\bm R^K$ such that $\bm R^k(\bm R^k)^T=1$ and that $\bm\Lambda^k=\bm\Lambda'^k\bm R^k$, and $\bm g^k(\bm u) = (\bm R^k)^T\bm g'^k(\bm u)$ for all $\bm u$.
    \item (strict identifiability of the residual variance) The covariance matrix $\bm\Sigma$ is unique.
\end{itemize}
\end{theorem}

\subsection{Strict identifiability after post-processing}
 It is standard practice in the Bayesian factor analysis literature to apply post-processing to account for identifiability issues in inferring the factor loadings matrix. We build on such approaches to develop a post-processing algorithm to infer parameters satisfying strict identifiability. After post-processing, we have strict identifiability for each parameter.
\begin{proposition}Suppose the data and model \eqref{eq: nifty+model} satisfy Assumptions \ref{assp: anchor}-\ref{assp: sigma}. After the post-processing steps described in Algorithm 1,  
 the model parameters $(\bm \Lambda,\bm g,\bm\Sigma)$ are strictly identifiable.
\end{proposition} 
   
 \subsection{Posterior consistency} 
With the identifiability results, we achieve guaranteed posterior consistency under suitable priors. We first show Bayesian posterior consistency for the model parameters, then develop a large support property for the induced density, in order to show the induced density also has posterior consistency in a strong sense.

\begin{theorem}(Posterior consistency of parameters)\label{thm: consistency_parameters}
Denote the collection of model parameters by $\Theta = (\bm\Lambda, \bm g, \bm\Sigma)$. Suppose the prior distribution for the parameters has full sup-norm 
 support around the true value $\Theta^0$ and the data satisfy Assumptions \ref{assp: anchor}-\ref{assp: sigma}. Let $H(\Theta)$ denote the post-processed parameters under 
 Algorithm 1. Then for any 
 $\epsilon$-neighborhood around $\Theta^0$, $\mathcal N_\epsilon(\Theta^0)$, 
$$\text{pr}(H(\Theta) \in \mathcal N_\epsilon(\Theta^0)\mid\bm x_1,\ldots,\bm x_N)\to 1\quad P^\infty_0\text{almost surely}.$$
\end{theorem}   

Next, we establish the posterior consistency of the induced density of $\bm x_i$. As a mild condition, the true density generating the data should be close to a latent factor structure described by model \ref{eq: nifty+model}, which is formulated as follows.
 \begin{assumption}
     \label{assp: truedensity}
There exists $\bm \Lambda^0\in\mathbb R^{P\times H}$, $\bm g^0=(g_1^0,\ldots,g_h^0)$, and $\bm\Sigma^0=\text{diag}\{(\sigma^0_1)^2,\ldots,(\sigma^0_P)^2\}$ such that the $P$-dimensional multivariate density can be represented as  
	\begin{equation}
		f^0(\bm x_i) =  \int_{[0,1]^H}\phi_{\bm\Sigma}[\bm x_i-{\bm\Lambda}^0\bm g^0(\bm u_i)]d\bm u_i,
	\end{equation}
 with $\sup|g_h^0(u)|<\infty$ for any $u\in[0,1]$, and $f^0$   strictly positive and finite.
 \end{assumption}  
  The next theorem states that the induced prior on the density $f$ assigns positive probability to arbitrary small neighborhoods of any $f_0$ satisfying Assumption \ref{assp: truedensity}. The neighborhoods are defined using Kullback-Leibler (KL) divergence. For two continuous distributions with densities $f_2$ and $f_2$, the KL divergence is defined to be 
\[KL(f_1, f_2) = \int f_1(\bm x)\log \frac{ f_1(\bm x)}{ f_2(\bm x)}d\bm x.\]
We denote an $\epsilon$-sized KL
neighborhood around a density $f^0$ as $KL_\epsilon(f^0)$.
\begin{theorem}\label{thm: consistency}
	Suppose that data are generated from model \eqref{eq: nifty+model}, with the true parameter $\Theta^0$. Let $f^0$ denote the corresponding density. Suppose that the prior distribution has full sup-norm support around the true value $\Theta^0$ and the data satisfy Assumption \ref{assp: anchor}-\ref{assp: sigma}, then $\Pi_{\bm f}[KL_\epsilon(f^0)]>0$ for all $\epsilon>0$. 
\end{theorem}

We conclude this section with a consistency result for density estimation.

\begin{proposition}\label{prop: consistency}
	Suppose that data are generated from model \eqref{eq: nifty+model} and satisfy Assumption \ref{assp: anchor}. Let $f^0$ denote the corresponding density. If the priors for $\bm\Lambda,\bm g,\bm\Sigma$ satisfy the conditions in Theorem \ref{thm: consistency}, then the posterior distribution of the induced $f$ converges as follows:
	\[\Pi[\mathcal U_\epsilon(f^0)\mid \bm x_1,\ldots,\bm x_N]\to 1\quad \mathbb P^\infty_0\text{-almost surely,}\] 
	where $U_\epsilon(f^0):=\{f:\int |f-f_0|\text{d}\bm x<\epsilon\}$ is an $\epsilon$-ball around the true density, and $\mathbb P^N_0$ denotes the distribution of $\bm x_1,\ldots \bm x_N$ under $f^0$.
\end{proposition} 

\section{Numerical Experiments} \label{sec: numerical}
\subsection{Density estimation and factor analysis}
\begin{figure}[H]
        \centering   
        \begin{subfigure}[t]{0.45\textwidth}
                \centering \includegraphics[width=.5\textwidth]{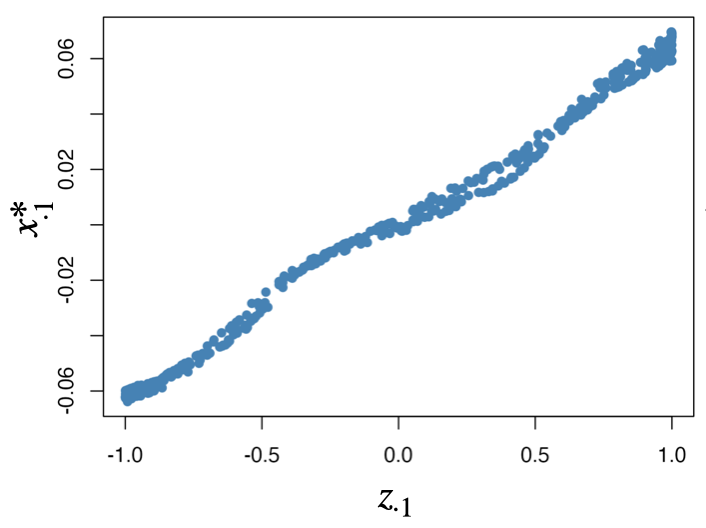}
                \caption{}
        \end{subfigure} \quad 
        \begin{subfigure}[t]{0.45\textwidth}
                \centering \includegraphics[width=.5\textwidth]{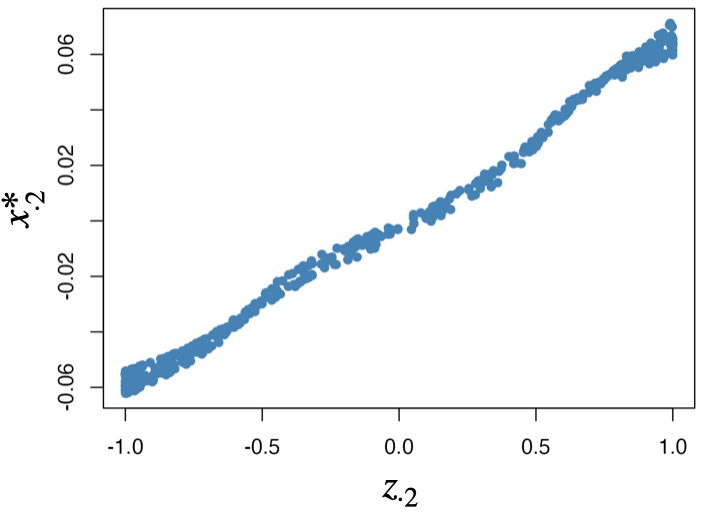}  
                \caption{}
        \end{subfigure} 
        \caption{\label{fig: 2-curve-diffusion}Anchor data of setting 1 learned from diffusion maps, plotted against true latent variables. }
\end{figure}

We used three different simulated settings to assess the performance in density estimation and parameter estimation. We compare with Gaussian linear factor models [using cumulative shrinkage prior CUSP \citep{legramanti2020bayesian}], Bayesian mixture of factor analyzers [BMFA, \citep{papastamoulis2020clustering}], Bayesian Gaussian copula factor analysis [BFA, \citep{murray2013bayesian}] GP-LVMs, and variational auto-encoders in terms of density estimation accuracy and latent variable interpretability. We consider the following settings: 
\begin{enumerate} 
	\item Independent but non-Gaussian marginal distribution in $\mathbb R^2$:
	$\bm x_i = N[(z_{i1}, z_{i2})^T,\sigma^2 \bm I_{2}], z_{i1} \sim \text{Beta}(0.4, 0.4), z_{i2} \sim\text{Gamma}(1,1),\sigma^2= 0.01, z_{i1} {\perp\!\!\!\!\perp}  z_{i2}.$ 
	\item Gaussian linear factor in $\mathbb R^{20}$: $\bm x_i = \bm\Lambda\bm\eta_i+\bm\varepsilon_i,\quad\bm\eta_i\sim N_2(0,\bm I_2),\sigma^2= 0.01, \lambda_{jk}\stackrel{\text{iid}}{\sim} N(0,1).$
	\item Curved shape in $\mathbb R^{10}$:
	$\bm x_i = N[(2z_{i1},2z_{i1}^2,2z_{i2},2z_{i2}^2,0,\ldots,0)^T,\sigma^2 \bm I_{10}]$ with $\sigma^2=0.01$ and $z_{ij}\stackrel{\text{iid}}\sim U(0,1).$  
\end{enumerate}
	In each experiment, we generated $n$ data points and varied the training size $n$ from 100 to 800 to examine the impact of sample size on model performance. We use the diffusion-map-based algorithm to pretrain a set of anchor data for each case. 
    For implementing NIFTY, we adopt the following setup conventions:
\begin{itemize}
    \item The number of anchor dimensions \( K \) and the augmented data \( \bm{x}_i^* \) are obtained via a diffusion map–based pre-training procedure. These are treated as fixed inputs to the NIFTY model.
    \item Residual variances for the anchor dimensions ($\sigma^{*2}_i$) are pre-estimated during pre-training, while residual variances  ($\sigma^2_i$) for the non-anchor dimensions are inferred via MCMC.
    \item We specify an overfitted number of factors \( \bar{H} = KL \) (with a default of \( L = 5 \)), and rely on sparsity-inducing priors to infer the effective number of factors \( H \).
    \item During MCMC (10,000 total iterations with a burn-in of 5,000), we sample \( \bm{\Lambda} \), \( \sigma_i^2 \), \( u_i \), and the slopes ($\alpha_{lh}$) of the transformation functions \( g_h \).
    \item After MCMC, we apply post-processing (Algorithm~\ref{algo: matchalign}) to align and summarize the posterior samples \((\bm{\Lambda}^{(1)}, \bm{g}^{(1)}, \bm{\Sigma}^{(1)}), \ldots, (\bm{\Lambda}^{(M)}, \bm{g}^{(M)}, \bm{\Sigma}^{(M)})\).
\end{itemize}
    For all three settings, the pre-training yields $K=2$ anchor dimensions, which coincides with the true number of independent latent variables. Figure \ref{fig: 2-curve-diffusion} displays a sanity check for the augmented data in setting 1. The anchor data recover the true independent factors that generate the dataset.

\begin{figure}[H]
        \centering   
        \begin{subfigure}[t]{0.25\textwidth}
                \centering \includegraphics[width=.45\textwidth]{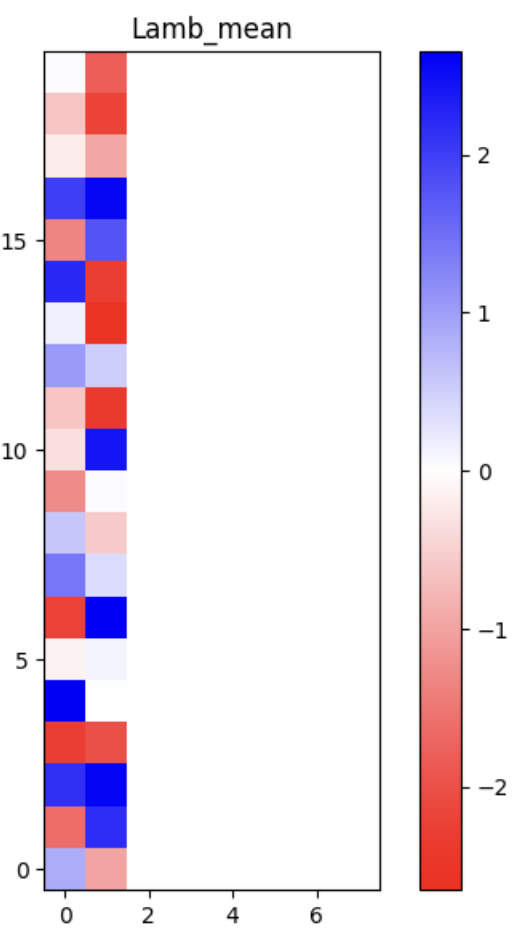}
                \caption{Posterior mean of $\bm\Lambda$ for $n=100$.}         \end{subfigure} \quad 
        \begin{subfigure}[t]{0.25\textwidth}
                \centering \includegraphics[width=.45\textwidth]{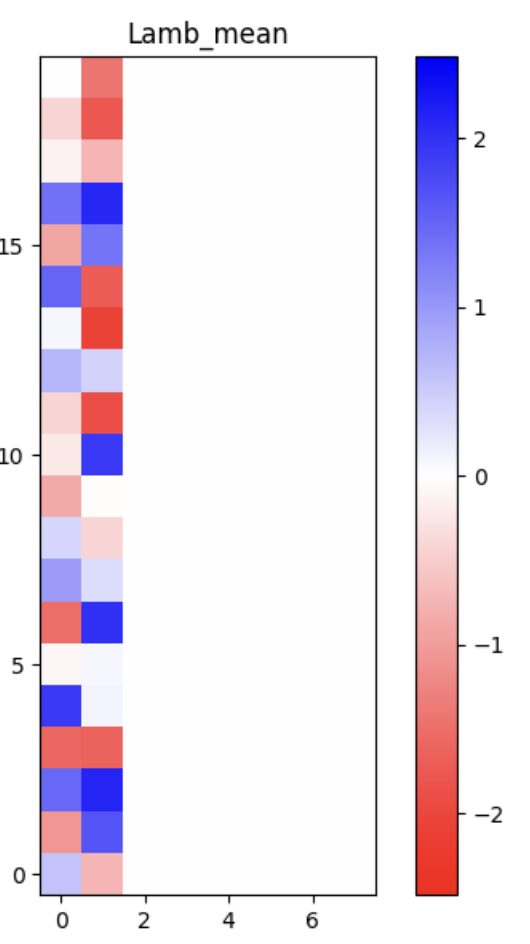}  
                \caption{Posterior mean of $\bm\Lambda$ for $n=800$.}
        \end{subfigure}   
        \quad 
        \begin{subfigure}[t]{0.25\textwidth}
                \centering \includegraphics[width=.25\textwidth]{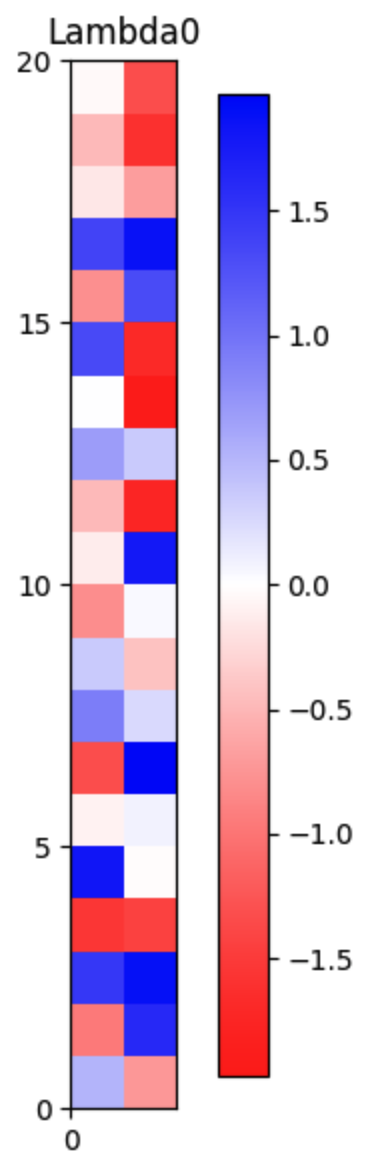}                \caption{True $\bm \Lambda_0$.}
        \end{subfigure}  \\ 
        \begin{subfigure}[t]{0.35\textwidth}
                \centering \includegraphics[width=.75\textwidth]{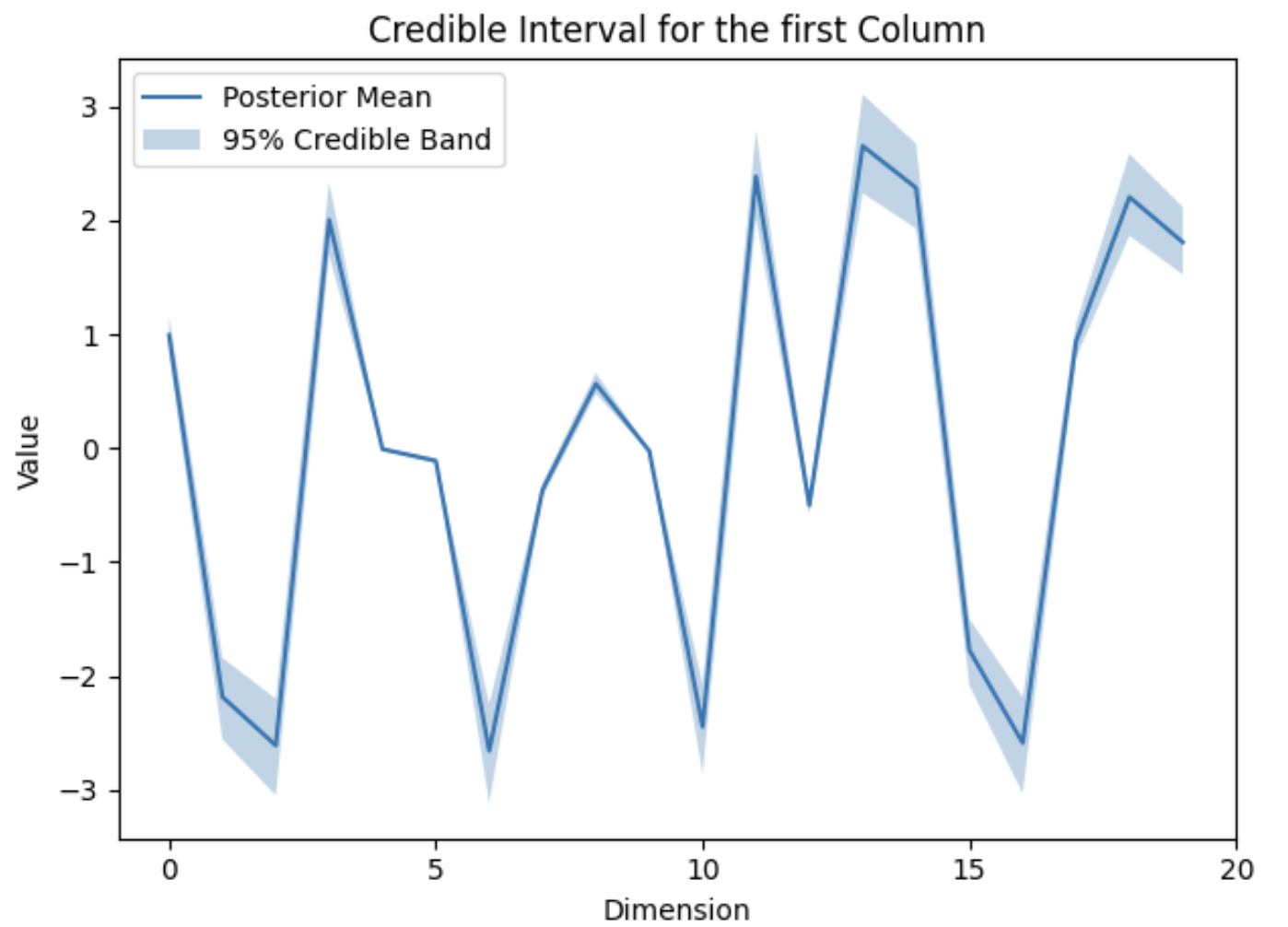}  \caption{The 95\% credible band for the nonzero values in $\bm\Lambda$ for $n=100$.}
        \end{subfigure}  \quad
        \begin{subfigure}[t]{0.35\textwidth}
                \centering \includegraphics[width=.75\textwidth]{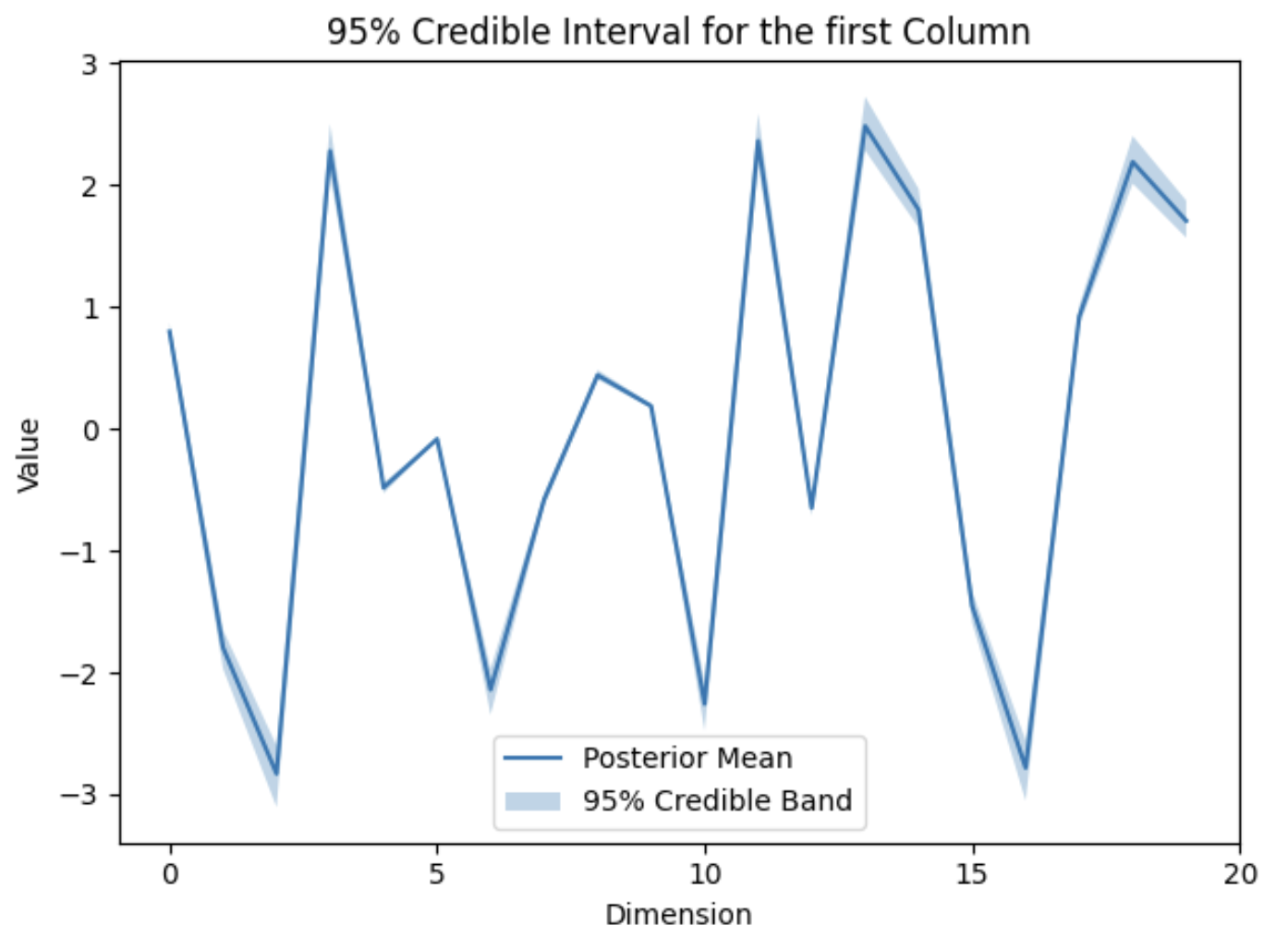}      \caption{The 95\% credible band for the nonzero values in $\bm\Lambda$ for $n=800$.}
        \end{subfigure}  \\ 
        \caption{\label{fig: numerical_loadings} Example results for simulation case 2. The upper bound on the number of factors is chosen as $H=8$. The posterior mean of the loading matrix for $n=100$ and $n=800$ is plotted in (a) and (b), the credible interval non-zero values are plotted in (d) and (e), with 2-dimensional ground truth in (c).} 
\end{figure}

We show estimated loading matrices for different sample sizes in simulation case 2 in Figure \ref{fig: numerical_loadings}. As $n$ increases, the posterior mean $\bm\Lambda$ converges to the truth, which supports our strict identifiability and posterior consistency results. 

We evaluate accuracy in characterizing the unknown density of the data using sliced Wasserstein (SW) distance \citep{bonneel2015sliced} between
test data and data sampled from the posterior predictive. We estimate SW distance using the 
\texttt{pot} python package. We fixed the test sample size to $1,000$ to limit Monte Carlo errors in SW distance estimation. As shown in Figure \ref{fig: numerical_loadings}, the posterior estimation for the loadings is close to the true parameter. As $n$ increases, the posterior mean approaches closer to the truth and the credible band becomes narrower. 

\subsubsection*{Comparison with other latent factor models}
We implemented GP-LVMs, variational autoencoders (VAEs), and Gaussian linear factor models (PPCA) in the same simulated settings and compare performance with NIFTY. Further implementation details are in the Supplementary Materials, Section 3. 

\begin{figure}[H]
        \centering   
        \begin{subfigure}[t]{0.25\textwidth}
                \centering \includegraphics[width=1\textwidth]{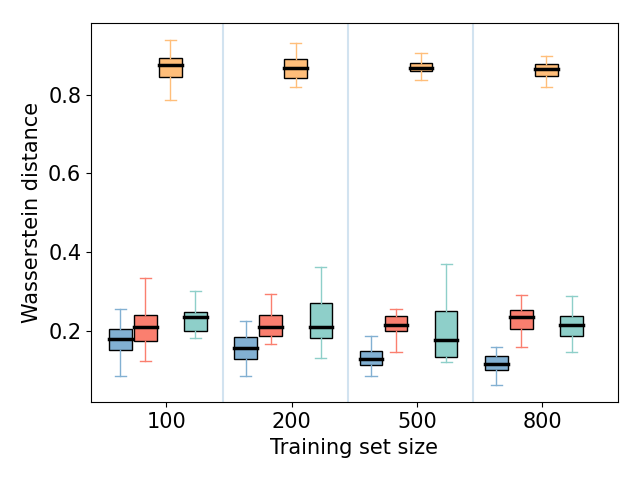}
                \caption{2-dimensional data generated from independent Gamma and Beta distribution.}         \end{subfigure} \quad 
        \begin{subfigure}[t]{0.25\textwidth}
                \centering \includegraphics[width=1\textwidth]{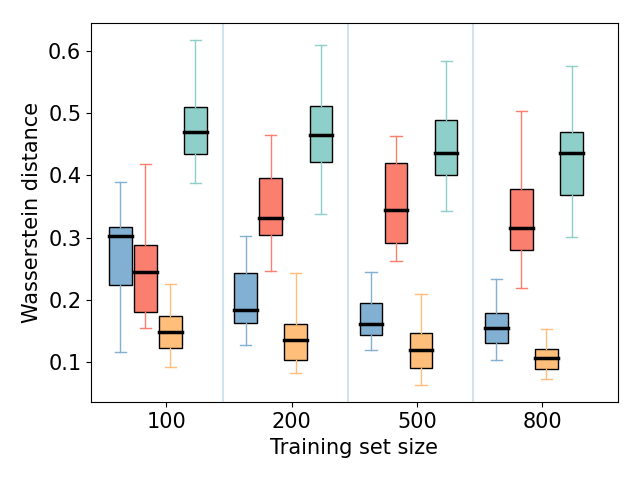}  
                \caption{20-dimensional data generated from a Gaussian linear factor.}
        \end{subfigure}   
        \quad 
        \begin{subfigure}[t]{0.25\textwidth}
                \centering \includegraphics[width=1\textwidth]{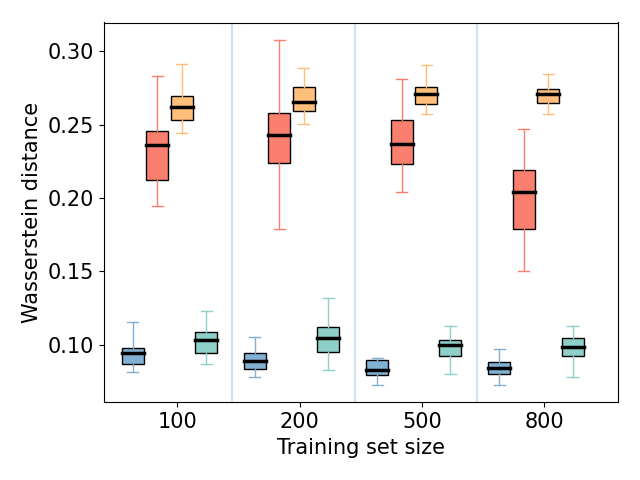}                \caption{10-dimensional data generated from two latent curves.}
        \end{subfigure} 
        \begin{subfigure}[t]{0.08\textwidth}
                \centering \includegraphics[width=1\textwidth]{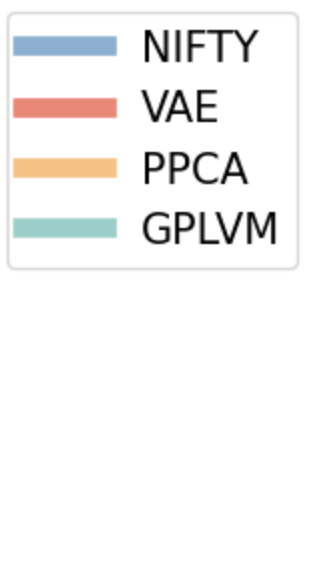} 
        \end{subfigure}   
        \caption{\label{fig: simu-res-Wasserstein} Box plots of the out-of-sample sliced Wasserstein distance. The three panels respectively display results from the three simulated data sets, and boxes are filled with different colors according to the method.} 
\end{figure}

\paragraph{Point estimation: } Figure \ref{fig: simu-res-Wasserstein} shows that NIFTY achieves the smallest out-of-sample sliced Wasserstein distance in most experiments and has the smallest variance across replicates. Due to Monte Carlo error, the minimum-achievable distance is not zero. Meanwhile, PPCA performs remarkably well when data are generated from a Gaussian linear factor model but poorly in non-Gaussian cases. GP-LVM effectively captures non-Gaussianity in panels (a) and (c), and is the second best among the methods. VAE performs comparably with GP-LVM and NIFTY for the 2-dimensional independent data but fails on the curved data. 

\begin{figure}[H]
        \centering   
        \begin{subfigure}[t]{0.2\textwidth}
                \centering \includegraphics[width=1\textwidth]{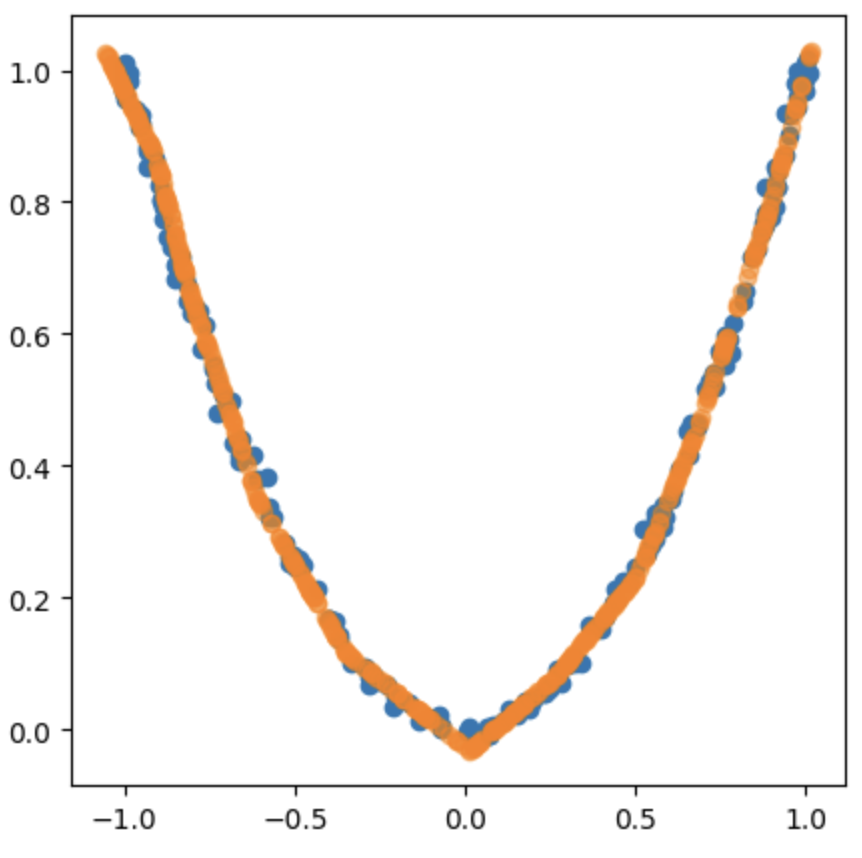}
                \caption{NIFTY}
        \end{subfigure} \quad 
        \begin{subfigure}[t]{0.2\textwidth}
                \centering \includegraphics[width=1\textwidth]{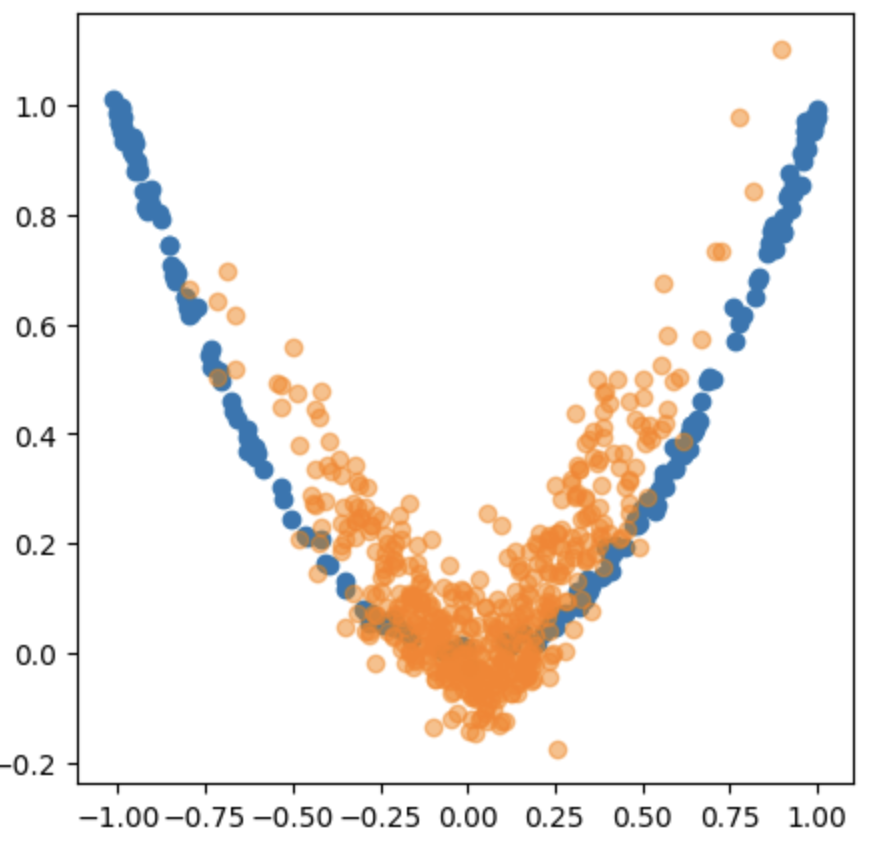}  
                \caption{VAE}
        \end{subfigure}   
        \quad 
        \begin{subfigure}[t]{0.2\textwidth}
                \centering \includegraphics[width=1\textwidth]{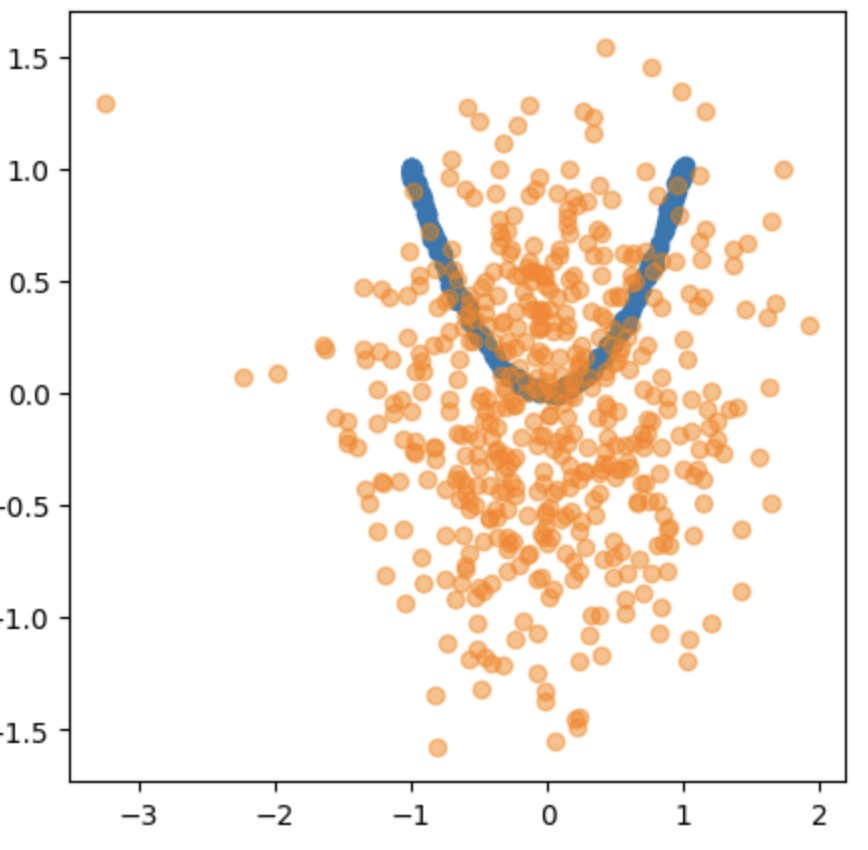}                \caption{PPCA}
        \end{subfigure} \quad 
        \begin{subfigure}[t]{0.2\textwidth}
                \centering \includegraphics[width=1\textwidth]{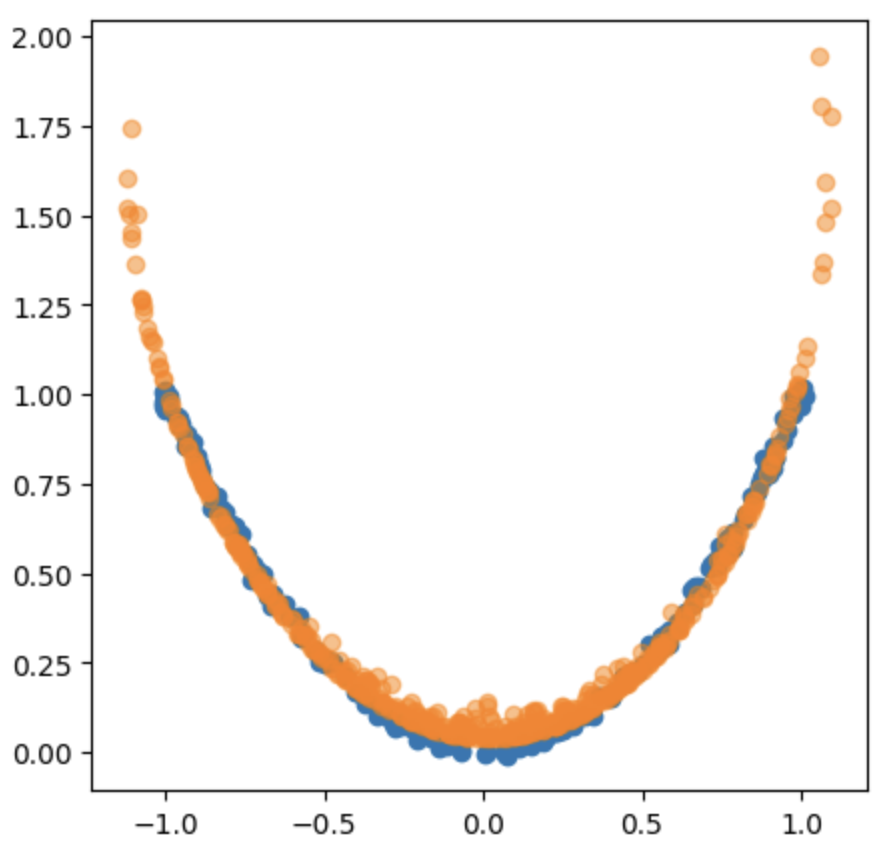} 
                \caption{GPLVM}
        \end{subfigure} 
        \caption{\label{fig: 2-curve-compare}Comparing out-of-sample data generation in the curved data experiment. The blue curves are from the ground truth, and the orange curves are data points generated by each method. Only the first two dimensions are plotted.} 
\end{figure}

The latent factors learned by VAE exhibit strong nonlinear relationships, deviating significantly from a standard normal distribution. To provide better visual insight into the results, we plot the first two dimensions of the generated data versus the ground truth in the curved data experiments, shown in Figure \ref{fig: 2-curve-compare}. PPCA generates Gaussian noise, while VAE and GP-LVM capture the nonlinear relationship between the two dimensions but fail to recover the spread and variance of the distribution accurately. In contrast, NIFTY accurately recovers the underlying curve and distribution of the data.  

\begin{table}[htbp]
\centering
\caption{Frequentist coverage of 95\% credible intervals for out-of-sample covariance matrix \(\text{Cov}(\bm x_i)\). The brackets show the mean width of the element-wise credible intervals.}
\begin{tabular}{lcccc}
\hline
\textbf{Method} & \textbf{Simulation 1} & \makecell{\textbf{Simulation 2} \\$(N=100, P=20)$}  & \makecell{\textbf{Simulation 2} \\$(N=500, P=50)$ }& \textbf{Simulation 3} \\
\hline
NIFTY & 0.95 (0.027) & 0.91 (0.342) & 0.89 (0.289) & 0.91(0.069) \\
Gaussian Factor & 0.88 (0.022) & 0.45 (0.188)& 0.15 (0.090) & 0.54(0.161) \\
Copula Factor & 0.94 (0.024) & 0.41(0.208)& 0.39(0.267) & 0.15(0.018) \\
\hline
\end{tabular}
\label{tab:coverage}
\end{table}

\paragraph{Uncertainty quantification:} 
We assess the uncertainty quantification performance of three Bayesian models: NIFTY, Bayesian Gaussian linear factor models, and Bayesian copula factor models. We estimated 95\% credible intervals for each element of the covariance and calculated the empirical coverage of the true covariance values. For Gaussian linear factor models, posterior samples of the covariance matrix are obtained as $\bm\Lambda^{(m)} {\bm\Lambda^{(m)}}^T + \bm\Sigma^{(m)}$, while for NIFTY we used $\bm\Lambda^{(m)}\text{Cov}(\bm\eta_i^{(m)}){\bm\Lambda^{(m)}}^T + \bm\Sigma^{(m)}$. Since the copula factor model does not explicitly estimate the covariance, we generate 500 synthetic data points from the estimated model for each posterior draw and compute the empirical covariance from the simulated data. 

Table~\ref{tab:coverage} reports the coverage rates, showing that NIFTY demonstrates substantially better coverage compared to the other methods. In Simulation 1, where $p = 20$ but only 2 dimensions have nonzero signal, all methods successfully capture the zero covariances in the remaining 18 dimensions, resulting in reasonably high average coverage. In Simulation 2, where the data are generated exactly from a Gaussian linear factor model, both the Gaussian and Copula-based methods tend to underestimate uncertainty. Although both models are well specified in this case, the large number of parameters combined with modest sample sizes and the use of shrinkage priors makes the results unsurprising. In Simulation 3, where the data are generated from two latent nonlinear curves, the Gaussian and Copula models fail to provide reliable coverage, while NIFTY maintains strong performance. We also considered two different dimensional settings for the Gaussian data cases: $N=100, P=20$ and $N=500, P=50$, and observed that NIFTY remains robust as the dimension increases.

\subsection{Data visualization with dimension reduction\label{sec: data-vis}}
The latent locations can be viewed as coordinates of the data embedded in low-dimensional unit cubes. When the dimension of the cube is 2 or 3, the latent locations can be naturally used to visualize high-dimensional data. There is a rich literature on unsupervised dimension reduction algorithms for data visualization, ranging from diffusion maps to the popular t-distributed stochastic
neighbor embedding (t-SNE) \citep{van2008visualizing}].

Many of the current dimension reduction algorithms for data visualization are uninterpretable black boxes, with an unclear relationship between the low-dimensional coordinates and the original high-dimensional data. In fact, it is typically not possible to map back from the low-dimensional coordinates to the observed data space. In contrast, NIFTY has a transparent generative probability model with a loadings matrix that concretely shows the relationship between the low-dimensional factors and the original data. In addition, NIFTY provides uncertainty quantification through a Bayesian approach and we can verify that dimensionality reduction is not leading to a significant loss of information by comparing the empirical distribution of the data to the posterior predictive distribution.


\subsubsection*{A Swiss roll embedded in ten-dimensional space}
\begin{figure}[H]
\centering 
\begin{subfigure}[t]{0.2\textwidth}
	\centering
	\includegraphics[width=.9\textwidth]{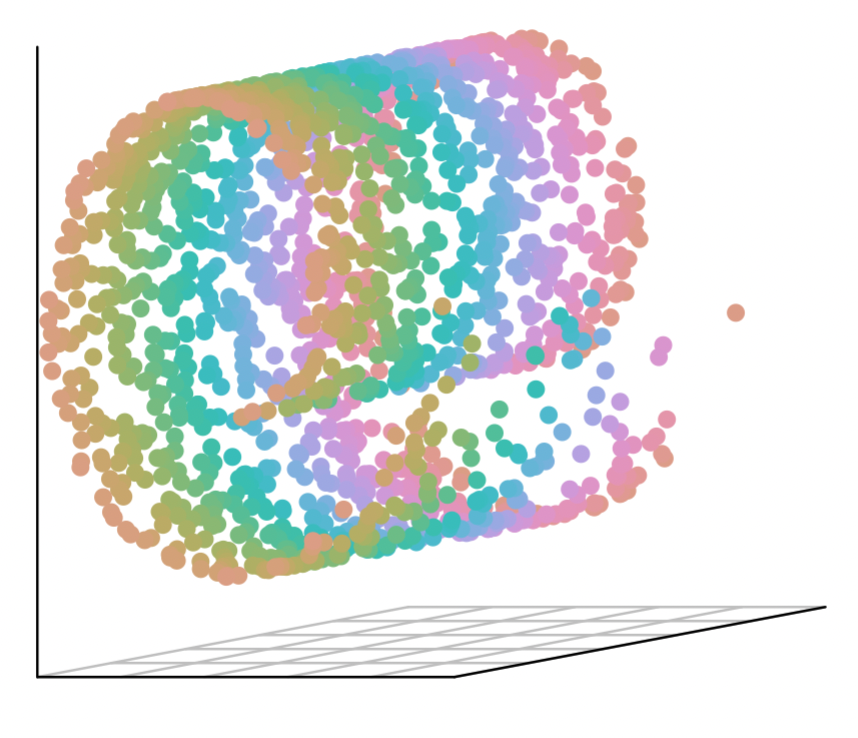}
 \caption{The three-dimensional Swiss roll generated from uniform variables $u_i$ and $v_i$.}
\end{subfigure}  \quad
\begin{subfigure}[t]{0.2\textwidth}
	\centering
	\includegraphics[width=.9\textwidth]{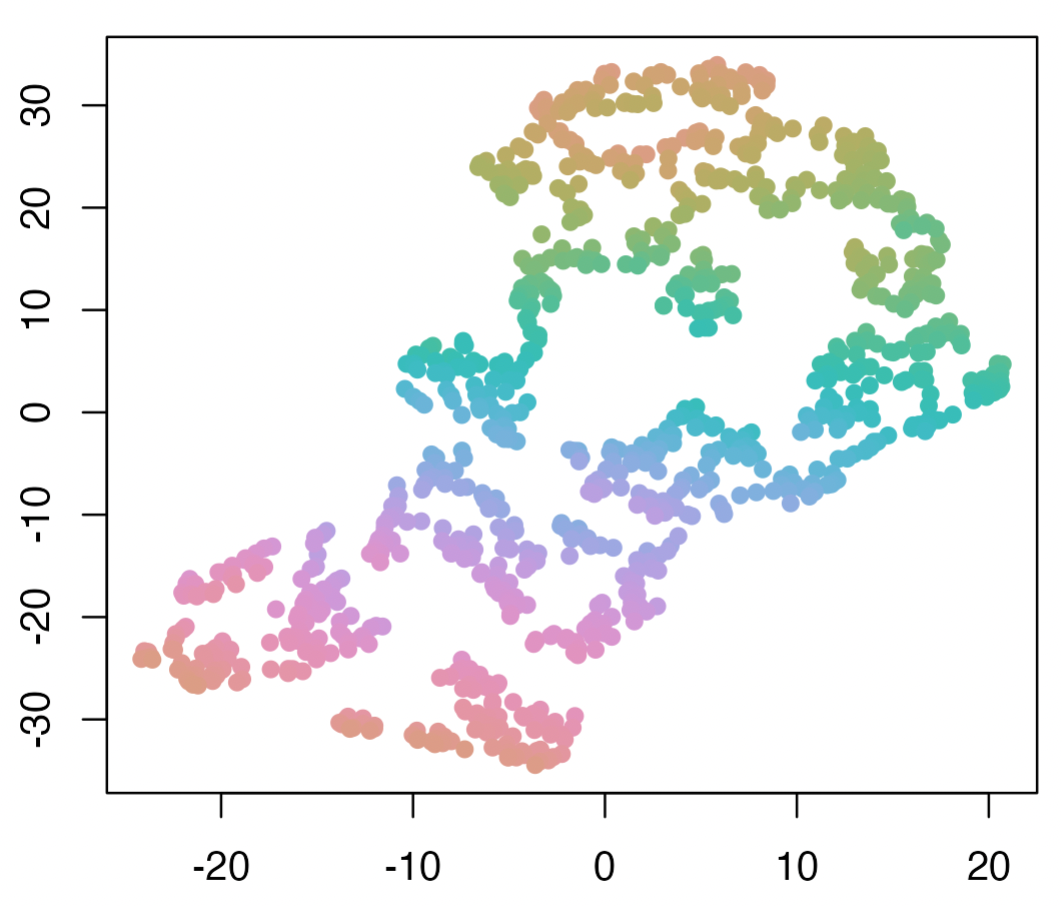}
  \caption{2D visualization from diffusion maps.}
\end{subfigure}  \quad
\begin{subfigure}[t]{0.2\textwidth}
	\centering
	\includegraphics[width=.9\textwidth]{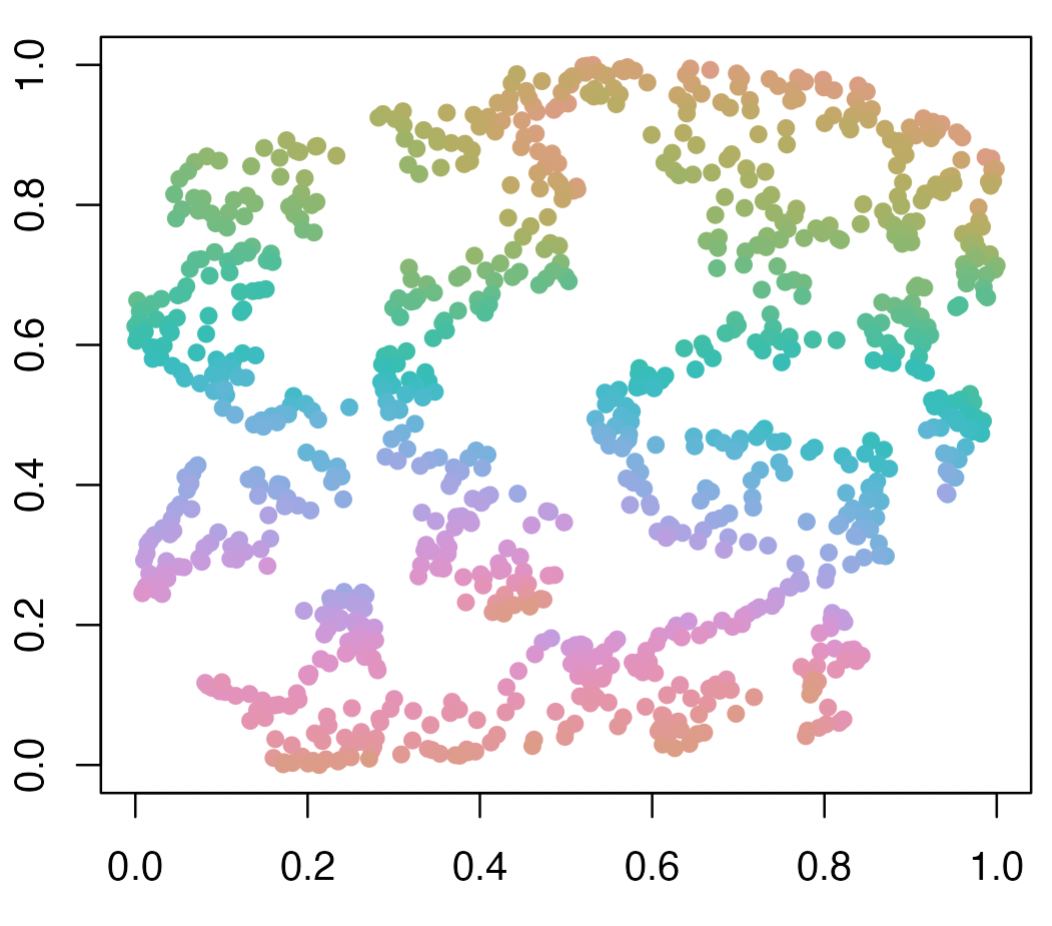}
  \caption{2D visualization from NIFTY latent locations.}
\end{subfigure} \quad
\begin{subfigure}[t]{0.24\textwidth}
	\centering
	\includegraphics[width=.9\textwidth]{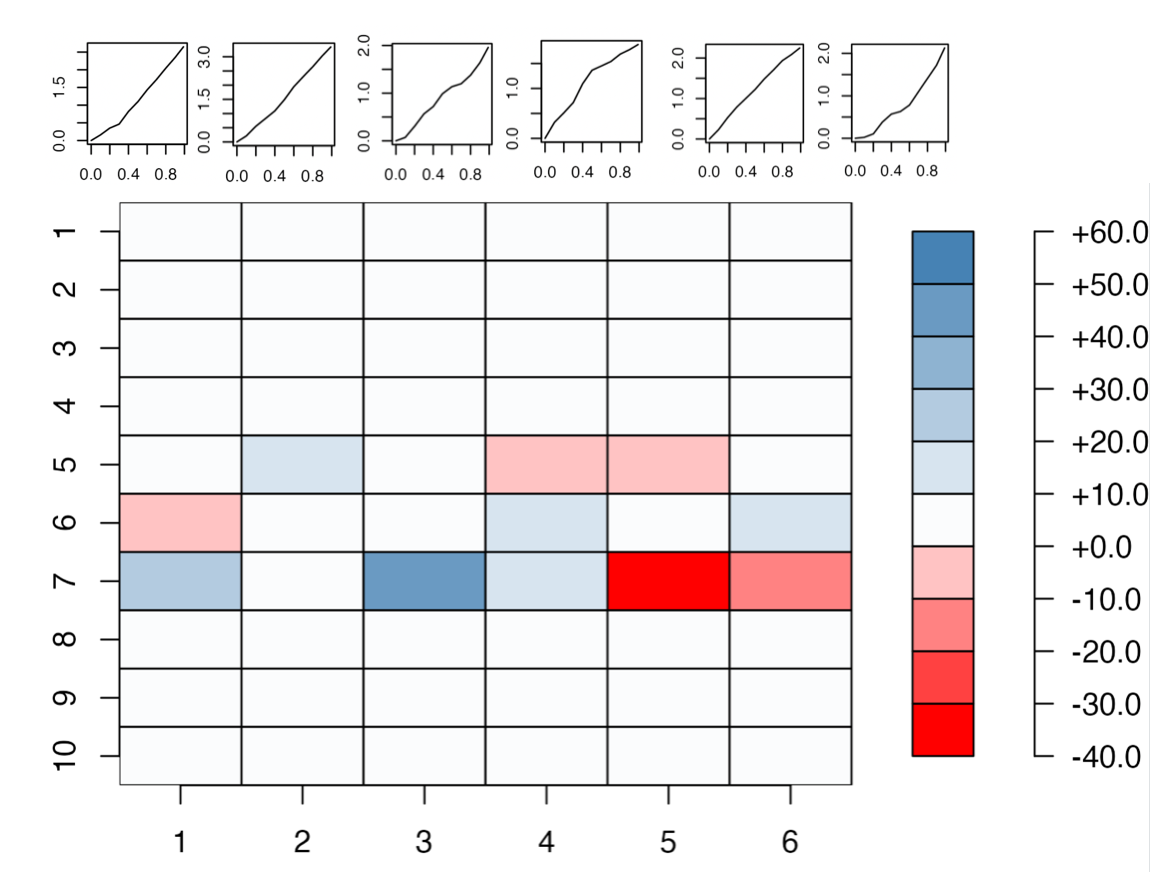}
  \caption{Loading matrix and latent mappings.}
\end{subfigure}
	\caption{\label{fig: swiss_visualization}Visualization of a Swiss roll embedded in a ten-dimensional space. The scatterplots in panels a-c are colored according to $v_i$. We use diffusion maps to generate anchor data and apply NIFTY with $K=2$ and $H=6$. NIFTY provides an unfolded view of the two latent factors, which is close to the truth. The mapping from latent factors to the data is also accessible.}
\end{figure}

In this example, a three-dimensional Swiss roll is first generated from two independent variables and then embedded in a ten-dimensional space. Starting with $u_i\sim \text{U}(0,1)$ and $v_i\sim \text{U}(0,1)$, we let $y_i\sim N(\bm\mu,0.01I_{10})$ with $\bm\mu= [0,0,0,0,(3\pi u_i+\frac{3}{2}\pi)\sin(3\pi u_i+\frac{3}{2}\pi)),(3\pi u_i+\frac{3}{2}\pi)\cos(3\pi u_i+\frac{3}{2}\pi)),v_i,0,0,0]^T.$ Figure \ref{fig: swiss_visualization}(a) plots the data on the three-dimensional Swiss roll. Figure \ref{fig: swiss_visualization}(b) visualizes the two-dimensional representation from diffusion maps. 
Figure \ref{fig: swiss_visualization}(b) shows the inferred latent locations from NIFTY. Both algorithms unfold the roll to a two-dimensional surface, revealing the true generating factors. With NIFTY, we are able to track the transformation from latent factors to data.

\subsubsection*{Gaussian clusters with heteroscedasticity}
We demonstrate a case where NIFTY 
clearly improves data visualization over the pre-training method. The data are sampled from five Gaussian distributions with different means and variances. The Gaussian means are equally spaced along one axis in a 20-dimensional space, and the variances of each cluster increase from $1^2$ to $5^2$. 

\begin{figure}[H]
\centering 
\begin{subfigure}[t]{0.2\textwidth}
	\centering
	\includegraphics[width=.9\textwidth]{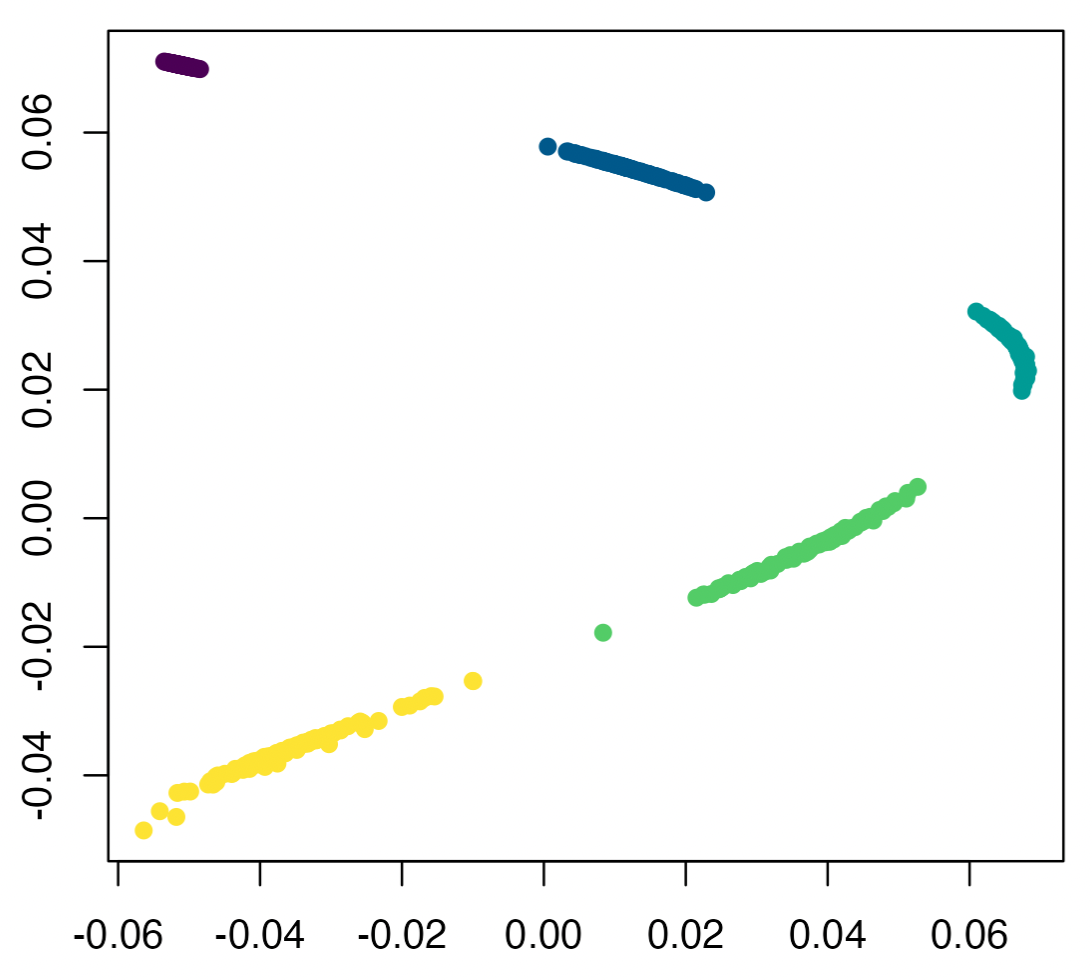}
  \caption{2D visualization from diffusion maps.}
\end{subfigure}  \quad
\begin{subfigure}[t]{0.2\textwidth}
	\centering
	\includegraphics[width=.9\textwidth]{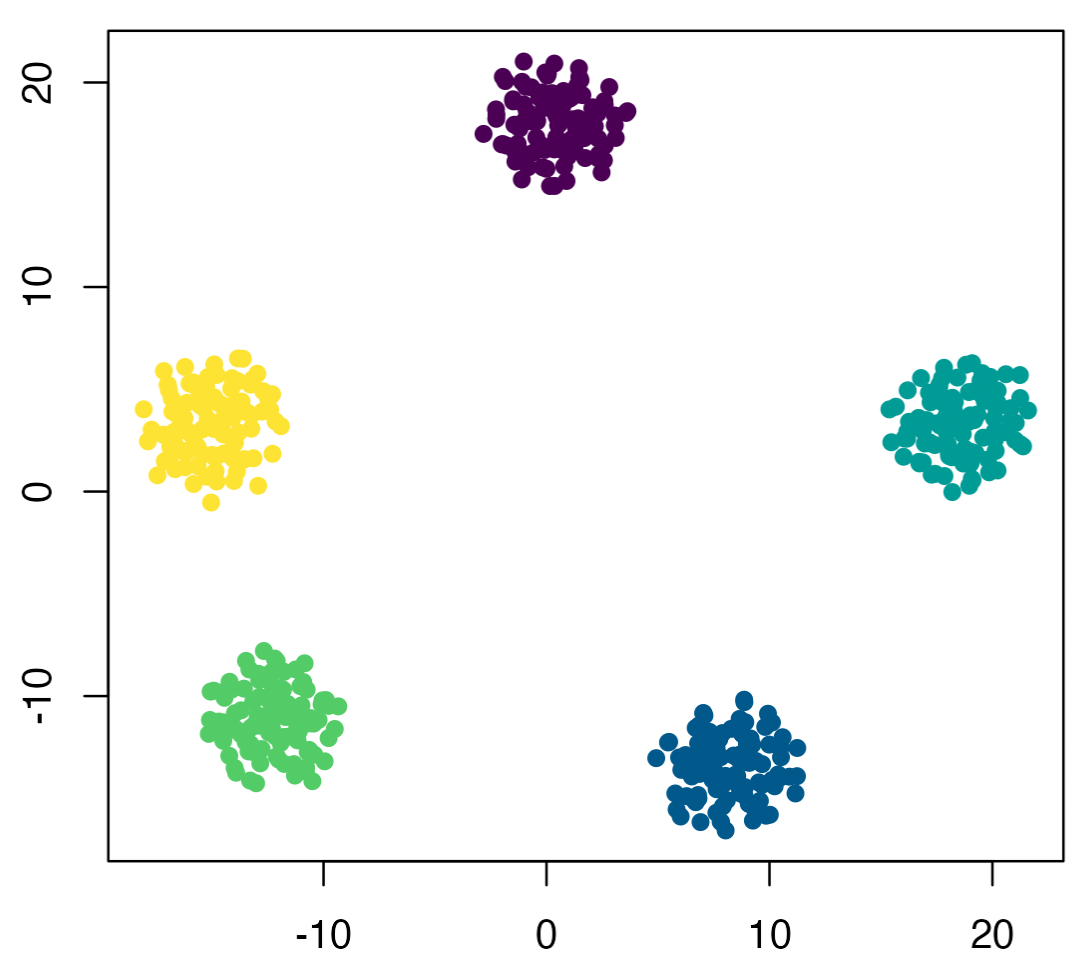}
  \caption{2D visualization from t-SNE.}
\end{subfigure}    \quad
\begin{subfigure}[t]{0.2\textwidth}
	\centering
	\includegraphics[width=.9\textwidth]{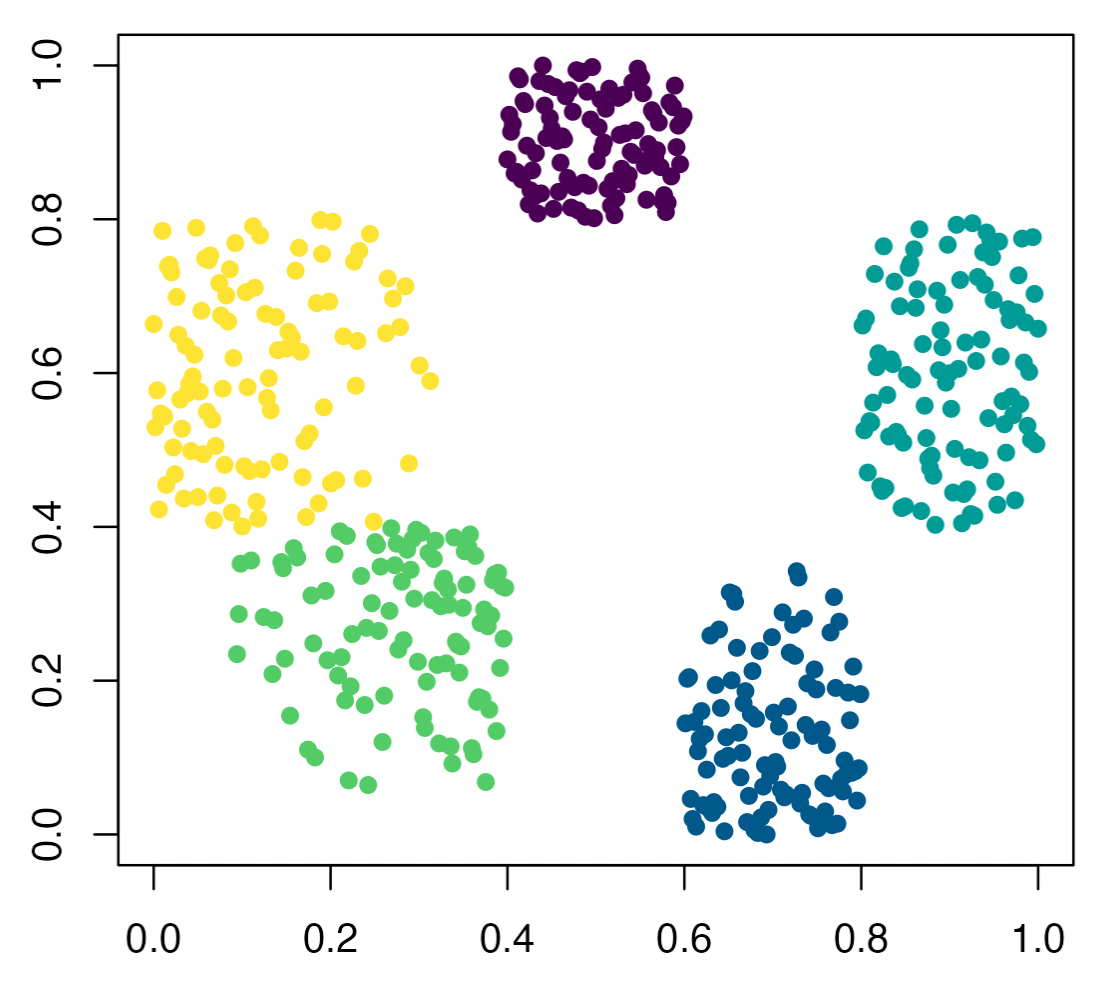}
  \caption{2D visualization from NIFTY latent locations.}
\end{subfigure}\quad
\begin{subfigure}[t]{0.2\textwidth}
	\centering
	\includegraphics[width=.7\textwidth]{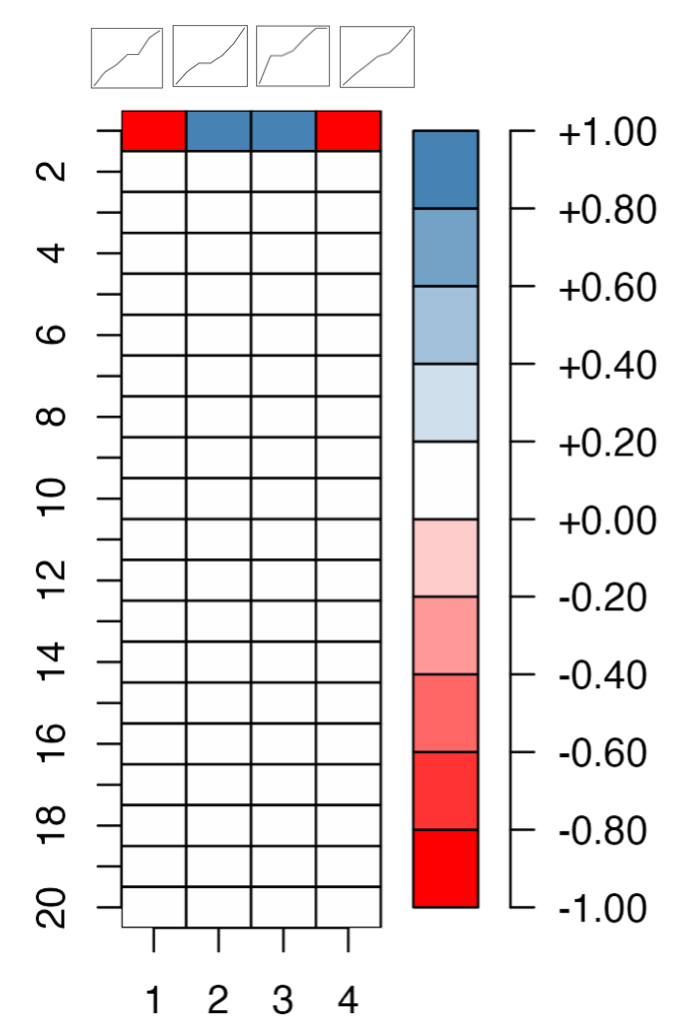}
  \caption{Loading matrix and latent mappings.}
\end{subfigure}
	\caption{\label{fig: cluster_visualization}Visualization of 500 points sampled from five 20-dimensional Gaussian with standard deviation being 1, 2, 3, 4, 5, colored according to cluster label (1 to 5 from dark purple to light yellow). Latent locations from NIFTY arguably provide the best visual results, with a clear presentation of cluster patterns and heteroscedasticity between clusters. Besides, the projection from the 2D representation to the data is tractable.}
\end{figure}

From Figure \ref{fig: cluster_visualization}(a), we find that diffusion maps are not ideal for visualizing the cluster pattern in a 2D space.  As shown in Figure \ref{fig: cluster_visualization}(b), t-SNE also does not have ideal performance in this case, preserving only compact clusters and not capturing heteroskedasticity. To obtain better visual results, we change the pre-training method from diffusion maps to t-SNE. From panel (c) of Figure \ref{fig: cluster_visualization}, the latent locations in NIFTY not only display compact cluster patterns but also provide a visual guide of the variance within each cluster. Moreover, the contribution of the dimensions in the data to the low-dimensional representation can be tracked via the loading matrix and nonlinear mappings (panel d).  

\section{Audio Classification of Bird Species\label{sec: bird}}

In monitoring bird migration and population dynamics, machine learning has become popular for species identification based on bird vocalization recordings \citep{tolkova2019feature}. The state of the art is based on deep neural network (DNN) classifiers, which can perform excellently when a large number of labeled audio recordings are available for each species \citep{lehikoinen2023successful}.
 Unfortunately, accurate labels can be difficult to obtain, particularly for rare species. To address the problem of limited labels in DNN classifier training, it is common to augment observed data with fake data \citep{lauha2022domain}. 
For example, one can take the original labeled spectrograms of the bird audio recordings, and for each observed spectrogram generate multiple fake spectrograms that are perturbations. Ideally, these perturbations would mimic real-world variation in bird vocalizations, but in practice this process tends to be quite ad hoc.

We apply NIFTY to a Finland bird species monitoring study \citep{lehikoinen2023successful} with the goals of improving data augmentation for downstream DNN classifiers or alternatively directly using NIFTY for classification based on limited training data. By training NIFTY on available spectrogram data from a species of interest, we can learn a lower-dimensional structure in the data and exploit this structure in generating new spectrograms representative of the variation in calls from a given species. This should provide a more accurate approach to generate new spectrograms that are close to the limited available training data for data augmentation. Current practice adds an arbitrary amount of noise, slightly shifts the spectrogram image, or masks some of the data \citep{salamon2017deep}. 

\begin{figure}[H]
\centering  
	  \begin{subfigure}[t]{0.45\textwidth}
                \centering 
	\includegraphics[width=.7\textwidth]{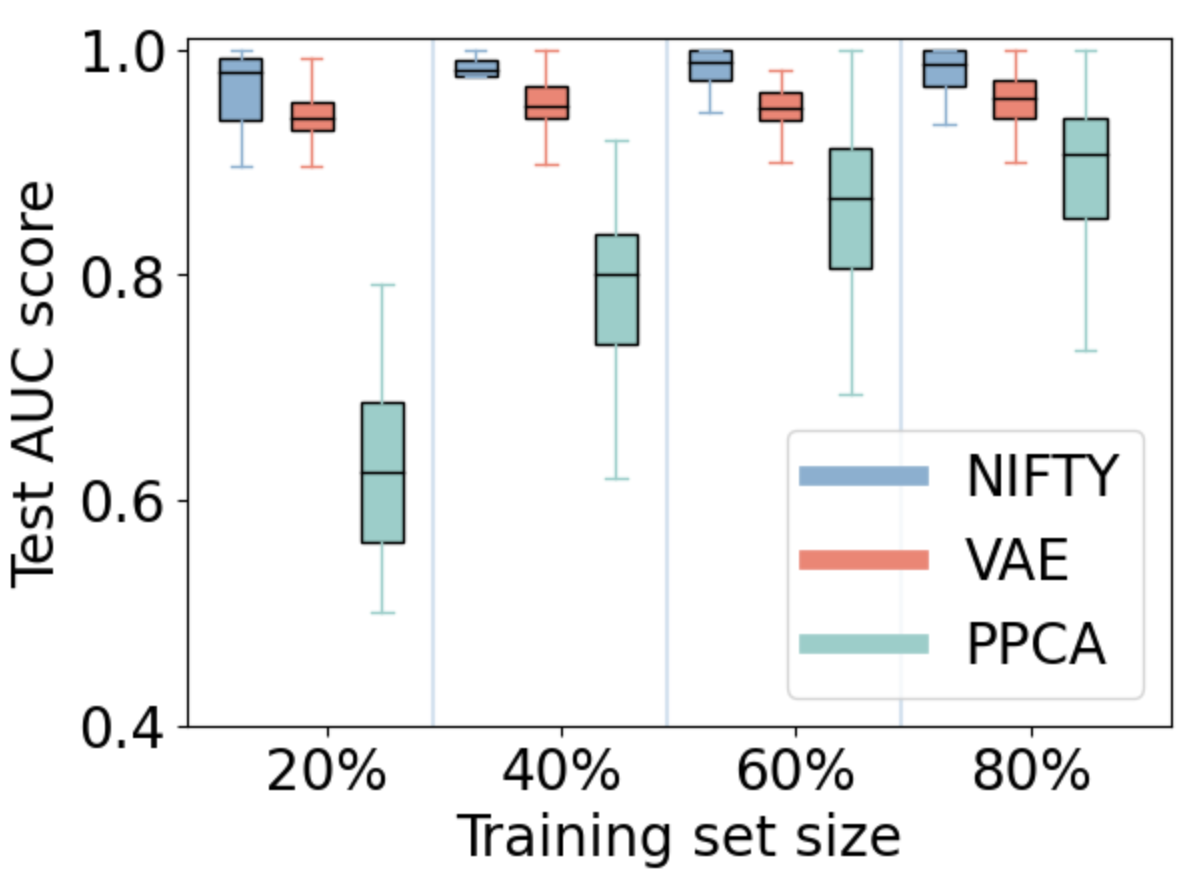}
 \caption{Comparing classification accuracy in terms of AUC using latent factors learned from NIFTY, VAE, and PPCA against using an SVM with the vectorized spectrogram data.}
        \end{subfigure} \quad
	  \begin{subfigure}[t]{0.45\textwidth}
                \centering 
	\includegraphics[width=.95\textwidth]{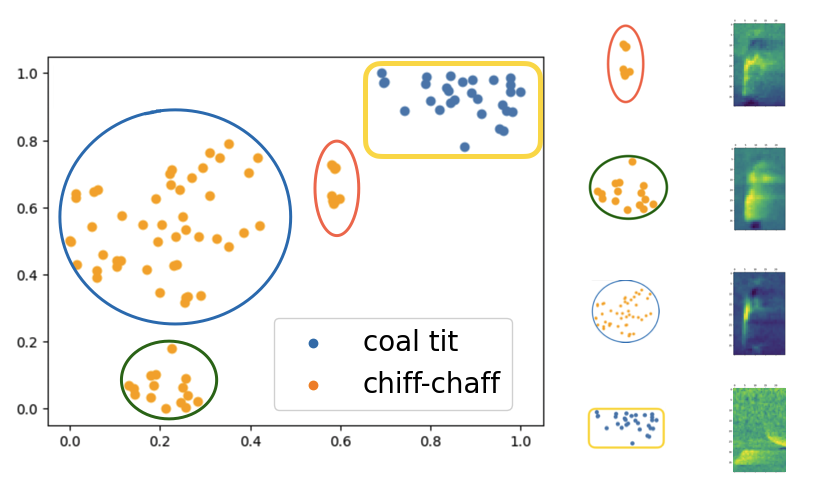}
 \caption{2D visualization from the NIFTY latent locations and a zoom-in view of the sub-clusters within common chiffchaff.}
        \end{subfigure} 
        \\
	\caption{\label{fig: bird_classification}Left: classification accuracy using latent factors learned from NIFTY, VAE, and using an SVM with the vectorized spectrogram data. Test classification accuracy in terms of AUC under the ROC curve is displayed as the training set varies from 20\% to 80\% of the dataset. We have 102 data points with labels, consisting of 73 chiffchaff calls and 29 coal tit calls. The number of latent factors is five for both NIFTY and VAE.  Right: 2D visualization of NIFTY latent locations. Locations are colored according to the true species identity - either chiffchaff or coal tit. There are distinct clusters for the species. For the common chiffchaff, we also see sub-clusters. The zoom-in view of each sub-cluster provides information about different call types within one species.}\end{figure}

\subsection{Data augmentation to increase training samples}

We pre-process the audio files by clipping the songs into syllables according to manual labels, then convert them to spectrograms by applying windowing and the fast Fourier transform (FFT). Each call is treated as an independent data point, and we do not model time-ordering or temporal dynamics across calls. As a result, each clip of data $\bm S_{i}$ is a $M\times T$ spectrogram, with each entry $S_{i[mt]}$ encoding power within the $m$th frequency band at a discrete time point $t$. In our examples, $M=120$ and $T=25$. From Figure \ref{fig: bird_classification}(b), bird calls within a species share similar shapes or vary from several call types. That allows us to represent the $120\times 25$-dimensional spectrogram by mapping from a few factors.

\begin{figure}[H]
\centering 
	\includegraphics[width=.6\linewidth]{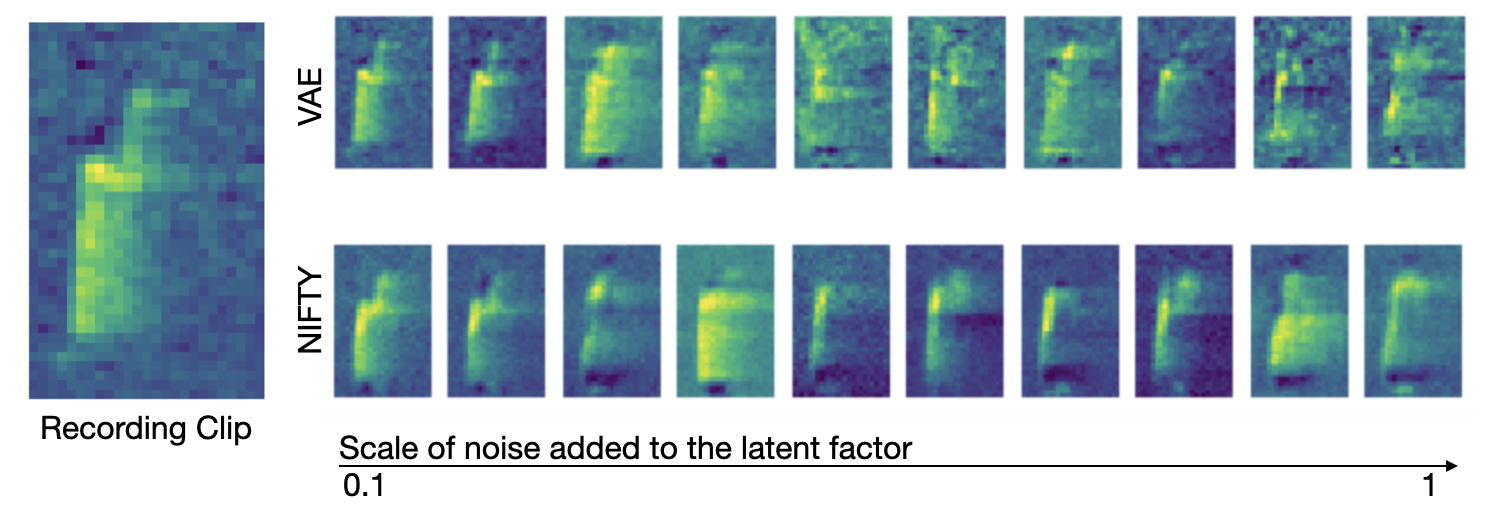}   	\caption{\label{fig: bird_aug}Comparing data generation performance between NIFTY and VAE for the common chiffchaff. VAE's output becomes more noisy as added noise increases, but NIFTY's output remains near the underlying manifold. }
\end{figure}

We apply NIFTY to calls from the common chiffchaff, a migratory warbler that lives in Europe, Asia, and northern Africa. The data include $N=73$ collected samples of spectrograms, and each vectorized spectrogram has $P=3000$. Through pre-training using diffusion maps, we choose five anchor dimensions. After running NIFTY, we can generate new latent locations from Unif$[0,1]^5$ and apply the learned mappings and loading matrix. Figure \ref{fig: bird_aug} shows new generated data around one clip of a call from the common chiffchaff obtained by adding a small random perturbation to the latent location inferred for that clip. One can tune the deviation of generated data from the original data by varying the magnitude of added noise to make it closer or further from a specific bird call. The figure shows what happens to the spectrogram as we increase the amount of perturbation of the latent location for VAE and for NIFTY. We find that NIFTY produces spectrograms that maintain the key attributes of the original bird call, whereas VAEs become degraded by noise. 

We also generate 100 new song clips using NIFTY and VAE, by sampling latent factors from the assumed distribution and applying the transformation. The audio clips, along with the original calls, are available on \url{https://nifty-master.github.io/}. There are occasionally strong noises in the VAE generated calls, and some clips are unlike bird calls (e.g. the call at the 9th second), while NIFTY generates more realistic variations.


\subsection{Classification based on latent factors}
Most existing bird acoustic monitoring frameworks are built on large-scale data and neural network-based classifiers trained on labeled bird song recordings, such as BirdNet \citep{kahl2021birdnet}. 
Such algorithms can be very accurate but require lots of labeled data for each bird species, collected under different conditions in terms of distance from the microphone, background noise, etc. There is a need for accurate classifiers that are less data-hungry, motivating the use of NIFTY in this context.

We demonstrate the potential of NIFTY by focusing on binary classification between the common chiffchaff and the coal tit. The coal tit is a slightly larger bird that inhabits an overlapping range. Letting $y_i\in\{-1,1\}$ denote the label of data $\bm x_i$, we apply a simple support vector machine (SVM) classifier \citep{cortes1995support}, which learns a hyperplane $\bm w\bm x +\bm b=0$ to separate the two species. Formally, we solve an optimization problem $\min_{\bm w,b} \|\bm w\|$ such that $y_i(\bm w\bm x_i +\bm b)>0$ for all $i=1,\ldots, N$. As shown in Figure \ref{fig: bird_classification}(a), fitting such a classifier on high-dimensional data results in low accuracy in predicting the label. Therefore, we used the low-dimensional factors $\bm\eta_i$ learned from NIFTY to fit an SVM instead of $\bm x_i$, which drastically enhances accuracy, especially when the sample size is small. Additionally, we compare with using a VAE to generate low-dimensional features for the classification model. Applying first-stage dimension reduction with either NIFTY or VAE leads to much better performance in small training-size cases. Across the different training sample sizes, NIFTY performs consistently better than VAEs.
The 2D visualization (panel b) of the common chiffchaff vocalizations displays subcluster patterns, which are as expected given the variation in vocalization types. The audio clips of each sub-cluster are also available at \url{https://nifty-master.github.io/}. 

Overall, we find that NIFTY generates more realistic data and improves classification accuracy when the number of labels is limited, bypassing the need for a much more complex and data-hungry DNN.

\section{Discussion} 

The focus of this article is on introducing a relatively simple latent variable modeling framework for identifiable and interpretable dimensionality reduction, with sufficient flexibility to characterize complex data. We illustrate this flexibility in building realistic generative models for bird vocalization data. The proposed framework represents an attractive competitor to popular GP-LVM and VAE approaches. Building on our initial developments for NIFTY, there are several natural next steps.
First, we can accommodate observed data ${\bm x}_i$ that have a variety of measurement scales, including binary, categorical, count, and continuous, through one of two simple modifications. One direction is to define a Gaussian linear factor model for the underlying data ${\bm x}_i^*$ and then let $x_{ij}=h_j(x_{ij}^*)$ with $h_j(\cdot)$ an appropriate link function for the $j$th variable type. Another is to define each $x_{ij}$ as belonging to an exponential family that includes latent factors as predictors in a generalized linear model (GLM). In both cases, it is of interest to consider alternatives to the MCMC-based Bayesian inference approach including pure optimization approaches for rapid dimension reduction and variational approximations to speed up Bayesian analyses.

Another interesting direction forward is to carefully consider the role of pre-training in the practical performance and theoretical properties of NIFTY. While we focused on manifold learning algorithms for pre-training, such as diffusion maps, it would be appealing to be able to adaptively select the type of pre-training most appropriate for the data at hand. For example, some data may have a lower-dimensional manifold structure, while other data may be characterized more appropriately as a stratified space.

One contribution in this paper that is of considerable independent interest beyond the specific factor structure of NIFTY is our highlighting of the important problem of {\em latent posterior shift}. We proposed a particular strategy to deal with this problem in our inferences, but a careful study of when and why such shifts occur and how they can be optimally handled remains an open problem.

There are many additional natural extensions of NIFTY. For example, there has been considerable interest in recent work in using factor models for multi-study \citep{de2021bayesian} and multi-type data. Even in the Gaussian linear factor case, interesting challenges arise in inferring study-specific versus shared factors \citep{chandra2023inferring}. By extending the NIFTY structure to such settings, we enable considerable gains in flexibility over the state of the art, while maintaining identifiability.

\bibliographystyle{chicago}
\bibliography{ref.bib}

\begin{thebibliography}{}

\bibitem[\protect\citeauthoryear{Allman, Matias, and Rhodes}{Allman
  et~al.}{2009}]{allman2009identifiability}
Allman, E.~S., C.~Matias, and J.~a. Rhodes (2009).
\newblock {Identifiability of Parameters in Latent Structure Models with Many
  Observed Variables}.
\newblock {\em The Annals of Statistics\/}~{\em 37}, 3099–3132.

\bibitem[\protect\citeauthoryear{Arminger and Muth{\'e}n}{Arminger and
  Muth{\'e}n}{1998}]{arminger1998bayesian}
Arminger, G. and B.~O. Muth{\'e}n (1998).
\newblock {A Bayesian Approach to Nonlinear Latent Variable Models Using the
  Gibbs Sampler and the Metropolis-hastings Algorithm}.
\newblock {\em Psychometrika\/}~{\em 63}, 271--300.

\bibitem[\protect\citeauthoryear{Arora, Ge, Halpern, Mimno, Moitra, Sontag, Wu,
  and Zhu}{Arora et~al.}{2013}]{arora2013practical}
Arora, S., R.~Ge, Y.~Halpern, D.~Mimno, A.~Moitra, D.~Sontag, Y.~Wu, and M.~Zhu
  (2013).
\newblock {A Practical Algorithm for Topic Modeling with Provable Guarantees}.
\newblock In {\em {International Conference on Machine Learning}}, pp.\
  280--288. PMLR.

\bibitem[\protect\citeauthoryear{Bing, Bunea, Ning, and Wegkamp}{Bing
  et~al.}{2020}]{bing2020adaptive}
Bing, X., F.~Bunea, Y.~Ning, and M.~Wegkamp (2020).
\newblock {Adaptive Estimation in Structured Factor Models with Applications to
  Overlapping Clustering}.
\newblock {\em The Annals of Statistics\/}~{\em 48\/}(4), 2055 -- 2081.

\bibitem[\protect\citeauthoryear{Bonneel, Rabin, Peyr{\'e}, and
  Pfister}{Bonneel et~al.}{2015}]{bonneel2015sliced}
Bonneel, N., J.~Rabin, G.~Peyr{\'e}, and H.~Pfister (2015).
\newblock {Sliced and Radon Wasserstein Barycenters of Measures}.
\newblock {\em Journal of Mathematical Imaging and Vision\/}~{\em 51}, 22--45.

\bibitem[\protect\citeauthoryear{Carvalho, Polson, and Scott}{Carvalho
  et~al.}{2009}]{carvalho2009handling}
Carvalho, C.~M., N.~G. Polson, and J.~G. Scott (2009).
\newblock {Handling Sparsity Via the Horseshoe}.
\newblock In {\em {Artificial Intelligence and Statistics}}, pp.\  73--80.
  PMLR.

\bibitem[\protect\citeauthoryear{Chandra, Canale, and Dunson}{Chandra
  et~al.}{2023}]{chandra2023escaping}
Chandra, N.~K., A.~Canale, and D.~B. Dunson (2023).
\newblock {Escaping the Curse of Dimensionality in Bayesian Model-based
  Clustering.}
\newblock {\em Journal of Machine Learning Research\/}~{\em 24}, 144--1.

\bibitem[\protect\citeauthoryear{Chandra, Dunson, and Xu}{Chandra
  et~al.}{2023}]{chandra2023inferring}
Chandra, N.~K., D.~B. Dunson, and J.~Xu (2023).
\newblock {Inferring Covariance Structure From Multiple Data Sources Via
  Subspace Factor Analysis}.
\newblock {\em Arxiv Preprint Arxiv:2305.04113\/}.

\bibitem[\protect\citeauthoryear{Coifman and Lafon}{Coifman and
  Lafon}{2006}]{coifman2006diffusion}
Coifman, R.~R. and S.~Lafon (2006).
\newblock {Diffusion Maps}.
\newblock {\em Applied and Computational Harmonic Analysis\/}~{\em 21\/}(1),
  5--30.

\bibitem[\protect\citeauthoryear{Comon}{Comon}{1994}]{comon1994independent}
Comon, P. (1994).
\newblock {Independent Component Analysis, a New Concept?}
\newblock {\em Signal Processing\/}~{\em 36\/}(3), 287--314.

\bibitem[\protect\citeauthoryear{Cortes and Vapnik}{Cortes and
  Vapnik}{1995}]{cortes1995support}
Cortes, C. and V.~Vapnik (1995).
\newblock {Support-vector Networks}.
\newblock {\em Machine Learning\/}~{\em 20}, 273--297.

\bibitem[\protect\citeauthoryear{Creswell, White, Dumoulin, Arulkumaran,
  Sengupta, and Bharath}{Creswell et~al.}{2018}]{creswell2018generative}
Creswell, A., T.~White, V.~Dumoulin, K.~Arulkumaran, B.~Sengupta, and A.~a.
  Bharath (2018).
\newblock {Generative Adversarial Networks: an Overview}.
\newblock {\em IEEE Signal Processing Magazine\/}~{\em 35\/}(1), 53--65.

\bibitem[\protect\citeauthoryear{De~Vito, Bellio, Trippa, and
  Parmigiani}{De~Vito et~al.}{2021}]{de2021bayesian}
De~Vito, R., R.~Bellio, L.~Trippa, and G.~Parmigiani (2021).
\newblock {Bayesian Multistudy Factor Analysis for High-throughput Biological
  Data}.
\newblock {\em The Annals of Applied Statistics\/}~{\em 15\/}(4), 1723--1741.

\bibitem[\protect\citeauthoryear{Duan, Young, Nishimura, and Dunson}{Duan
  et~al.}{2020}]{duan2020bayesian}
Duan, L.~L., A.~L. Young, A.~Nishimura, and D.~B. Dunson (2020).
\newblock {Bayesian Constraint Relaxation}.
\newblock {\em Biometrika\/}~{\em 107\/}(1), 191--204.

\bibitem[\protect\citeauthoryear{Dunson and Wu}{Dunson and
  Wu}{2021}]{dunson2021inferring}
Dunson, D.~B. and N.~Wu (2021).
\newblock {Inferring Manifolds From Noisy Data Using Gaussian Processes}.
\newblock {\em Arxiv Preprint Arxiv:2110.07478\/}.

\bibitem[\protect\citeauthoryear{El~Karoui and Wu}{El~Karoui and
  Wu}{2016}]{el2016graph}
El~Karoui, N. and H.-t. Wu (2016).
\newblock {Graph Connection Laplacian Methods Can Be Made Robust to Noise}.

\bibitem[\protect\citeauthoryear{Fruchter}{Fruchter}{1954}]{fruchter1954introduction}
Fruchter, B. (1954).
\newblock {\em {Introduction to Factor Analysis.}}
\newblock {Van Nostrand Series in Psychology}.

\bibitem[\protect\citeauthoryear{Fr{\"u}hwirth-Schnatter, Hosszejni, and
  Lopes}{Fr{\"u}hwirth-Schnatter et~al.}{2024}]{fruhwirth2024sparse}
Fr{\"u}hwirth-Schnatter, S., D.~Hosszejni, and H.~F. Lopes (2024).
\newblock {Sparse Bayesian Factor Analysis When the Number of Factors Is
  Unknown}.
\newblock {\em Bayesian Analysis\/}~{\em 1\/}(1), 1--31.

\bibitem[\protect\citeauthoryear{Ghahramani, Hinton, and Others}{Ghahramani
  et~al.}{1996}]{ghahramani1996algorithm}
Ghahramani, Z., G.~E. Hinton, and Others (1996).
\newblock {The EM Algorithm for Mixtures of Factor Analyzers}.
\newblock Technical report, Technical Report CRG-TR-96-1, University of
  Toronto.

\bibitem[\protect\citeauthoryear{Ghosal, Ghosh, and Ramamoorthi}{Ghosal
  et~al.}{1999}]{ghosal1999posterior}
Ghosal, S., J.~K. Ghosh, and R.~Ramamoorthi (1999).
\newblock {Posterior Consistency of Dirichlet Mixtures in Density Estimation}.
\newblock {\em The Annals of Statistics\/}~{\em 27\/}(1), 143--158.

\bibitem[\protect\citeauthoryear{Ghosal and Van Der~Vaart}{Ghosal and Van
  Der~Vaart}{2017}]{ghosal2017fundamentals}
Ghosal, S. and A.~Van Der~Vaart (2017).
\newblock {\em {Fundamentals of Nonparametric Bayesian Inference}}, Volume~44.
\newblock Cambridge University Press.

\bibitem[\protect\citeauthoryear{Harman}{Harman}{1976}]{harman1976modern}
Harman, H.~H. (1976).
\newblock {\em {Modern Factor Analysis}}.
\newblock University of Chicago Press.

\bibitem[\protect\citeauthoryear{Ho, Jain, and Abbeel}{Ho
  et~al.}{2020}]{ho2020denoising}
Ho, J., A.~Jain, and P.~Abbeel (2020).
\newblock {Denoising Diffusion Probabilistic Models}.
\newblock {\em Advances in Neural Information Processing Systems\/}~{\em 33},
  6840--6851.

\bibitem[\protect\citeauthoryear{Hoffman and Johnson}{Hoffman and
  Johnson}{2016}]{hoffman2016elbo}
Hoffman, M.~D. and M.~J. Johnson (2016).
\newblock {ELBO Surgery: yet Another Way to Carve up the Variational Evidence
  Lower Bound}.
\newblock In {\em {Workshop in Advances in Approximate Bayesian Inference,
  Nips}}, Volume~1.

\bibitem[\protect\citeauthoryear{J{\"o}reskog}{J{\"o}reskog}{1967}]{joreskog1967some}
J{\"o}reskog, K.~G. (1967).
\newblock {Some Contributions to Maximum Likelihood Factor Analysis}.
\newblock {\em Psychometrika\/}~{\em 32\/}(4), 443--482.

\bibitem[\protect\citeauthoryear{Kahl, Wood, Eibl, and Klinck}{Kahl
  et~al.}{2021}]{kahl2021birdnet}
Kahl, S., C.~M. Wood, M.~Eibl, and H.~Klinck (2021).
\newblock {Birdnet: a Deep Learning Solution for Avian Diversity Monitoring}.
\newblock {\em Ecological Informatics\/}~{\em 61}, 101236.

\bibitem[\protect\citeauthoryear{Kim and Mnih}{Kim and
  Mnih}{2018}]{kim2018disentangling}
Kim, H. and A.~Mnih (2018).
\newblock {Disentangling by Factorising}.
\newblock In {\em {International Conference on Machine Learning}}, pp.\
  2649--2658. PMLR.

\bibitem[\protect\citeauthoryear{Kingma and Welling}{Kingma and
  Welling}{2014}]{kingma2014stochastic}
Kingma, D.~P. and M.~Welling (2014).
\newblock {Stochastic Gradient VB and the Variational Auto-encoder}.
\newblock In {\em {Second International Conference on Learning Representations,
  Iclr}}, Volume~19, pp.\  121.

\bibitem[\protect\citeauthoryear{Kobyzev, Prince, and Brubaker}{Kobyzev
  et~al.}{2020}]{kobyzev2020normalizing}
Kobyzev, I., S.~J. Prince, and M.~a. Brubaker (2020).
\newblock {Normalizing Flows: an Introduction and Review of Current Methods}.
\newblock {\em IEEE Transactions on Pattern Analysis and Machine
  Intelligence\/}~{\em 43\/}(11), 3964--3979.

\bibitem[\protect\citeauthoryear{Kolouri, Park, Thorpe, Slepcev, and
  Rohde}{Kolouri et~al.}{2017}]{kolouri2017optimal}
Kolouri, S., S.~R. Park, M.~Thorpe, D.~Slepcev, and G.~K. Rohde (2017).
\newblock {Optimal Mass Transport: Signal Processing and Machine-learning
  Applications}.
\newblock {\em IEEE Signal Processing Magazine\/}~{\em 34\/}(4), 43--59.

\bibitem[\protect\citeauthoryear{Kundu and Dunson}{Kundu and
  Dunson}{2014}]{kundu2014latent}
Kundu, S. and D.~B. Dunson (2014).
\newblock {Latent Factor Models for Density Estimation}.
\newblock {\em Biometrika\/}~{\em 101\/}(3), 641--654.

\bibitem[\protect\citeauthoryear{Lauha, Somervuo, Lehikoinen, Geres, Richter,
  Seibold, and Ovaskainen}{Lauha et~al.}{2022}]{lauha2022domain}
Lauha, P., P.~Somervuo, P.~Lehikoinen, L.~Geres, T.~Richter, S.~Seibold, and
  O.~Ovaskainen (2022).
\newblock {Domain-specific Neural Networks Improve Automated Bird Sound
  Recognition Already with Small Amount of Local Data}.
\newblock {\em Methods in Ecology and Evolution\/}~{\em 13\/}(12), 2799--2810.

\bibitem[\protect\citeauthoryear{Legramanti, Durante, and Dunson}{Legramanti
  et~al.}{2020}]{legramanti2020bayesian}
Legramanti, S., D.~Durante, and D.~B. Dunson (2020).
\newblock {Bayesian Cumulative Shrinkage for Infinite Factorizations}.
\newblock {\em Biometrika\/}~{\em 107\/}(3), 745--752.

\bibitem[\protect\citeauthoryear{Lehikoinen, Rannisto, Camargo, Aintila, Lauha,
  Piirainen, Somervuo, and Ovaskainen}{Lehikoinen
  et~al.}{2023}]{lehikoinen2023successful}
Lehikoinen, P., M.~Rannisto, U.~Camargo, A.~Aintila, P.~Lauha, E.~Piirainen,
  P.~Somervuo, and O.~Ovaskainen (2023).
\newblock {A Successful Crowdsourcing Approach for Bird Sound Classification}.
\newblock {\em Citizen Science\/}~{\em 8\/}(1).

\bibitem[\protect\citeauthoryear{Li and Chen}{Li and Chen}{2016}]{li2016review}
Li, P. and S.~Chen (2016).
\newblock {A Review on Gaussian Process Latent Variable Models}.
\newblock {\em Caai Transactions on Intelligence Technology\/}~{\em 1\/}(4),
  366--376.

\bibitem[\protect\citeauthoryear{Makalic and Schmidt}{Makalic and
  Schmidt}{2015}]{makalic2015simple}
Makalic, E. and D.~F. Schmidt (2015).
\newblock {A Simple Sampler for the Horseshoe Estimator}.
\newblock {\em IEEE Signal Processing Letters\/}~{\em 23\/}(1), 179--182.

\bibitem[\protect\citeauthoryear{Martin and Mcdonald}{Martin and
  Mcdonald}{1975}]{martin1975bayesian}
Martin, J.~K. and R.~P. Mcdonald (1975).
\newblock {Bayesian Estimation in Unrestricted Factor Analysis: a Treatment for
  Heywood Cases}.
\newblock {\em Psychometrika\/}~{\em 40\/}(4), 505--517.

\bibitem[\protect\citeauthoryear{Mclachlan, Peel, and Bean}{Mclachlan
  et~al.}{2003}]{mclachlan2003modelling}
Mclachlan, G.~J., D.~Peel, and R.~W. Bean (2003).
\newblock {Modelling High-dimensional Data by Mixtures of Factor Analyzers}.
\newblock {\em Computational Statistics \& Data Analysis\/}~{\em 41\/}(3-4),
  379--388.

\bibitem[\protect\citeauthoryear{Moran, Sridhar, Wang, and Blei}{Moran
  et~al.}{2022}]{moran2022identifiable}
Moran, G.~E., D.~Sridhar, Y.~Wang, and D.~Blei (2022).
\newblock {Identifiable Deep Generative Models Via Sparse Decoding}.
\newblock {\em Transactions on Machine Learning Research\/}~(2835-8856).

\bibitem[\protect\citeauthoryear{Murray, Dunson, Carin, and Lucas}{Murray
  et~al.}{2013}]{murray2013bayesian}
Murray, J.~S., D.~B. Dunson, L.~Carin, and J.~E. Lucas (2013).
\newblock {Bayesian Gaussian Copula Factor Models for Mixed Data}.
\newblock {\em Journal of the American Statistical Association\/}~{\em
  108\/}(502), 656--665.

\bibitem[\protect\citeauthoryear{Papastamoulis}{Papastamoulis}{2020}]{papastamoulis2020clustering}
Papastamoulis, P. (2020).
\newblock {Clustering Multivariate Data Using Factor Analytic Bayesian Mixtures
  with an Unknown Number of Components}.
\newblock {\em Statistics and Computing\/}~{\em 30\/}(3), 485--506.

\bibitem[\protect\citeauthoryear{Poworoznek, Ferrari, and Dunson}{Poworoznek
  et~al.}{2021}]{poworoznek2021efficiently}
Poworoznek, E., F.~Ferrari, and D.~Dunson (2021).
\newblock {Efficiently Resolving Rotational Ambiguity in Bayesian Matrix
  Sampling with Matching}.
\newblock {\em Arxiv Preprint Arxiv:2107.13783\/}.

\bibitem[\protect\citeauthoryear{Rezende, Mohamed, and Wierstra}{Rezende
  et~al.}{2014}]{rezende2014stochastic}
Rezende, D.~J., S.~Mohamed, and D.~Wierstra (2014).
\newblock {Stochastic Backpropagation and Variational Inference in Deep Latent
  Gaussian Models}.
\newblock In {\em {International Conference on Machine Learning}}, Volume~2,
  pp.\ ~2.

\bibitem[\protect\citeauthoryear{Roberts and Stramer}{Roberts and
  Stramer}{2002}]{roberts2002langevin}
Roberts, G.~O. and O.~Stramer (2002).
\newblock {Langevin Diffusions and Metropolis-hastings Algorithms}.
\newblock {\em Methodology and Computing in Applied Probability\/}~{\em 4},
  337--357.

\bibitem[\protect\citeauthoryear{Roberts and Tweedie}{Roberts and
  Tweedie}{1996}]{roberts1996exponential}
Roberts, G.~O. and R.~L. Tweedie (1996).
\newblock {Exponential Convergence of Langevin Distributions and Their Discrete
  Approximations}.
\newblock {\em Bernoulli\/}, 341--363.

\bibitem[\protect\citeauthoryear{Salamon and Bello}{Salamon and
  Bello}{2017}]{salamon2017deep}
Salamon, J. and J.~P. Bello (2017).
\newblock {Deep Convolutional Neural Networks and Data Augmentation for
  Environmental Sound Classification}.
\newblock {\em IEEE Signal Processing Letters\/}~{\em 24\/}(3), 279--283.

\bibitem[\protect\citeauthoryear{Shan and Daubechies}{Shan and
  Daubechies}{2022}]{shan2022diffusion}
Shan, S. and I.~Daubechies (2022).
\newblock {Diffusion Maps: Using the Semigroup Property for Parameter Tuning}.
\newblock In {\em {Theoretical Physics, Wavelets, Analysis, Genomics: an
  Indisciplinary Tribute to Alex Grossmann}}, pp.\  409--424. Springer.

\bibitem[\protect\citeauthoryear{Shen and Wu}{Shen and
  Wu}{2022}]{shen2022scalability}
Shen, C. and H.-t. Wu (2022).
\newblock {Scalability and Robustness of Spectral Embedding: Landmark Diffusion
  Is All You Need}.
\newblock {\em Information and Inference: a Journal of the Ima\/}~{\em
  11\/}(4), 1527--1595.

\bibitem[\protect\citeauthoryear{Titsias and Lawrence}{Titsias and
  Lawrence}{2010}]{titsias2010bayesian}
Titsias, M. and N.~D. Lawrence (2010).
\newblock {Bayesian Gaussian Process Latent Variable Model}.
\newblock In {\em {Proceedings of the Thirteenth International Conference on
  Artificial Intelligence and Statistics}}, pp.\  844--851. JMLR Workshop and
  Conference Proceedings.

\bibitem[\protect\citeauthoryear{Tolkova}{Tolkova}{2019}]{tolkova2019feature}
Tolkova, I. (2019).
\newblock {Feature Representations for Conservation Bioacoustics: Review and
  Discussion}.
\newblock {\em Harvard University\/}.

\bibitem[\protect\citeauthoryear{Tomczak and Welling}{Tomczak and
  Welling}{2018}]{tomczak2018vae}
Tomczak, J. and M.~Welling (2018).
\newblock {VAE with a Vampprior}.
\newblock In {\em {International Conference on Artificial Intelligence and
  Statistics}}, pp.\  1214--1223. PMLR.

\bibitem[\protect\citeauthoryear{Van Der~Maaten and Hinton}{Van Der~Maaten and
  Hinton}{2008}]{van2008visualizing}
Van Der~Maaten, L. and G.~Hinton (2008).
\newblock {Visualizing Data Using T-sne.}
\newblock {\em Journal of Machine Learning Research\/}~{\em 9\/}(11).

\bibitem[\protect\citeauthoryear{Wang, Blei, and Cunningham}{Wang
  et~al.}{2021}]{wang2021posterior}
Wang, Y., D.~Blei, and J.~P. Cunningham (2021).
\newblock {Posterior Collapse and Latent Variable Non-identifiability}.
\newblock {\em Advances in Neural Information Processing Systems\/}~{\em 34},
  5443--5455.

\bibitem[\protect\citeauthoryear{Wellner and Others}{Wellner and
  Others}{2013}]{wellner2013weak}
Wellner, J. and Others (2013).
\newblock {\em {Weak Convergence and Empirical Processes: with Applications to
  Statistics}}.
\newblock Springer Science \& Business Media.

\bibitem[\protect\citeauthoryear{Xu, Schmidt, Makalic, Qian, and Hopper}{Xu
  et~al.}{2016}]{xu2016bayesian}
Xu, Z., D.~F. Schmidt, E.~Makalic, G.~Qian, and J.~L. Hopper (2016).
\newblock {Bayesian Grouped Horseshoe Regression with Application to Additive
  Models}.
\newblock In {\em {AI 2016: Advances in Artificial Intelligence: 29th
  Australasian Joint Conference, Hobart, Tas, Australia, December 5-8, 2016,
  Proceedings 29}}, pp.\  229--240. Springer.

\bibitem[\protect\citeauthoryear{Yalcin and Amemiya}{Yalcin and
  Amemiya}{2001}]{yalcin2001nonlinear}
Yalcin, I. and Y.~Amemiya (2001).
\newblock {Nonlinear Factor Analysis as a Statistical Method}.
\newblock {\em Statistical Science\/}, 275--294.

\bibitem[\protect\citeauthoryear{Zheng, Ng, and Zhang}{Zheng
  et~al.}{2022}]{zheng2022identifiability}
Zheng, Y., I.~Ng, and K.~Zhang (2022).
\newblock {On the Identifiability of Nonlinear ICA: Sparsity and Beyond}.
\newblock {\em Advances in Neural Information Processing Systems\/}~{\em 35},
  16411--16422.

\end{thebibliography}
\appendix  
\section{Proofs}
\subsection*{Proof of Proposition \ref{proposition: CoRe}} 
\begin{proof}
Let $F_k^0(u)$ denote the CDF of the prior for $u_{ik}$ defined in \eqref{eq: prior_u} and $F_k^N(u)$ denote the corresponding empirical CDF with $N$ data points. 
	By the definition of a Wasserstein-2 distance, when $\nu\to\infty$, 
	 the $i$th order statistics of $\bm u_{\cdot k}$ converge in probability to $i/N$. Hence for any $\epsilon>0$ and any $u\in[0,1]$, there exists an $M$ large enough, such that
	 $$\text{pr}(|F_k^M(u)-u|\ge \epsilon)=\text{pr}(|i/M-u|\ge \epsilon)\to 0, \text{ as } \nu\to\infty,$$
with $i$ chosen such that 
$u_{(i-1)k} < u < u_{(i)k}$ with 
$u_{(1)k},\ldots,u_{(M)k}$ the order statistics of 
$u_{1k},\ldots,u_{Mk}$.
	 As $N\to\infty$, the distance between empirical CDF and true CDF converges to zero. Therefore, under prior specification \eqref{eq: prior_u},
	 $$F_k^0(u) \to u.$$ 
\end{proof}

\subsection*{Proof of Theorem \ref{thm: main_identify}}
\begin{proof} 
We prove the identifiability results in three parts: first the strict identifiability of latent locations $u_{ik}$s, then the strict identifiability of $\bm\Sigma$, and finally the generic identifiability of $\bm\Lambda$ and $\bm g$.

\subsubsection*{Identifiability of latent locations}
Under Assumption \ref{assp: sigma}, the value of $\sigma_{j_1}^2,\ldots,\sigma_{j_K}^2$ are known or estimated in advance. In  Assumption \eqref{assp: anchor}, 
the $k$th anchor dimension depends only the factor $h_k$.
Identifiability of $u_{ik}$ can be described as:  for each $k=1,\ldots,K$, if two sets of parameters $(u_{ik}, g_{h_k}, \lambda_{jh_k})$ and $(u_{ik}',g_{h_k}',\lambda_{jh_k}')$ yield the same likelihood for $x_{ij_k}$, then we must have $u_{ik} = u_{ik}'$ or $u_{ik} = 1 - u_{ik}'$. Further, there is a constant $c$ such that $g_{h_k}(u) = c g_{h_k}'(u)$ for any $u\in[0,1]$ and $\lambda_{jh_k} = 1/c \lambda_{jh_k}'$. 

Since the log-likelihood of the two sets of parameters should be equal, we have
$$ [x_{ij_k} -  \lambda_{jh_k}g_{hk}(u_{ik})]^2/\sigma^2_{j_k} = [x_{ij_k} -  \lambda_{jh_k}'g_{hk}'(u_{ik}')]^2/\sigma^2_{j_k}.$$ 
Furthermore, $$\lambda_{jh_k}^2g_{h_k}(u_{ik})^2-(\lambda_{jh_k}')^2g'_{h_k}(u_{ih_k}')^2 -  x_{ij_k} [\lambda_{jh_k}g_{h_k}(u_{ik})-\lambda_{jh_k}'g'_{hk}(u_{ik}')] = 0$$
 holds for any value of $x_{ij_k}$, and therefore we must have  $\lambda_{jh_k}g_{h_k}(u_{ik})=\lambda_{jh_k}'g'_{hk}(u_{ik}')$.
Since $g_h$ are monotonely increasing functions from $[0,1]$ to $\mathbb R$, the inverse exists and $$u_{ik}=g_{h_k}^{-1}[\lambda_{jh_k}^{-1}\lambda_{jh_k}'g'_{h_k}(u'_{ik})].$$  
If $\lambda_{jh_k}\lambda_{jh_k}'>0$, since both $g^{-1}_h$ and $g'_h$ are monotonely non-decreasing functions for every $h$, $g^{-1}_{h_k}\circ \lambda_{jh_k}^{-1}\lambda_{jh_k}'g'_{h_k}$ is also a non-decreasing function. Actually, it is an identity function, because
$$\begin{aligned}
	t = \text{pr}(u_{ik}\le t) &=\text{pr}(g_{h_k}^{-1}[\lambda_{jh_k}^{-1}\lambda_{jh_k}'g'_{h_k}(u'_{ik})]\le t) \\
	& = \text{pr}(u'_{ik}\le (g^{-1}_{h_k}\circ \lambda_{jh_k}^{-1}\lambda_{jh_k}'g'_{h_k})^{-1}(t))=(g^{-1}_{h_k}\circ \lambda_{jh_k}^{-1}\lambda_{jh_k}'g'_{h_k})^{-1}(t).
\end{aligned}$$
Similarly, when $\lambda_{jh_k}\lambda_{jh_k}'<0$, we have  $u_{ik}=1-u_{ik}'$. Furthermore, we show that $g^{-1}_{h_k}\circ \lambda_{jh_k}^{-1}\lambda_{jh_k}'g'_{h_k}$ is an identity function or  $-g^{-1}_{h_k}\circ \lambda_{jh_k}^{-1}\lambda_{jh_k}'g'_{h_k}$ is an identity function.

\subsubsection*{Identifiability of residual covariance}
As $\Sigma$ is diagonal, it suffices to show identifiability of the diagonal elements $\sigma_j^2$. For ease of notation, we denote the $j$th element in $\bm\Lambda\bm\eta_i$ being a map from $\bm u_i$ as $m(\bm u_i)$. Since we have shown identifiability of $\bm u_i$, it suffices to show: if there exists $m, m'$ and $\sigma_j^2, (\sigma_j')^2$ such that
\[\frac{[x_{ij} - m(\bm u_i)]^2}{\sigma_j^2} =  \frac{[x_{ij} - m'(\bm u_i)]^2}{(\sigma_j')^2}, \forall x_{ij},\] 
then one must have 
$x_{ij} = \left(\frac{1}{\sigma_j^2}-\frac{1}{(\sigma_j')^2}\right)^{-1}\left(\frac{m(\bm u_i)}{\sigma_j^2}-\frac{m(\bm u_i)}{(\sigma_j')^2}\right)\ \text{or}\ x_{ij} = \left(\frac{1}{\sigma_j^2}+\frac{1}{(\sigma_j')^2}\right)^{-1}\left(\frac{m(\bm u_i)}{\sigma_j^2}+\frac{m(\bm u_i)}{(\sigma_j')^2}\right).$ With $\sigma_j^2\neq (\sigma_j')^2$, either equation will give a contradiction because $\bm u_i$ is a latent location vector specified by the anchor dimensions, but $x_{ij}$ can take any value. Therefore, we must have $\sigma_j^2 = (\sigma_j')^2$.

\subsubsection*{Identifiability of loadings and factors}
Without loss of generality, we assume that $\lambda_{jh_k}\lambda_{jh_k}'>0$ and hence $u_{ik}$ are identifiable for all $k$. Supposing there is another set of parameters $(\bm\Gamma,\bm e)$ yielding the same likelihood as $(\bm\Lambda,\bm g)$, we will show that there is a linear transformation between $\bm\Gamma$ and $\bm\Lambda$ within the $k$th block. 

We first show there exists a reversible linear transformation $\bm T\in\mathbb R^{H\times H}$, such that $\bm\Gamma = \bm\Lambda \bm T$ and $\bm e(\bm u) = T^{-1}\bm g(\bm u)$ for any value of $\bm u\in[0,1]^K$. 
Recall that in our model, columns in $\bm\Gamma$ and functions in $\bm e$ can be rearranged so that they can be written in partition $\bm\Gamma^1,\ldots,\bm\Gamma^K$ and $\bm e^1,\ldots,\bm e^K$ using the same criteria when partitioning $\bm\Lambda$ and $\bm g$. Since $u_{ik}$ are uniquely determined by the anchor dimension, we have 
	$$[\bm\Gamma^1 \bm e^1(u_{i1}),\ldots,\bm\Gamma^K \bm e^K(u_{iK})] = [\bm\Lambda^1 \bm g^1(u_{i1}),\ldots,\bm\Lambda^K \bm g^K(u_{iK})],$$ or omit some subscripts, $\bm\Gamma\bm e(\bm u) = \bm\Lambda\bm g(\bm u).$ Since both matrices are of full column rank, there exists a $\bm B$ matrix to make $\bm B\bm\Lambda = \bm I$ and therefore $\bm g(\bm u)=\bm B\bm\Gamma\bm e(\bm u)$ for every $\bm u$.  Let $T = \bm B\bm\Lambda$ denote the transformation between $\bm g$ and $\bm e$, then we also have $\bm\Gamma=\bm\Lambda \bm T$. 

 Next, we prove that $\bm T$ is a block diagonal matrix, with blocks divided by the $K$ partitions. Note that 
	$\bm T [e_1(u_{ik_1}), \ldots,e_H(u_{ik_H})]^T = [g_1(u_{ik_1}), \ldots,g_H(u_{ik_H})]^T.$   If $\bm T$ is not a block diagonal matrix, assume that $\bm T_{h_1h_2} \neq 0$ and that $k_{h_1} = 1$, $k_{h_2} = 2$. Then $g_{h_1}(u_{i1})$ must be a linear combination of $e_{h_1}(u_{i1})$ and $e_{h_2}(u_{i2})$, for any $u_{i1}$ and $u_{i2}$, which violates the definition of $g_{h_1}$. Therefore $T_{h_1h_2}=0$ if $k_{h_1} \neq  k_{h_2}$; hence $\bm T$ is a block diagonal matrix. 
\end{proof}

\subsection*{Proof of Theorem \ref{thm: consistency_parameters}}
\begin{proof}
	Let $L_{\Theta^0}^N(\bm x)$ denote the likelihood of $(\bm x_1,\ldots,\bm x_N)$ under parameters $\Theta^0 = \Lambda^0,g^0,\Sigma^0$, and $L_{\Theta}^N(\bm x)$ be the likelihood under parameters $\Theta = \Lambda,g,\Sigma$.  By generic identifiability, the subset of $\Theta$ satisfying $L_{\Theta}^N(\bm x)=L_{\Theta^0}^N(\bm x)$ has measure zero. Therefore, for any $\epsilon>0$, there must exist a $\delta_{\epsilon,N}>0$ such  that the prior $\Pi$ for $\Theta$ satisfies 
	$$\lim_{N \to \infty}\Pi\left[\mathcal N^C_\epsilon(\Theta^0)\cap \{|L_{\Theta^0}^N(\bm x) - L_{\Theta}^N(\bm x)|<\delta_{\epsilon,N}\}\right]= 0.$$
	We can define tests $\psi^N:= 1\{|L_{\Theta^0}^N(\bm x) - L_{\Theta}^N(\bm x)|\ge \delta_{\epsilon,N}\}$ with $\mathbb P_0^N\psi^N\to 0$ and $\sup_{\Theta^0\in\mathcal N^C_\epsilon(\Theta^0)}\mathbb P_\Theta^N(1-\psi^N)\to 0$. Applying Schwartz's theorem (theorem 6.16 in \cite{ghosal2017fundamentals}), the posterior of the model parameters is strongly consistent at $\Theta^0$.
\end{proof} 
\subsection*{Proof of Theorem \ref{thm: consistency}}
\begin{proof}
Let $f_{{\bm\Lambda},\bm g,{\bm\Sigma}}$ denote the induced density of $\bm x_i$ given $({\bm\Lambda}, \bm g, {\bm\Sigma})$ and $f^0$ the true data-generating density having parameters $({\bm\Lambda^0}, \bm g^0, {\bm\Sigma}^0)$.
	By definition, the KL divergence between $f_{{\bm\Lambda},\bm g,{\bm\Sigma}},f^0$ is
	\begin{equation}\label{step: KL}
		KL(f_{{\bm\Lambda},\bm g,{\bm\Sigma}},f^0)=\int f^0(\bm x)\log\frac{f^0(\bm x)}{f_{{\bm\Lambda},\bm g,{\bm\Sigma}}(\bm x)}d\bm x.
	\end{equation}

Suppose $\bm\Sigma^0=\text{diag}[(\sigma^0_1)^2,\ldots,(\sigma^0_P)^2]$ and $\bm\Sigma=\text{diag}(\sigma_1^2,\ldots,\sigma_P^2)$. Then, we introduce diagonal matrix $\bm \Gamma = \text{diag}\left[\frac{\sigma_1^2}{(\sigma^0_1)^2},\ldots,\frac{\sigma_P^2}{(\sigma^0_P)^2}\right]$, and represent $\bm\Sigma = \bm \Gamma^{1/2} \bm\Sigma^0 \bm \Gamma^{1/2}.$
We have 
\begin{equation*}
	\begin{aligned}
		\frac{f_{{\bm\Lambda}^0,\bm g^0,{\bm\Sigma^0}}(\bm x_i)}{f_{{\bm\Lambda}^0,\bm g^0,{\bm\Sigma}}(\bm x_i)}&= \frac{ \int_{[0,1]^H}\phi_{\bm\Sigma^0}[\bm x_i-{\bm\Lambda}^0\bm g^0(\bm u_i)]d\bm u_i}{ \int_{[0,1]^H}\phi_{\bm\Sigma}[\bm x_i-{\bm\Lambda}^0\bm g^0(\bm u_i)]d\bm u_i} \\
		&= \exp\left\{\frac{1}{2}(\bm x_i\bm\Gamma^{-1/2}-\bm x_i)^T(\bm\Sigma^0)^{-1}(\bm x_i\bm\Gamma^{-1/2}-\bm x_i)\right\},\\
	\end{aligned}
\end{equation*}
which goes to 1 as $\|\bm\Sigma-\bm\Sigma^0\|_\infty\to 0$. On the other hand, 
	\begin{equation*}
	\begin{aligned}
		\frac{f_{{\bm\Lambda}^0,\bm g^0,{\bm\Sigma}}(\bm x_i)}{f_{{\bm\Lambda},\bm g,{\bm\Sigma}}(\bm x_i)}&=\frac{ \int_{[0,1]^H}\phi_{\bm\Sigma}[\bm x_i-{\bm\Lambda}^0\bm g^0(\bm u_i)]d\bm u_i}{\int_{[0,1]^H}\phi_{\bm\Sigma}[\bm x_i-{\bm\Lambda}\bm g(\bm u_i)]d\bm u_i}\\
		&\le \sup_{\bm u_i\in[0,1]^H}\exp\left\{\frac{1}{2}[{\bm\Lambda}\bm g(\bm u_i)-{\bm\Lambda}^0\bm g^0(\bm u_i)]^T\bm \Sigma^{-1}[{\bm\Lambda}\bm g(\bm u_i)-{\bm\Lambda}^0\bm g^0(\bm u_i)]\right.\\
		&\left.\quad -\frac{1}{2}[\bm x - {\bm\Lambda}^0\bm g^0(\bm u_i)]^T\bm\Sigma^{-1}[{\bm\Lambda}\bm g(\bm u_i)-{\bm\Lambda}^0\bm g^0(\bm u_i)]\right\},
	\end{aligned}
	\end{equation*}
	which goes to 1 as $\|{\bm\Lambda}^0\bm g^0-{\bm\Lambda}\bm g\|_\infty\to 0.$ 
	With similar derivation on $\frac{f_{{\bm\Lambda},\bm g,{\bm\Sigma}}(\bm x)}{f_{{\bm\Lambda}^0,\bm g^0,{\bm\Sigma}}(\bm x)}$, we have $\log \frac{f^0(x)}{f_{{\bm\Lambda},\bm g,{\bm\Sigma}}(\bm x)}\to 0$ as $\|{\bm\Lambda}^0\bm g^0-{\bm\Lambda}\bm g\|_\infty \to 0.$
		Under Assumption \ref{assp: truedensity},
		$\frac{f^0(\bm x)}{f_{{\bm\Lambda},\bm g,{\bm\Sigma}}(\bm x)}$ has a finite upper bound. Therefore we can apply the dominated convergence theorem, and obtain $$KL(f_{{\bm\Lambda},\bm g,{\bm\Sigma}},f^0)=\int f^0(\bm x)\log\frac{f^0(\bm x)}{f_{{\bm\Lambda},\bm g,{\bm\Sigma}}(\bm x)}d\bm x\to 0$$ as $\|{\bm\Lambda}^0\bm g^0-{\bm\Lambda}\bm g\|_\infty\to 0$ and $\|\bm\Sigma-\bm\Sigma^0\|_\infty\to 0$.
		
		 Therefore, for an arbitrary $\epsilon>0$ bounding the KL divergence from $f^0$, we can find corresponding scalars $\delta_{\bm\Lambda,\epsilon},\delta_{\bm g,\epsilon},\delta_{\bm\Sigma,\epsilon}>0$ bounding sup-norm neighborhoods for the respective model parameters. As long as 
$\bm\Lambda$ is in the $\delta_{\bm\Gamma,\epsilon}$-sup-norm neighborhood of $\bm\Lambda^0$, $\bm g$ is in the $\delta_{\bm g,\epsilon}$-sup-norm neighborhood of $\bm g^0$ and  $\bm\Sigma$ is in the $\delta_{\bm\Sigma,\epsilon}$-sup-norm neighborhood of $\bm\Sigma^0$
		then $KL(f_{{\bm\Lambda},\bm g,{\bm\Sigma}},f^0)<\epsilon.$
		Therefore $\Pi_f[KL_\epsilon(f^0)]>0$. 
		\end{proof}
\subsection*{Proof of Proposition \ref{prop: consistency}}

Let $U_\epsilon(f^0):=\{f:\int |f-f_0|\text{d}\bm x<\epsilon\}$.
Our proof uses Theorem 2 of \cite{ghosal1999posterior}, which
provides sufficient conditions under which the posterior probability assigned to strong neighborhoods of the true data-generating density converge to one almost surely.
Their result involves conditions on the size of the parameter space in terms of $L_1$ metric entropy. Before proceeding, we review $L_1$ metric entropy and Theorem 2 of \cite{ghosal1999posterior}.

\begin{lemma}\label{lemma: ghosal}
    [Theorem 2 in \cite{ghosal1999posterior}]  Let $\Pi$ be a prior on $\mathcal{F}$. Suppose $f_0 \in \mathcal{F}$ is in the Kullback-Leibler support of $\Pi$ and let $U=\left\{f: \int\left|f-f_0\right| d y<\epsilon\right\}$. If there is a $\delta<\epsilon / 4$, $c_1, c_2>0, \beta<\epsilon^2 / 8$ and $\mathcal{F}_n \subset \mathcal{F}$ such that for all large $n$ :

(1) $\Pi\left(\mathcal{F}_n^c\right)<c_1 \exp \left(-n c_2\right)$, and,

(2) The $L_1$ metric entropy, $J\left(\delta, \mathcal{F}_n\right)<n \beta$, then $\Pi\left(U \mid Y_1, Y_2, \ldots, Y_n\right) \rightarrow 1$ a.s. $P_{f_0}$.
\end{lemma}

The $L_1$ metric entropy $J(\delta, \mathcal G)$ for $\mathcal G\subset \mathcal F$ is defined as the minimum of $$\log \left(k: \mathcal{G} \subset \bigcup_{i=1}^k\left\{f: \int\left|f-f_i\right| d y<\delta, f_1, f_2, \ldots, f_k \in \mathcal{F}\right\}\right).$$ 

\begin{proof}
We verify that our proposed prior framework satisfies the two conditions. 
Let $\mathcal H_n$ denote a subset of the parameter space and adopt the decomposition such that $\mathcal H_n=\mathcal H_{1n}\otimes \mathcal H_{2n}$, where  $\mathcal H_{1n}=\{(\bm\Lambda,\bm g):\|\bm\Lambda\bm g\|_\infty\le M_n\}$ and $\mathcal H_{2n}= [L_n,\infty)^{p}$ where $M_n =O(\sqrt n)$ and $L_n\to 0$. 

Since the prior for each element of $\bm \Lambda$ and each spline slope characterizing $\bm g$ is Gaussian, we obtain $\Pi(\mathcal H_{1n}^c)\le A_1\exp(-B_1M_n^2)$. For $\mathcal H_{2n}$, we can choose $L_n$ small enough such that $\nu(0,L_n)\le A_2\exp(-B_2n)$, where $\nu$ is the inverse-Gamma prior.   

By Theorem 2.7.1 in \cite{wellner2013weak}, if $\mathcal{X}$ is a bounded, convex subset of $\mathbb{R}^d$ with nonempty interior, and $m$ is a 1-Lipschitz mapping from $\mathcal{X}$ to $\mathbb R$, then there exists a constant $C_1$  such that the $L_\infty$ entropy of $m(\mathcal{X})\le C_1\lambda(\mathcal X^1)\delta^{-d}$
for every $\delta>0$, where $\lambda\left(\mathcal{X}^1\right)$ is the Lebesgue measure of the set $\{x: \| x-$ $\mathcal{X} \|<1\}$. In the NIFTY setting, $\mathcal X$ is a $K$-dimensional unit cube. Since $\bm g$ is modeled by linear splines, $(\Lambda g)_j$ is 1-Lipschitz for any $j=1,\ldots,p$. Therefore, we have bounded the $L_1$ metric entropy as $J(\delta, H_{1n})<C_1M_n/(3\delta^K)$.

This implies there are $C^* = [\exp(C_1M_n/(3\delta^K))]$ elements $\bm\mu_1, \bm\mu_2, \ldots, \bm\mu_{C^*}$ such that
 $$\mathcal{H}_{1 n} \subset \bigcup_{i=1}^{C^*}\left\{(\bm\Lambda,\bm g): \max_{j}\int_0^1\left|[\bm\Lambda\bm g(\bm u)]_j -[\bm \mu_i(\bm u)]_j\right| d \bm u<\delta\right\}.$$

We now consider the sieve 
$$\mathcal F_n = \{f_{(\bm\Lambda,\bm g,\bm\Sigma)}:(\bm\Lambda,\bm g)\in\mathcal H_{1n}, \text{diag} (\bm\Sigma)\in \mathcal H_{2n}\}.$$
Clearly $\mathcal F_n \subseteq F$ and $\mathcal F_n \uparrow F$. Observing that $\mathcal H_n\subset\mathcal F_n$ implies 
$\Pi\left(\mathcal{F}_n^c\right)=\left(\Pi^* \otimes \nu\right)\left(\mathcal{H}_n^c\right)=\left(\Pi^* \otimes \nu\right)\left\{\left(\mathcal{H}_{1 n}^c \otimes \mathcal{H}_{2 n}\right) \cup\left(\mathcal{H}_{1 n} \otimes \mathcal{H}_{2 n}^c\right)\right\} \leq c_1 \exp \left(-c_2 n\right), c_1, c_2>0.$
This proves the first condition in Lemma \ref{lemma: ghosal}. 

We now show that $J(\delta, \mathcal F_n)$ can be controlled by $J(\delta, \mathcal H_n)$ times a constant. Given a fixed value of $\bm u$ in the $K$ unit cube, for simplicity of notation, we denote the $j$th entry in $\bm\mu_i(\bm u)$ as $\mu_{ij}$ and denote the $j$th entry in $\bm\Lambda\bm g(\bm u)$ as $\hat\mu_{j}$. It can be shown for $L_n$ small enough, if $\mu_{ij} < \hat\mu_j$, then
$$\begin{aligned}
& \int\left|\phi_\sigma(y-\hat\mu_{j})-\phi_{L_n}\left(y-\mu_{ij}\right)\right| d y \\
& \quad=\left(2 \pi \sigma^2\right)^{-1 / 2} \int_{G_y} \exp \left\{-(y-\hat\mu_{j})^2 /\left(2 \sigma^2\right)\right\}-\left(2 \pi L_n^2\right)^{-1 / 2} \int_{G_y} \exp \left\{-\left(y-\mu_{ij}\right)^2 /\left(2 L_n^2\right)\right\} \\
& \quad+\left(2 \pi L_n^2\right)^{-1 / 2} \int_{G_y^c} \exp \left\{-\left(y-\mu_{ij}\right)^2 /\left(2 L_n^2\right)\right\}-\left(2 \pi \sigma^2\right)^{-1 / 2} \int_{G_y^c} \exp \left\{-(y-\hat\mu_{j})^2 /\left(2 \sigma^2\right)\right\} \\
& \quad \leq 2\left(\hat\mu_{j}-\mu_{ij}\right) /\left(2 \pi \sigma^2\right)^{1 / 2}+2\left(\hat\mu_{j}-\mu_{ij}\right) /\left(2 \pi L_n^2\right)^{1 / 2} \leq 4\left(\hat\mu_{j}-\mu_{ij}\right) /\left(2 \pi L_n^2\right)^{1 / 2},
\end{aligned}
$$
where $G_y = \{y \in\mathbb R : y > (\mu_{ij} + \hat\mu_{ij})/2\}.$ Similarly, we have $\int\left|\phi_\sigma(y-\hat\mu_{j})-\phi_{L_n}\left(y-\mu_{ij}\right)\right| d y \leq 4\left(\hat\mu_{j}-\mu_{ij}\right) /\left(2 \pi L_n^2\right)^{1 / 2}$ when $\mu_{ij} \ge \hat\mu_j$. This implies that the sequence $f_j:=f_{\bm \mu_j,\bm\Sigma}$ is a $\delta$-cover of $\mathcal F_n$ when $\bm\mu_j$ is a $\delta$-cover of $\mathcal H_{1n},$ because whenever $\max_j\int^1_0|\mu_{ij}-\hat\mu_j| d u <\delta,$ 
$$\int|f_{\mu_i}- f_{\bm \Lambda,\bm g,\bm\Sigma}|d\bm y\le \prod_{j=1}^p(4\left(\hat\mu_{j}-\mu_{ij}\right) /\left(2 \pi L_n^2\right)^{1 / 2})\le C_2(\delta/L_n)^p.$$
This suggests that  
$J(\delta,\mathcal F_n) \le C_3M_n(L_n/\delta)^{pK}.$
Since $M_n=O(\sqrt{n})$, we can set $\delta<\epsilon/4$ and choose $\beta<\epsilon^2/8$ such that $J(\delta,\mathcal F_n)<n
\beta$, and the second condition in Lemma \ref{lemma: ghosal} is satisfied. 
\end{proof}

\section{Diffusion Map\label{sec: review-DM}}

In developing nonlinear factor models, a key motivation is accommodating data concentrated close to a lower-dimensional and potentially nonlinear subspace. This motivates pre-training based on diffusion maps \citep{coifman2006diffusion}, a popular nonlinear dimension reduction method. In later sections, we demonstrate the use of t-SNE \citep{van2008visualizing} as an alternative to diffusion maps.

The basic idea of diffusion-map-based pre-training is to automatically learn the intrinsic dimension $K$ and compute a $K$-dimensional coordinate that preserves the local geometry in the original data.
The pre-training can be summarized in the following steps:
\begin{enumerate}
	\item Given the $P$-dimensional data $\bm x_1,\ldots,\bm x_N$, use diffusion maps to learn a $Q$-dimensional representation, denoted by $\bm x^*_i =x^*_{i1},\ldots,x^*_{iQ}$. The diffusion map can be summarized as follows:
 \begin{itemize} 
        \item Define a distance matrix in $\mathbb R^p$ as $$\kappa\left(\bm x, \bm x^{\prime}\right)=\exp \left(-\frac{\left\|\bm x-\bm x^{\prime}\right\|_{\mathbb{R}^D}^2}{\epsilon_{D M}^2}\right),$$ where $\epsilon_{DM}$  is a tuning parameter. In practice, we follow the criteria in \cite{shan2022diffusion} to tune $\epsilon_{DM}$.
        \item Let $W_{i j}=\frac{\kappa\left(\bm x_i, \bm x_j\right)}{d\left(\bm x_i\right) d\left(\bm x_j\right)} \in \mathbb{R}^{N \times N}, 1 \leq i, j, \leq N$, where $d\left(\bm x_i\right)=\sum_{j=1}^N \kappa \left(\bm x_i, \bm x_j\right)$.
        \item Define an $N \times N$ diagonal matrix $D$ as $D_{i i}=\sum_{j=1}^N W_{i j}$, where $i=1, \ldots, N$.
        \item Define the normalized graph Laplacian $L$ as $L=\frac{D^{-1} W-I}{\epsilon_{D M}^2} \in \mathbb{R}^{N \times N}$.
 \item Let $\left(\mu_j, \bm v_j\right)_{j=0}^{N-1}$ denote the eigenpairs of $-L$ with $\mu_0 \leq \mu_1 \leq \cdots, \leq \mu_{N-1}$ and $\bm v_j$'s are unit $N$-vectors. Then $\mu_0=0$ and $\bm v_0$ is a constant vector. Take the first $Q$ eigenvectors corresponding to the $Q$ largest  eigenvalues, $\left(\bm v_1, \cdots, \bm v_{Q}\right)$, as coordinates for the data set $\left\{\bm x_i\right\}_{j=1}^N$ in a $Q$-dimensional space.    
 \end{itemize}
	\item Construct a \emph{local covariance matrix} (proposed in \cite{dunson2021inferring}) at each point $\bm x_i^*$ using the coordinates learned in step 1, which is defined as $$C_{N, \epsilon}\left(\bm x^*_i\right) =\frac{1}{N} \sum_{i'=1}^N\left(\bm x^*_i-\bm x^*_{i'}\right)\left(\bm x^*_i-\bm x^*_{i'}\right)^{\top} 1\left(\left\|\bm x^*_i-\bm x^*_{i'}\right\|_2\le \epsilon  \right)$$.
	\item Define $ \lambda_{N, \epsilon, m}\left(\bm x^*_i\right)$ to be the $m$th largest eigenvalue in $C_{N, \epsilon}\left(\bm x^*_i\right)$. 
	     Then define the mean of the $m$th eigenvalues of the local covariance matrices as $\bar{\lambda}_{\epsilon, m}=\frac{1}{N} \sum_{i=1}^N \lambda_{N, \epsilon, m}\left(\bm x^*_i\right).$ 
	         \item Determine the dimension to be $K$ with respect to a threshold $\delta$:
    $K=\max\{k: \bar{\lambda}_{\epsilon, k+1}/\bar{\lambda}_{\epsilon, k}\ge \delta\}.$ We take $\delta=0.5$ as a default choice. 
\item After determining the dimension $K$, we take the first $K$ dimensions $(x_{i1}^*,\ldots, x_{iK}^*)$ as the anchor data.  We henceforth use $\bm x^* \in \mathbb R^{N\times K}$ to denote the augmented anchor data.
\item Estimate the residual variances: for each $k$, let $u^*_{ik} = r / N$ if $x_{ik}^*$ is ranked the $r$th smallest among $x_{1k}^*, \ldots, x_{Nk}^*$. Fit a piecewise-linear regression model $x_{ik}^* = \alpha_0 + \sum_{l=1}^L \alpha_l(u^*_{ik} - \frac{l-1}{L})1_{\{u^*_{ik}\in[\frac{l-1}{L}, \frac{l}{L})\}}+\epsilon_i, \epsilon_i\sim N(0,\sigma_k^2)$ and denote the fitted value of the least square estimator as $\hat x_{1k}^*, \ldots, \hat x_{Nk}^*$. Set $\hat\sigma_k^2 =\frac{ \sum_{i=1}^N(\hat x_{ik}^* - x_{ik}^*)^2}{N-L-2}$ as the residual variance for the $k$th anchor feature in Assumption \ref{assp: sigma}.  
\end{enumerate}  
    Let $M$ be a $K$-dimensional smooth, closed and connected Riemannian manifold isometrically embedded in $\mathbb R^P$ through $\iota: M \to \mathbb R^P$. \cite{coifman2006diffusion} show that the negative of the graph Laplacian approximates the Laplace-Beltrami operator of $M$ pointwisely. By choosing an appropriate $\epsilon_{DM}$ and bandwidth in the local covariance matrix, the augmented data approximates the discretization of an embedding of $\iota(M)$ into $\mathbb R^P$. We numerically find this algorithm stable and useful for extracting latent features. As shown in Figure \ref{fig: 2-curve-diffusion}, the diffusion-map-based pre-training produces two augmented anchor dimensions that recover the true latent variables generating the data in example 2.  
\end{document}